\documentclass[journal]{new-aiaa} 

\usepackage[utf8]{inputenc}
\usepackage{graphicx}
\usepackage{multirow}
\usepackage{amsmath}
\usepackage{makecell}
\usepackage[version=4]{mhchem}
\usepackage{siunitx}
\usepackage{longtable,tabularx}
\graphicspath{{images/}}
\usepackage{caption}
\usepackage{natbib}
\usepackage{subcaption}
\usepackage{chngpage}
\usepackage{comment}

\title{Studies of Transonic Aircraft Flows and Prediction of Initial Buffet Onset Using Large-Eddy Simulations\footnote{Presented at AIAA AVIATION 2023 Forum, San Diego, CA, 06/12/2023-06/16/2023, Paper AIAA 2023-4338}}

\author{Konrad A. Goc\footnote{Aerodynamics Engineer, Flight Sciences. Member AIAA, konrad.a.goc@boeing.com}}
\affil{The Boeing Company, Seattle, WA 98124}

\author{Rahul Agrawal \footnote{Ph.D. candidate, Department of Mechanical Engineering. Student Member AIAA, rahul29@stanford.edu }}
\affil{Center for Turbulence Research, Stanford University, Stanford, CA 94305}

\author{Sanjeeb T. Bose\footnote{Distinguished Engineer, Cadence Design Systems. Adjunct Professor, Institute for Computational and Mathematical Engineering, Stanford University.  Senior Member AIAA, stbose@stanford.edu}}
\affil{Cadence Design Systems, San Jose, CA 95134 }
\affil{Institute for Computational and Mathematical Engineering, Stanford University, Stanford, CA 94305}

\author{Parviz Moin\footnote{Franklin P. and Caroline M. Johnson Professor, Department of Mechanical Engineering, Stanford University. Fellow AIAA, moin@stanford.edu}}
\affil{Center for Turbulence Research, Stanford University, Stanford, CA 94305}

\begin{document}

\maketitle

\begin{abstract}

This article utilizes the Large-Eddy Simulation (LES) paradigm with a physics-based turbulence modeling approach, including a dynamic subgrid-scale model and an equilibrium wall model, to examine the flow over the NASA transonic Common Research Model (CRM), a flow configuration that has been the focus of several AIAA Drag Prediction Workshops (DPW's). The current work explores sensitivities to laminar-to-turbulent transition, wind tunnel mounting system, grid resolution, and grid topology and makes suggestions for current best practices in the context of large-eddy simulations of transonic aircraft flows. It is found that promoting the flow transition to turbulence via an array of cylindrical trip dots, including the sting mounting system in the simulations, and leveraging stranded boundary layer grids all tend to improve the quality of the LES solutions. Non-monotonic grid convergence in the LES calculations is observed to be strongly sensitive to grid topology with stranded meshes rectifying this issue relative to their hexagonal close-packed (HCP) counterparts. The details of the boundary layer profiles both at the leading edge of the wing and within the shock-induced separation bubble are studied, with thicknesses and integral measures reported, providing details about the boundary layer characteristics to turbulence modelers not typically available from complex aircraft flows. Finally, an assessment of the initial buffet prediction capabilities of LES is made in the context of a simpler NACA 0012 flow, with computational predictions showing reasonable agreement with available experimental data for the angle of attack at initial buffet onset and shock oscillation frequency associated with sustained buffet.

\end{abstract}

\section{Nomenclature}

{\renewcommand\arraystretch{1.0}
\noindent\begin{longtable*}{@{}l @{\quad=\quad} l@{}}
$C_L$ & lift coefficient \\
$C_{L,max}$ & maximum lift coefficient \\
$C_D$ & drag coefficient \\
$C_M$ & moment coefficient \\
$C_p$ & pressure coefficient \\
$\tau_w$ & wall stress \\
$\alpha$ & angle of attack \\
$\rho$ & density \\
$T$ & temperature \\
$u_i$ & velocity components \\
$y^+$ & viscous length scale \\
$U_{\infty}$ & freestream velocity \\
$U_{||}$ & velocity component parallel to the local surface tangent \\
$U_{I}$ & inviscid velocity component parallel to the local surface tangent \\
$Re$ & Reynolds number \\
$M$ & Mach number \\
$\eta$ & semispan fraction \\
$\delta_{99}$ & boundary layer thickness based on 99\% of inviscid velocity profile \\
$H$ & shape factor \\
$\theta$ & momentum thickness \\
$h_{0,max}$ & maximum stagnation enthalpy
\end{longtable*}}


\section{Introduction}

\label{sec:intro}

\lettrine{T}{he} accurate simulation of an aircraft in transonic buffet conditions is a key pacing item towards the vision of Certification and Qualification by Analysis (CQbA) \citep{mauery2021guide}. The vision of CQbA is that as the predictive accuracy of high-fidelity simulations is improved and the methodologies become well-validated, these approaches are to be leveraged in place of, or in addition to, traditional wind tunnel and flight testing as part of the aircraft certification process. Although the transonic NASA CRM case has been a focus in the external aerodynamics community for many years, emerging paradigms such as LES have been used very sparsely to date on this configuration. It has been studied extensively since the early 2000s using Reynolds-Averaged Navier Stokes (RANS) techniques as part of American Institute of Aeronautics and Astronautics (AIAA) Drag Prediction Workshops 1-7 (DPW1-7) \citep{levy2003data,laflin2005data,vassberg2008abridged,vassberg2010summary,levy2013summary,tinoco2018summary,tinoco2023summary}, with no scale-resolving (Lattice Boltzmann Method - LBM, LES, Detached Eddy Simulation - DES, etc.) approaches having been attempted in DPW1-6 on this configuration and limited efforts made in DPW7 as part of the ``Beyond RANS'' test case. Apart from the workshop series, several groups have investigated this flow with scale-resolving methods on their own. The work of Lehmkuhl et al. \cite{lehmkuhl2016large} was among the first of such efforts, which showed promising initial results on a coarse grid at one angle of attack. More recent work by Ghate et al. \cite{ghate2021transonic} studied, among other things, the influence of the Vreman subgrid-scale model coefficient \citep{vreman2004eddy} on the shock location and found limited sensitivity to this choice. This work also concluded that Wall-Modeled LES (WMLES) on structured curvilinear overset grids up to approx. 700 million cells predicted a delayed onset of shock-induced separation (and the associated lift curve breakdown) relative to both RANS simulations of the same configuration and experimental data. It was clear from these early scale-resolving studies that further investigation into the treatment of turbulent transition, grid resolution, and subgrid-scale modeling approach is needed to validate the predictive capabilities of LES in this flow regime. This paper aims to address several of these items and to provide an initial assessment of LES when used for cruise aircraft aerodynamic predictions. This paper is an extension of work previously presented at AIAA Aviation 2023 \citep{goc2023studies} and its conclusions are largely consistent with what was presented there, except the findings about the impact of the symmetry plane boundary condition. At that time, it was reported that a significant sensitivity was observed depending on whether the simulations were run in ``full span'' or ``half span'' (i.e. using a symmetry plane boundary condition on the center plane) mode. This sensitivity was not found to be reproducible, and we no longer advocate that the simulations be run in full span mode as a general practice (the reader is referred to Appendix \ref{sec:aaooendix_span} for more details).

\begin{figure}[!ht]
\centering
    \subfloat[\label{fig:crm_ntf_exp}]{\includegraphics[width=0.795\textwidth]{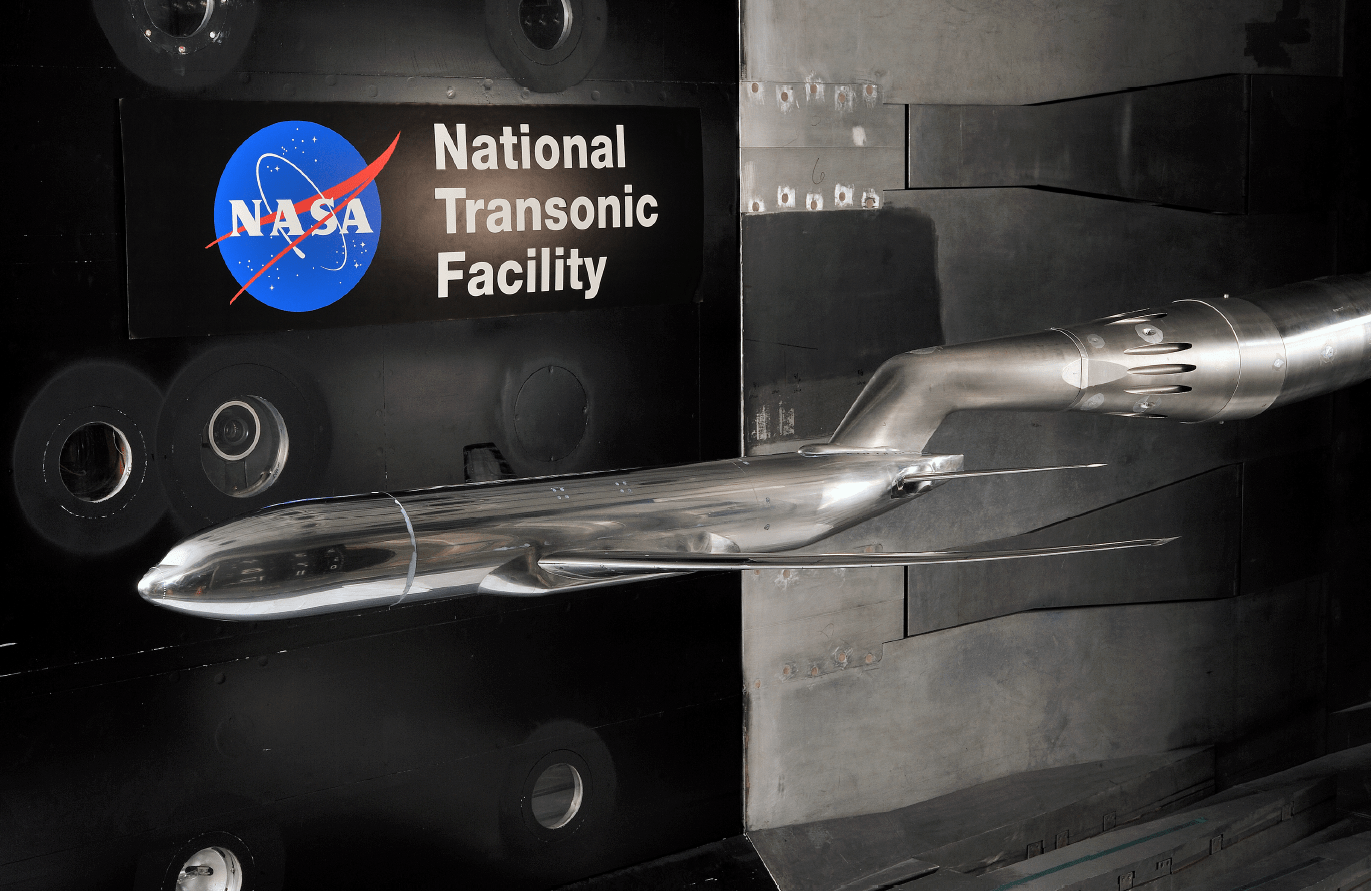}}\\
    \centering
    \subfloat[\label{fig:crm_ntf_cfd}]{\includegraphics[width=0.795\textwidth]{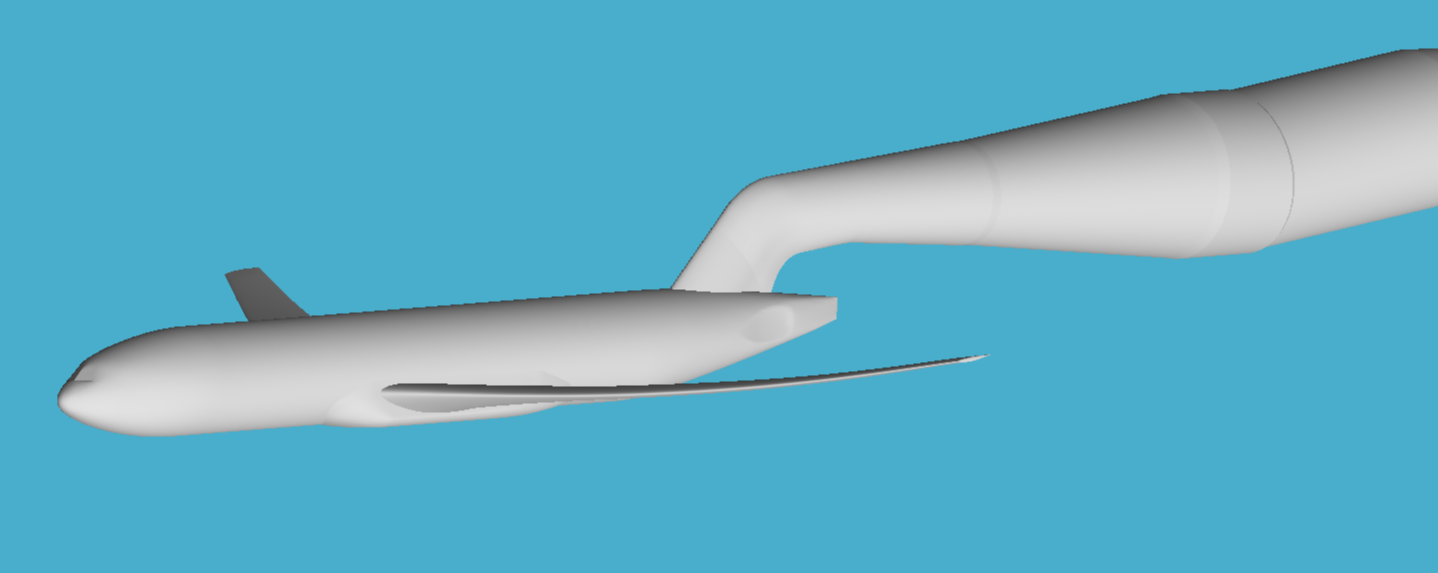}}
    \caption{Image of (a) the experimental apparatus from the NTF wind tunnel runs used as reference data in this article (reproduced from Ref. \cite{rivers2010experimental}) along with (b) the CFD geometry used in the calculations presented in this manuscript (mirrored about the symmetry plane for ease of viewing). The experimental image contains a horizontal stabilizer, but data from a stabilizer-off configuration was used as a reference to match the CFD geometry.  \label{fig:crm_geom}}
\end{figure}

Figure \ref{fig:crm_ntf_exp} shows an image of the CRM experiment performed at the NASA Langley National Transonic Facility (NTF), the data from which are used as a reference for this simulation campaign. The reference conditions for the calculations presented herein are $Re = 5.0 \times 10^6$, M = $0.85$ and $\alpha = 2.50^{\circ}-4.0^{\circ}$ spaced at $0.25^{\circ}$ increments. The corresponding experiments at these flow conditions were performed by Rivers et al. \cite{rivers2010experimental}. In all CRM calculations conducted in this paper\textsuperscript{\footnote{CRM geometries used were from DPW6: https://aiaa-dpw.larc.nasa.gov/Workshop6/DPW6-geom.html}}, the aeroelastic deflection reported by Tinoco et al. \cite{tinoco2018summary} is used for the baseline geometries just as it was in DPW6, resulting in new meshes and wing deformations for each angle of attack simulated, with up to $6.5''$ ($\approx 2.4\%$ of MAC) of wingtip deflection (at full scale) having been observed between the lowest angle of attack ($\alpha = 2.50^{\circ}$) and the highest angle of attack ($\alpha = 4.00^{\circ}$). At full scale, the aircraft mean aerodynamic chord has a length of $275.8''$, while the wingspan is $2313.5''$ ($\approx 8.4 \times$ MAC). The aspect ratio of the wing is $9$ and the leading edge is swept at an angle of $35^{\circ}$. The calculations described in this manuscript are run in a free-air setting because of the insufficient characterization of the NTF wind tunnel facility available in the open research domain at the time of executing the simulations. However, the sting mounting system visible at the tail of the aircraft in Fig. \ref{fig:crm_ntf_exp} is included in the simulations. As discussed in this paper, the inclusion of the sting mounting apparatus resulted in improvements to the shock location predictions on the wing in the simulations and therefore, unless otherwise noted, this piece of the geometry was retained for all simulations that were conducted in this manuscript.

The simulations presented below were performed using an explicit, unstructured, finite-volume solver for the compressible Navier-Stokes equations, charLES, developed by Cadence Design Systems. This code is 2\textsuperscript{nd}-order accurate in space, and 3\textsuperscript{rd}-order accurate in time, and utilizes Voronoi grids. More details about the governing equations and validation cases performed using charLES can be found in Refs.  \cite{bres2018large,goc2021large}. A combination of the dynamic Smagorinsky subgrid-scale model (DSM) \cite{moin1991dynamic} and the equilibrium wall model \cite{cabot2000approximate} were used based on experience in other external aerodynamic flow regimes \citep{agrawal2022non, goc2020sgs}. The solver also approximately preserves entropy in the adiabatic, inviscid limit and employs formally skew-symmetric operators. These choices are aimed at effectuating higher level conservation principles discretely. 

The remainder of this paper is organized as follows. Section \ref{sec:grids} details the grids used for the LES calculations of the flow over the CRM. In section \ref{sec:trip_dots_plate} we describe validation efforts for treatment of laminar-to-turbulent transition in LES calculations. Then, in section \ref{sec:trip_dots_crm} we describe the extension of these learnings to the full CRM aircraft configuration. The influence of the sting mounting system is explored in section \ref{sec:sting}. Sensitivities to the grid resolution and topology are established in section \ref{sec:topology}, including both isotropic (section \ref{sec:isotropic_strand}) and anisotropic (section \ref{sec:anisotropic_strand}) strand meshing approaches. The CRM investigations generally proceed chronologically in the direction of improving accuracy. The transition and sting mounting studies are meant to inform the subsequent investigations into grid topology effects, with the highest accuracy solutions being achieved on the final anisotropic strand meshes. An extension of the LES framework to the prediction of initial buffet onset on a NACA 0012 is made in section \ref{sec:buffet}. Finally, conclusions are drawn in section \ref{sec:summary}. 


\section{Description of Grids Used}
\label{sec:grids}

\begin{figure}[!ht]
    \subfloat[\label{fig:isostrand_schematic}]{\includegraphics[width=0.495\textwidth]{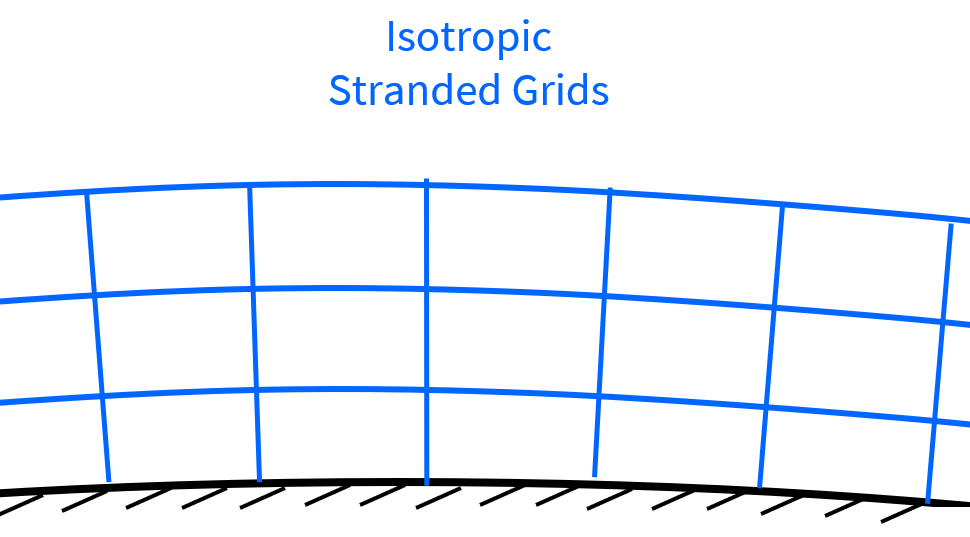}}
    \subfloat[\label{fig:anisostrand_schematic}]{\includegraphics[width=0.495\textwidth]{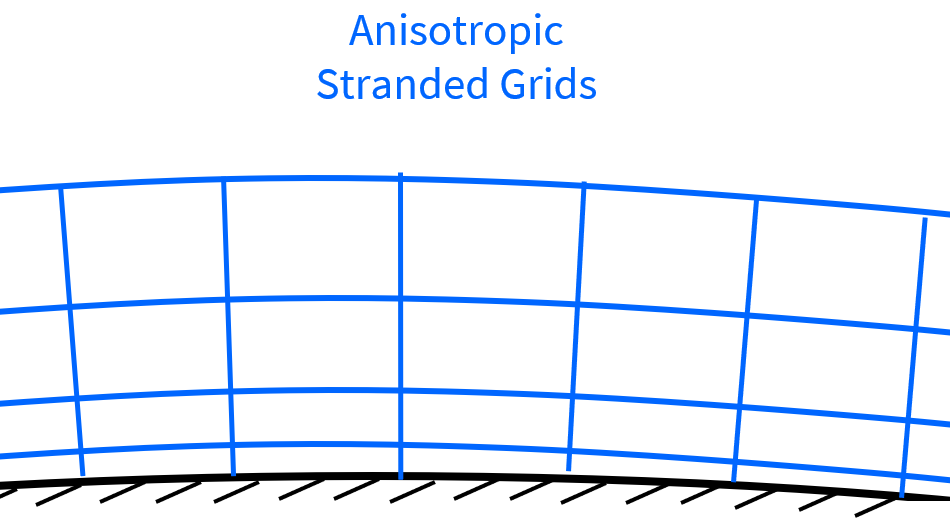}}\\
    \centering
    \subfloat[\label{fig:hcp_schematic}]{\includegraphics[width=0.495\textwidth]{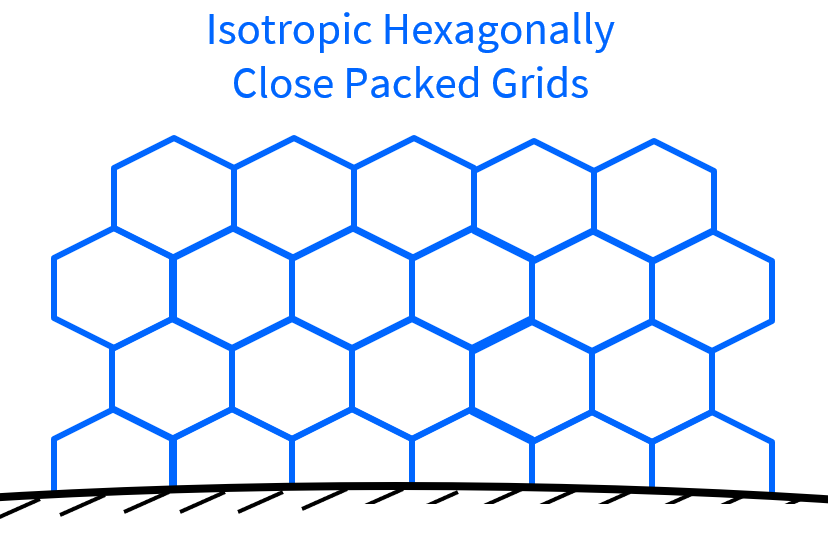}}
    \caption{Schematics of the three types of Voronoi meshes employed in this study: (a) Isotropic Stranded grids, (b) Anisotropic Stranded grids with a maximum 4:1 anisotropy (exaggerated in this schematic view for clarity), and (c) Isotropic Hexagonally Close Packed grids. Further details on the grids used in the CRM calculations are presented in Tables \ref{tab:hcp_meshes}-\ref{tab:aniso_strand_meshes} \label{fig:grid_schematics}}
\end{figure}

This brief section is meant to provide details on the meshes used in the subsequent sections of the manuscript, explicating the geometric features represented on each mesh and the finest near-wall resolution. The various mesh topologies explored are shown in the schematic in Fig. \ref{fig:grid_schematics}. All meshes considered are constructed using a Voronoi meshing approach \citep{fortune2017voronoi}, which offers numerical advantages over more typical (e.g. tetrahedral unstructured) meshes such as more precise control of the local grid spacing, more accurate central reconstruction, and improved numerical stability. Further details on recent investigations that leveraged this meshing approach with LES can be found in \citep{goc2021milestone,bres2018large,lozano2022performance}. Three types of Voronoi meshes were explored in this study: isotropic hexagonal close-packed (HCP) grids, isotropic stranded (locally conforming to the shape of the aerodynamic surface in the near-wall region) grids, and anisotropic stranded grids with near-wall anisotropy aspect ratios of $4$:$1$. It is known \citep{toosi2017anisotropic} that in simple flows, LES can achieve cost savings (while maintaining solution accuracy) from some near-wall anisotropy in the mesh due to the physical anisotropy present in near-wall turbulent structures, but that this anisotropy should not exceed certain thresholds (typically $\mathcal{O}(10)$) that would prohibit adequate grid support for the length scales associated with near-wall turbulence. Based on the findings in the literature and our investigations on a simple transonic NACA 0012 flow (not shown), an anisotropy ratio of $4$:$1$ (wall-normal: stream/span-wise) was selected as a reasonable compromise between the additional accuracy gained from improved wall-normal refinement and the increased cost associated with the time step drop incurred due to the decrease in the wall-normal length scale. Table \ref{tab:hcp_meshes} describes the details of the geometric features of the CRM represented on each of the three hexagonal close-packed (HCP) meshes, along with the associated grid counts and near-wall grid spacing. Tables \ref{tab:iso_strand_meshes} and \ref{tab:aniso_strand_meshes} detail similar descriptions for stranded meshes, isotropic and anisotropic variants respectively. The subsequent sections describe results from a combination of these meshes, as follows: Mesh HCP-F is used as the basis for the studies in Sections \ref{sec:trip_dots_crm} and \ref{sec:sting}, while meshes S-IC through S-IF and S-AXC through S-AF are used for simulations presented in Sections \ref{sec:isotropic_strand} and \ref{sec:anisotropic_strand}. Finally, section \ref{sec:bl_details} utilizes mesh S-AM.

\begin{table}[!ht]
    \caption{Details of the geometric features represented and near-wall resolution level on each HCP mesh and the associated grid count. The meshes are given a label HCP-C (HCP-Coarse) through HCP-XF (HCP-Extra Fine) and are isotropic in all three coordinate directions.}
    \begin{adjustwidth}{-1in}{-1in}
    \begin{center}
    \begin{tabular}{ccccc}
        &
        \textbf{\begin{tabular}[c]{@{}c@{}} 42 Mcv HCP\\Mesh (HCP-C)\end{tabular}} & 
        \textbf{\begin{tabular}[c]{@{}c@{}} 150 Mcv HCP\\Mesh (HCP-M)\end{tabular}} &  
        \textbf{\begin{tabular}[c]{@{}c@{}} 570 Mcv HCP\\Mesh (HCP-F)\end{tabular}} &
        \textbf{\begin{tabular}[c]{@{}c@{}} 2.2 Bcv HCP\\Mesh (HCP-XF)\end{tabular}}
        \\ \cline{2-5} \\
        Trip Dots & \begin{tabular}[c]{@{}c@{}} Included \\ (but sub-grid) \end{tabular} & \begin{tabular}[c]{@{}c@{}} Included \\ (but sub-grid) \end{tabular} & Varies & Included \\ \\
        Sting Mount & Included & Included & Varies & Included \\ \\
        Aeroelastic \\ Deflection & Included & Included & Included & Included \\  \\
        Near-Wall \\ Resolution (mm) &  13.7 &  6.8 & 3.4 & 1.7 \\ \\
        Points Per MAC &  $2^9$ &  $2^{10}$ & $2^{11}$ & $2^{12}$ \\ 
    \end{tabular}
    \label{tab:hcp_meshes}
    \end{center}
    \end{adjustwidth}
\end{table}

\begin{table}[!ht]
    \caption{Details of the geometric features represented and near-wall resolution level on each isotropic stranded mesh and the associated grid count. The meshes are given a label that starts with Strand-Isotropic (S-I...) and goes from S-IC through S-IF (Coarse, Medium, Fine) and are isotropic in all three coordinate directions.}
    \begin{adjustwidth}{-1in}{-1in}
    \begin{center}
    \begin{tabular}{cccc}
        &
        \textbf{\begin{tabular}[c]{@{}c@{}} 1:1 Isotropic \\ 45 Mcv\\ Strand Mesh\\ (S-IC)\end{tabular}} & 
        \textbf{\begin{tabular}[c]{@{}c@{}} 1:1 Isotropic \\ 170 Mcv\\ Strand Mesh\\ (S-IM)\end{tabular}} & 
        \textbf{\begin{tabular}[c]{@{}c@{}} 1:1 Isotropic \\ 670 Mcv\\ Strand Mesh\\ (S-IF)\end{tabular}}
        \\ \cline{2-4} \\
        Trip Dots & Excluded & Excluded & Excluded \\ \\
        Sting Mount & Included & Included & Included \\ \\
        Aeroelastic \\ Deflection & Included & Included & Included \\ \\
        Near-Wall \\ Resolution (mm) &  13.7 & 6.8 & 3.4 \\ \\
        Points Per MAC  &  $2^9$ &  $2^{10}$ &  $2^{11}$ \\ 
    \end{tabular}
    \label{tab:iso_strand_meshes}
\end{center}
\end{adjustwidth}
\end{table}

\begin{table}[!ht]
    \caption{Details of the geometric features represented and near-wall resolution level on each anisotropic stranded mesh and the associated grid count. The meshes are given a label that starts with Strand-Anisotropic (S-A...) and goes from S-AXC through S-AF (Extra Coarse, Coarse, Medium, Fine).}
    \begin{adjustwidth}{-1in}{-1in}
    \begin{center}
    \begin{tabular}{ccccc}
        &
        \textbf{\begin{tabular}[c]{@{}c@{}} 4:1 Anisotropic \\ 32.5 Mcv\\ Strand Mesh\\ (S-AXC)\end{tabular}} &  
        \textbf{\begin{tabular}[c]{@{}c@{}} 4:1 Anisotropic \\ 85 Mcv\\ Strand Mesh\\ (S-AC)\end{tabular}} &  
        \textbf{\begin{tabular}[c]{@{}c@{}} 4:1 Anisotropic \\ 310 Mcv\\ Strand Mesh\\ (S-AM)\end{tabular}} &
        \textbf{\begin{tabular}[c]{@{}c@{}} 4:1 Anisotropic \\ 1.1 Bcv\\ Strand Mesh\\ (S-AF)\end{tabular}}
        \\ \cline{2-5} \\
        Trip Dots  & Excluded & Excluded & Excluded & Excluded \\ \\
        Sting Mount & Included & Included & Included & Included \\ \\
        Aeroelastic \\ Deflection & Included & Included & Included & Included \\ \\
        Wall-Normal \\ Resolution (mm) & 6.8 & 3.4 & 1.7 & 0.85 \\ \\
        Stream/Span-wise \\ Resolution (mm) & 27.4 & 13.7 & 6.8 & 3.4 \\ \\
        Points Per MAC \\ Wall-Normal/ \\ Stream and Span-wise & $2^{8}$/$2^{10}$ & $2^{9}$/$2^{11}$ & $2^{10}$/$2^{12}$ & $2^{11}$/$2^{13}$ \\ 
    \end{tabular}
    \label{tab:aniso_strand_meshes}
\end{center}
\end{adjustwidth}
\end{table}

\section{Treatment of Turbulent Transition}

\subsection{Trip Dot Validation on a Flat Plate}
\label{sec:trip_dots_plate}

Unlike the case of true flight conditions where the flow typically transitions at or very near to the leading edge of the aircraft wing, the Reynolds numbers achieved as part of the reference NTF data for the CRM do not lead to a robust transition at the leading edge. As a consequence, leading-edge trip dots were installed on the aircraft to ensure transition by 10\% of the chord length \citep{rivers2010experimental}. In the context of LES of complex engineering flows, the role of these trip dots in transitioning the flow toward turbulence is not exactly understood. As a building block, a canonical flow setup was chosen to study this phenomenon. Namely, a flow involving a Blasius boundary layer encountering a single trip dot on a flat plate was chosen. This case was studied experimentally in Ref. \cite{ichimiya1993structure}. The reported wake-spreading angle (measured as the angle between the symmetry plane and the edge of the wake) behind the trip dot varied from $4.8^{\circ}-6.0^{\circ}$ depending on the height above the wall where the wake was measured at a roughness Reynolds number of $Re_k \sim 1000$ (based on the trip dot height). This Reynolds number is representative of the roughness Reynolds number encountered in wind tunnel conditions for the CRM \citep{rivers2010experimental, evans2020test} when $Re_k$ is computed based on the reported trip dot heights and freestream flow conditions.

\begin{figure}[!ht]
    \begin{center}
        \subfloat[\label{fig:single_trip_dot_grid_a}]{\includegraphics[width=0.33\textwidth]{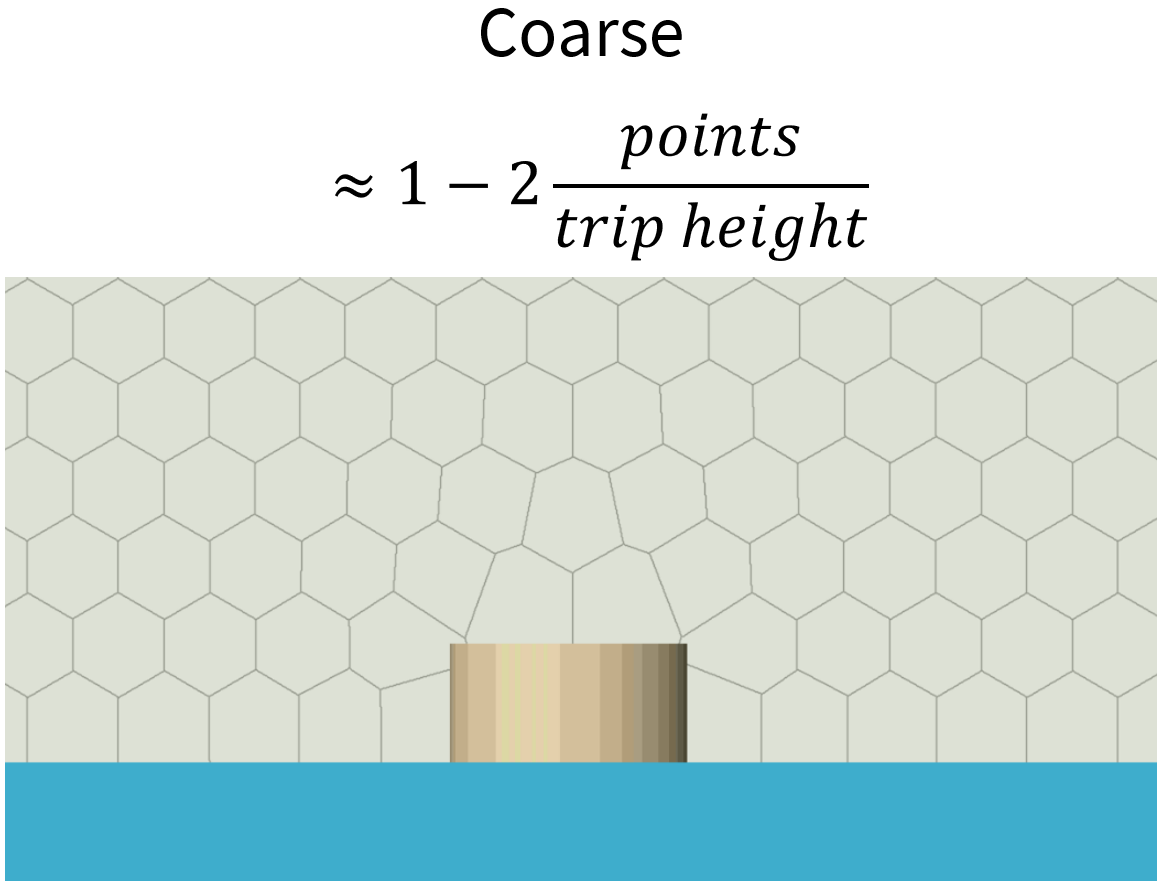}}
        \subfloat[\label{fig:single_trip_dot_grid_b}]{\includegraphics[width=0.33\textwidth]{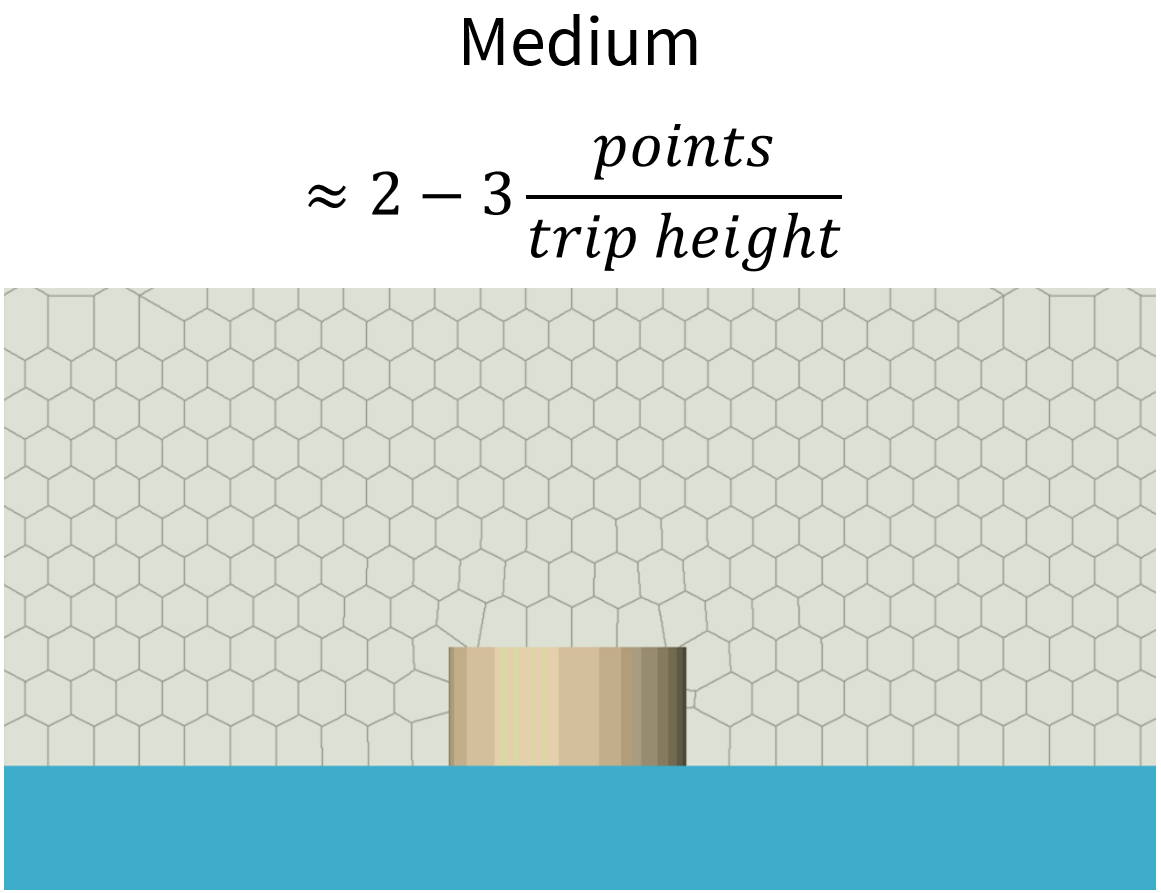}}
        \subfloat[\label{fig:single_trip_dot_grid_b}]{\includegraphics[width=0.33\textwidth]{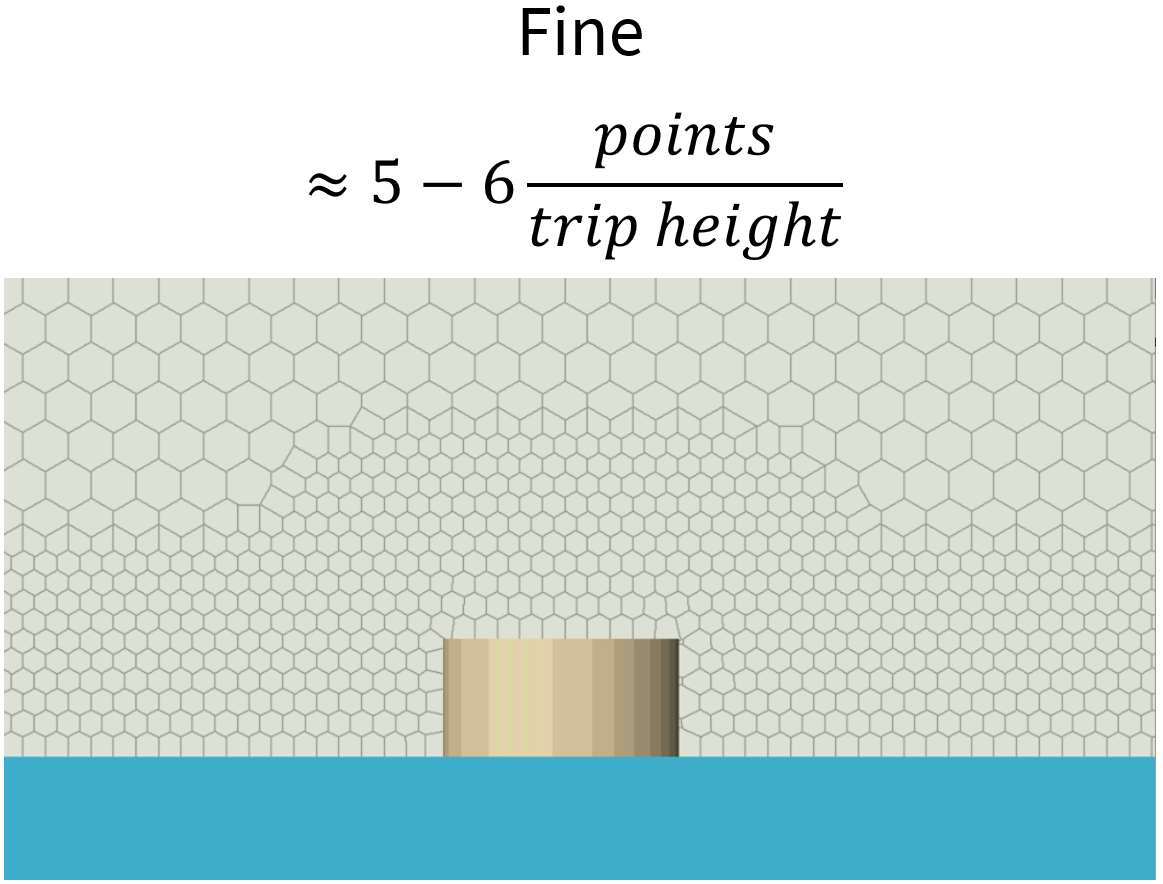}}
    \end{center}
     
    \caption{Spanwise slices of the three grids considered for the single trip dot calculations, ranging from (a) a coarse grid with just 1-2 points per trip dot height to (b) a medium grid with 2-3 points per trip height and (c) a fine grid with 5-6 points per trip dot height.  \label{fig:trip_dot_grids}}
\end{figure}

\begin{figure}[!ht]
\begin{center}
\includegraphics[width=0.85\textwidth]{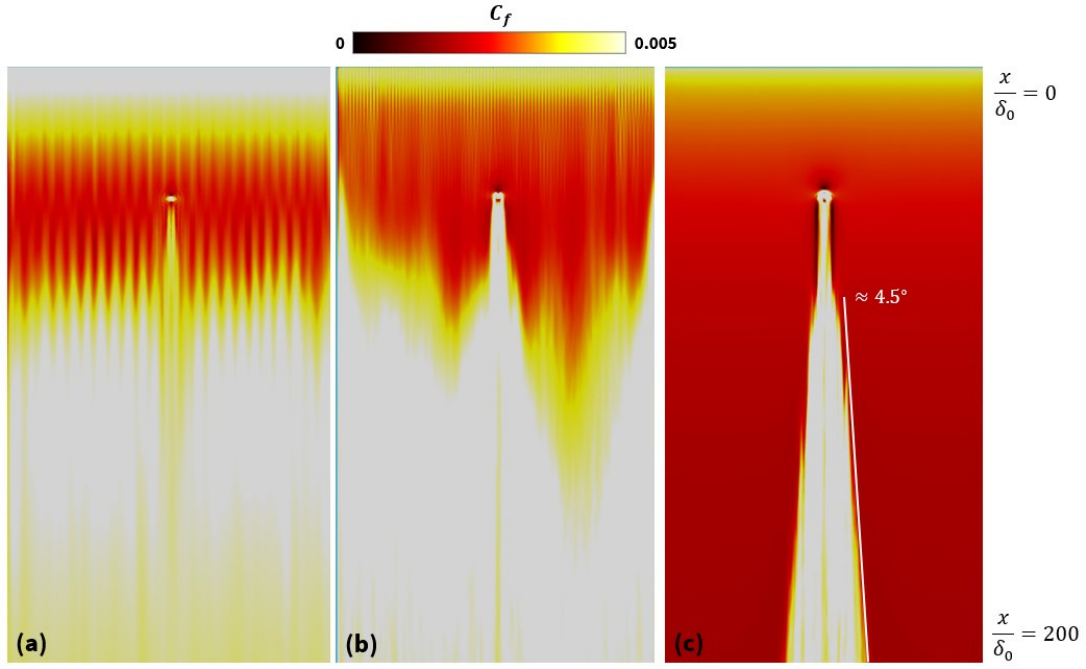}
\caption{Surface contours of the average skin friction for the single trip dot flat-plate calculations ranging from (a) a coarse grid with 1 point per trip dot height, (b) a medium grid with 2 points per trip dot height, and (c) a fine grid with 4 points per trip dot height. The flow is from top to bottom and extends approximately 200 inlet boundary-layer thicknesses in this visualization. \label{fig:cf_wake_spreading}}
\end{center}
\end{figure}

To faithfully replicate the laminar experimental inflow, a Blasius boundary-layer solution was provided at the inlet of the domain to an LES simulation. The inlet boundary layer thickness was tuned such that the extrapolated boundary layer grew to match the reported experimental boundary layer thickness ($\delta_{99} \sim 2 cm$) at the location of the trip dot. Since the DSM was employed for this study, it is expected that the model coefficient turns off in the laminar regions of the flow and no special turbulence modeling treatment was therefore employed to account for the laminar region of the flow, consistent with previous investigations of transitional boundary layer flows \cite{sayadi2012large}. A no-slip closure was applied on the wall in all cases. Three grid resolutions, as shown in Fig.\ref{fig:trip_dot_grids}, were examined. Note that the grids are uniformly refined by a factor of two in all three coordinate directions to up to $\approx 5-6$ points per trip dot height, using an HCP Voronoi meshing approach. The medium resolution on the flat plate was considered to be the upper limit of feasibility in accordance with the resulting grid counts that would be required to simulate the full aircraft geometry if that resolution per trip dot height were to be employed to resolve the trip dots on the CRM (which was the case on mesh HCP-XF). The fine grid resolution of the flat plate trip dot with 5-6 points per trip dot height would approximately scale to $\approx 8.5$ billion control volumes on a full aircraft simulation, near to or beyond the current limit of computational feasibility for realistic industrial configurations \cite{agrawalreynolds}. Studies in which only the immediate vicinity of the trip dots was refined were not pursued at this time, due to the additional validation effort required to understand the required de-refinement strategy in the wake of the trip dots that would lead to a reasonable compromise between solution accuracy and affordability. A ``blind'' gridding approach in which the entirety of the trip dot upstream/downstream and immediate vicinity regions were kept at the same resolution was considered instead. For this reason, even though finer resolutions were achievable in the context of this simple flow, they were not pursued in this effort because they would be impractical on full aircraft simulations. The wake-spreading angles for the three considered grids are shown in Fig. \ref{fig:cf_wake_spreading} as measured by the contours of surface skin friction coefficient, since this quantity is readily available in LES simulations. The region of elevated skin friction magnitude downstream of the trip dot is identified as the turbulent wake. It is clear that the coarsest grid fails to capture any meaningful impact associated with the trip dot, as patterned bands of transition appear at regularly spaced intervals that approximately match the grid spacing, meaning that the transition is almost entirely grid-induced in this case. On the medium mesh with 2-3 points per trip dot height, some impact of the trip dot is visible, with the transition occurring earlier than in the untripped regions of the flow. Also of note is the interaction of the grid with the periodic boundary plane, which injects disturbances from the spanwise boundaries that contaminate the rest of the flow field, making this solution of limited use at the spanwise boundaries of the domain. Finally, with 5-6 points per trip dot height, a clean wake with a spreading angle of $\approx 4.5^{\circ}$ is achieved (as measured using the contours of skin friction on the surface), which is in reasonable agreement with the experimental results \citep{ichimiya1993structure}, especially when considering the measurements nearest to the wall reported in that paper showed a spreading angle of $\approx 4.8^{\circ}$. Based on these findings, a grid resolution of 5-6 grid points has been identified as an entry point into trip dot wake-resolving simulations, while with 2-3 points, some disturbance in the skin friction due to the presence of the trip dots can be captured.

\subsection{Impact of Laminar-to-Turbulent Transition on the Transonic CRM}
\label{sec:trip_dots_crm}

\begin{figure}[!ht]
    \subfloat[\label{fig:crm_trip_dot_grid_a}]{\includegraphics[width=0.495\textwidth]{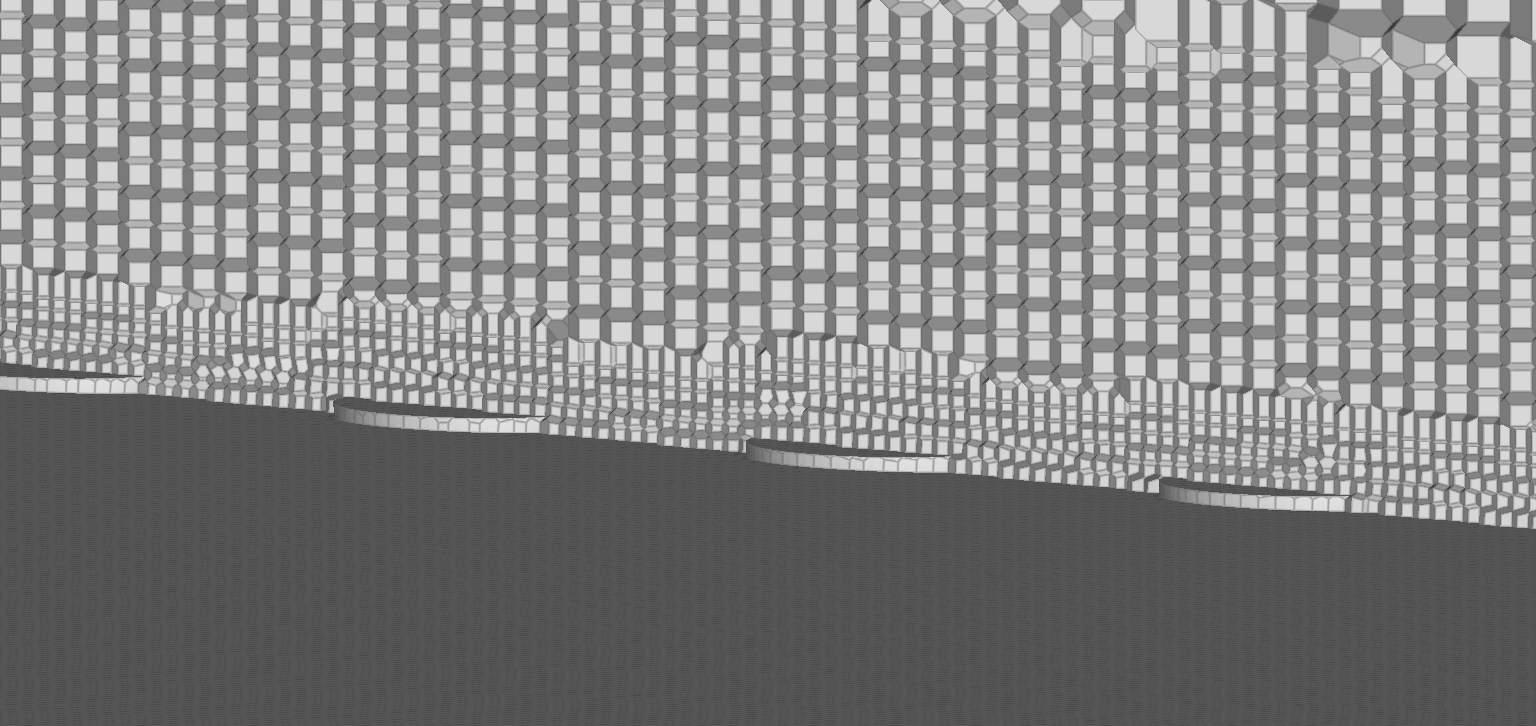}}
    \subfloat[\label{fig:crm_trip_dot_grid_b}]{\includegraphics[width=0.495\textwidth]{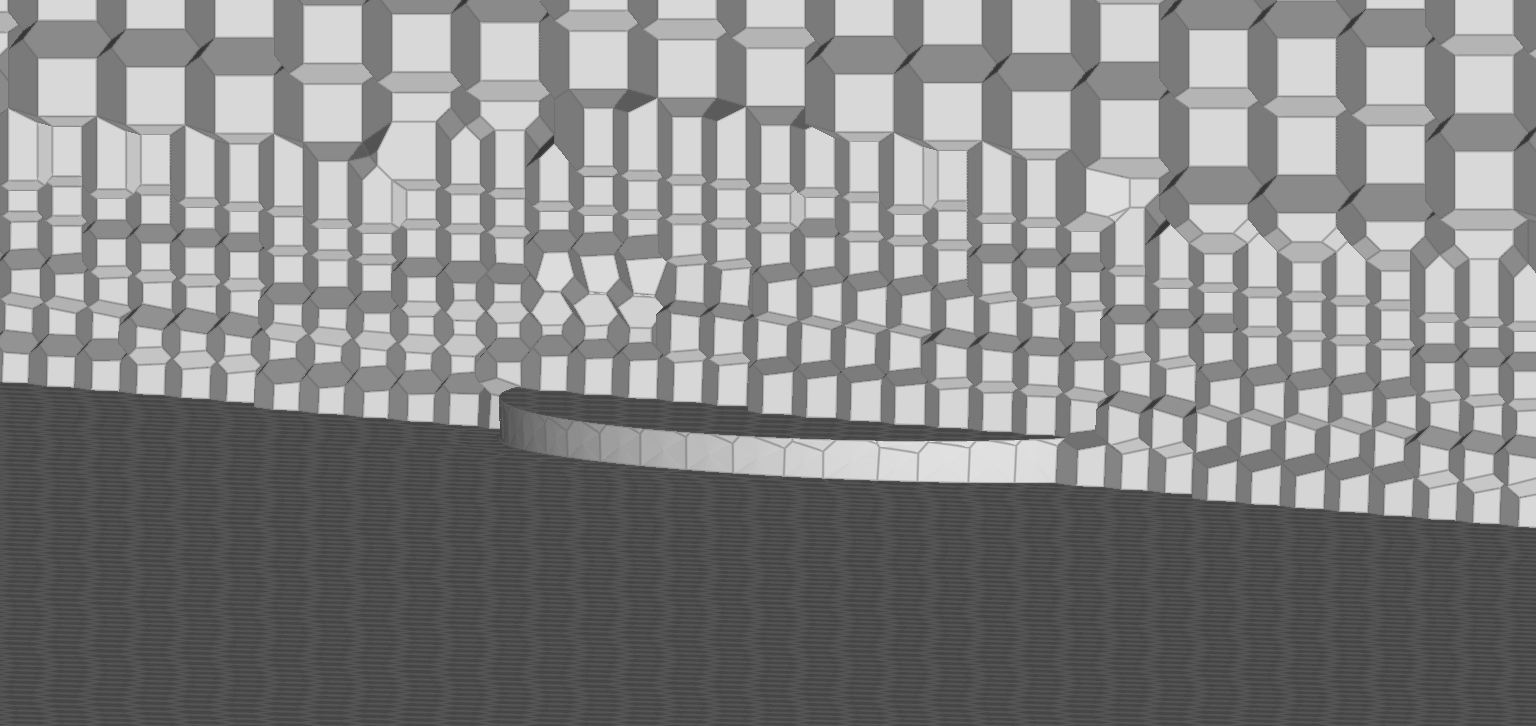}}
     
    \caption{Crinkle-cut view from mesh HCP-F of the grid in the vicinity of the geometric trip dots, including (a) a view of 4 trip dots and (b) a zoomed-in view of a single trip dot. Only 1 point per trip dot height is achieved on Mesh HCP-F. A finer calculation on mesh HCP-XF had a resolution of 2 points per trip dot. \label{fig:crm_trip_dot_grid}}
\end{figure}

\begin{figure}[!ht]
    \subfloat[\label{fig:yehudi_hcpL11_untripped}]{\includegraphics[width=0.495\textwidth]{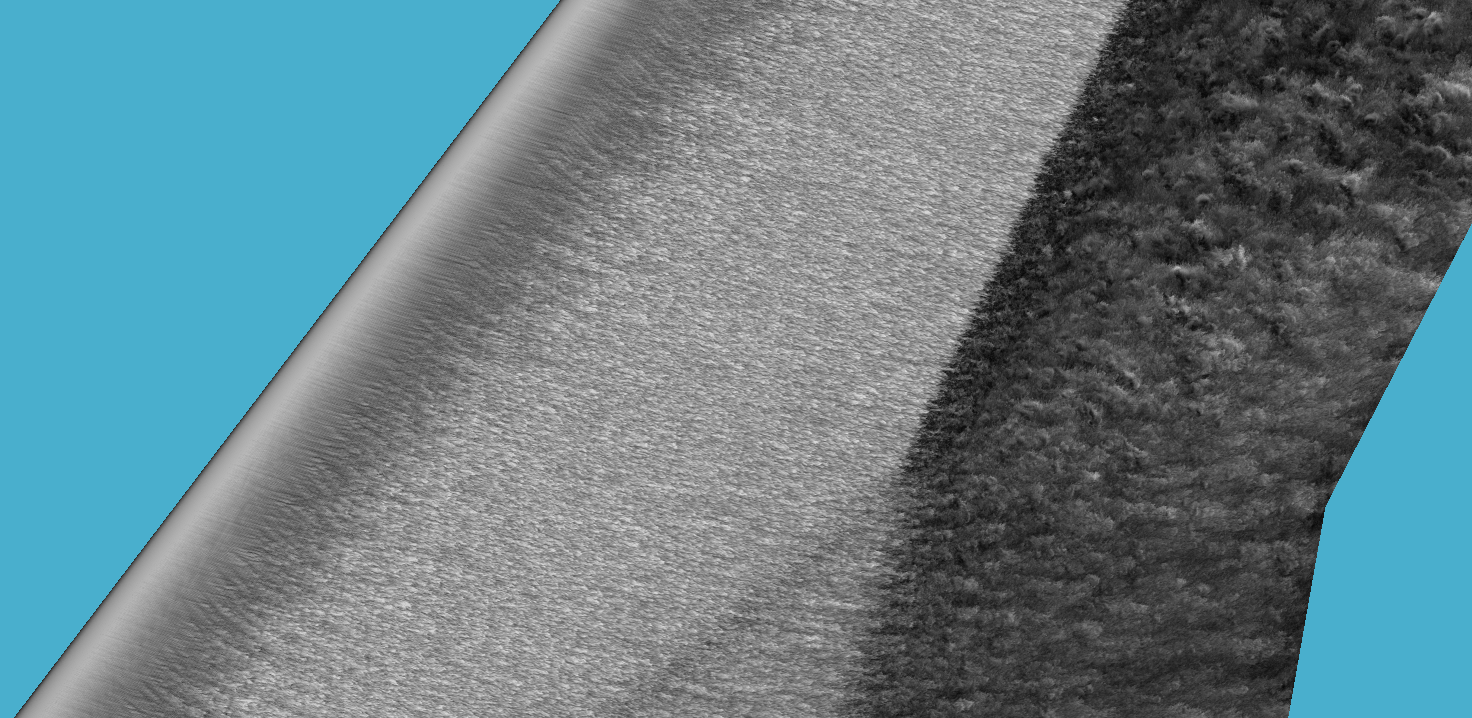}}
    \subfloat[\label{fig:yehudi_hcpL11_tripdot}]{\includegraphics[width=0.495\textwidth]{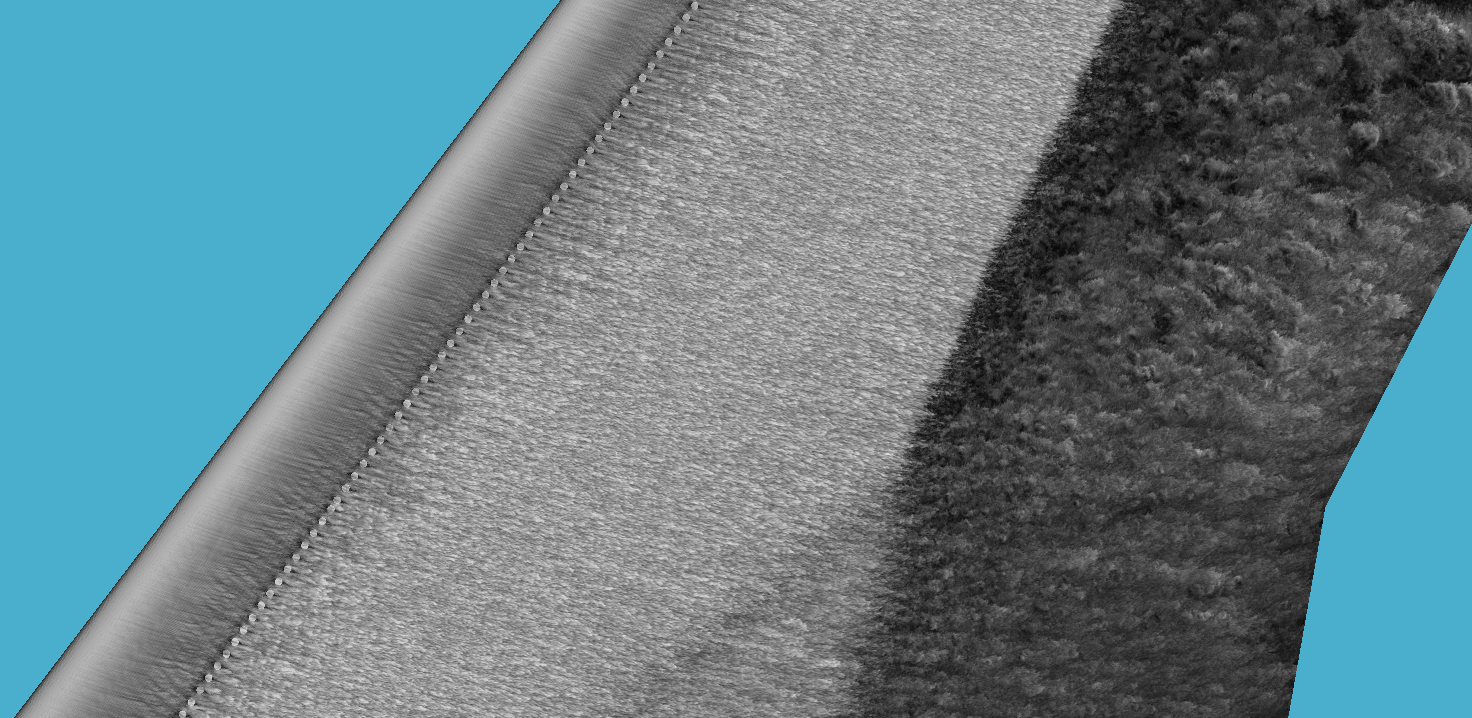}}\\ 
    \centering
    \subfloat[\label{fig:yehudi_hcpL11_num_trip01}]{\includegraphics[width=0.495\textwidth]{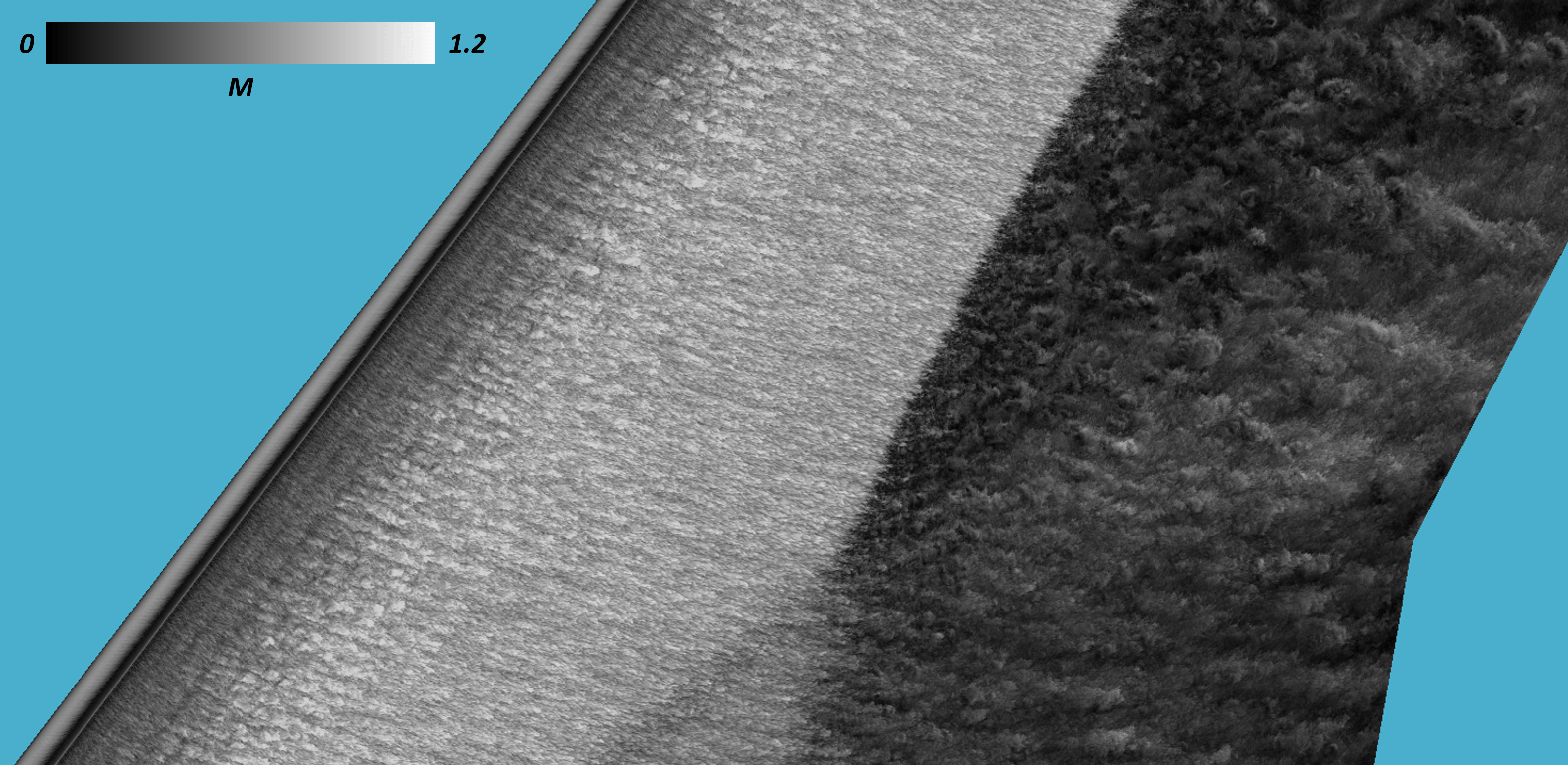}}\
    
    \subfloat[\label{fig:yehudi_hcpL11_untripped_zoom}]{\includegraphics[width=0.495\textwidth]{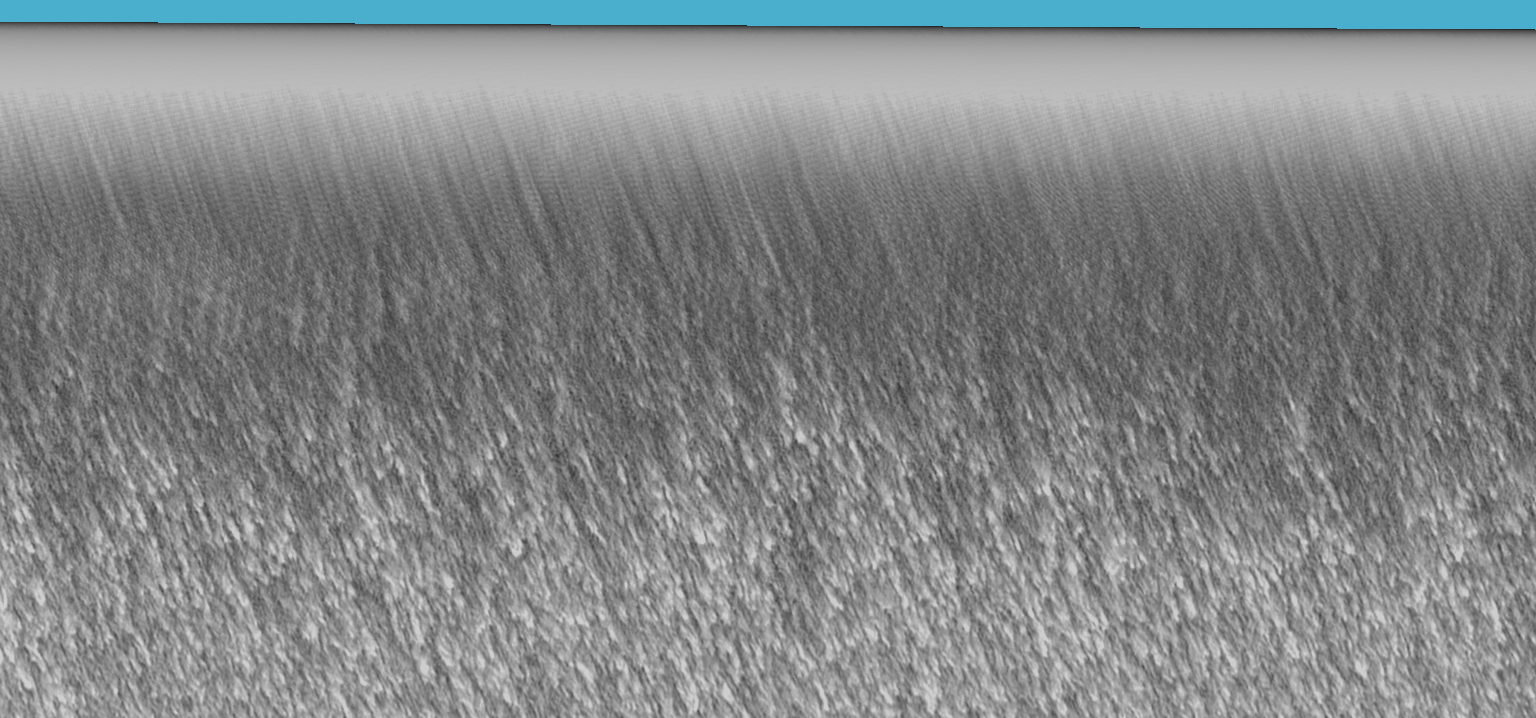}}
    \subfloat[\label{fig:yehudi_hcpL11_tripdot_zoom}]{\includegraphics[width=0.495\textwidth]{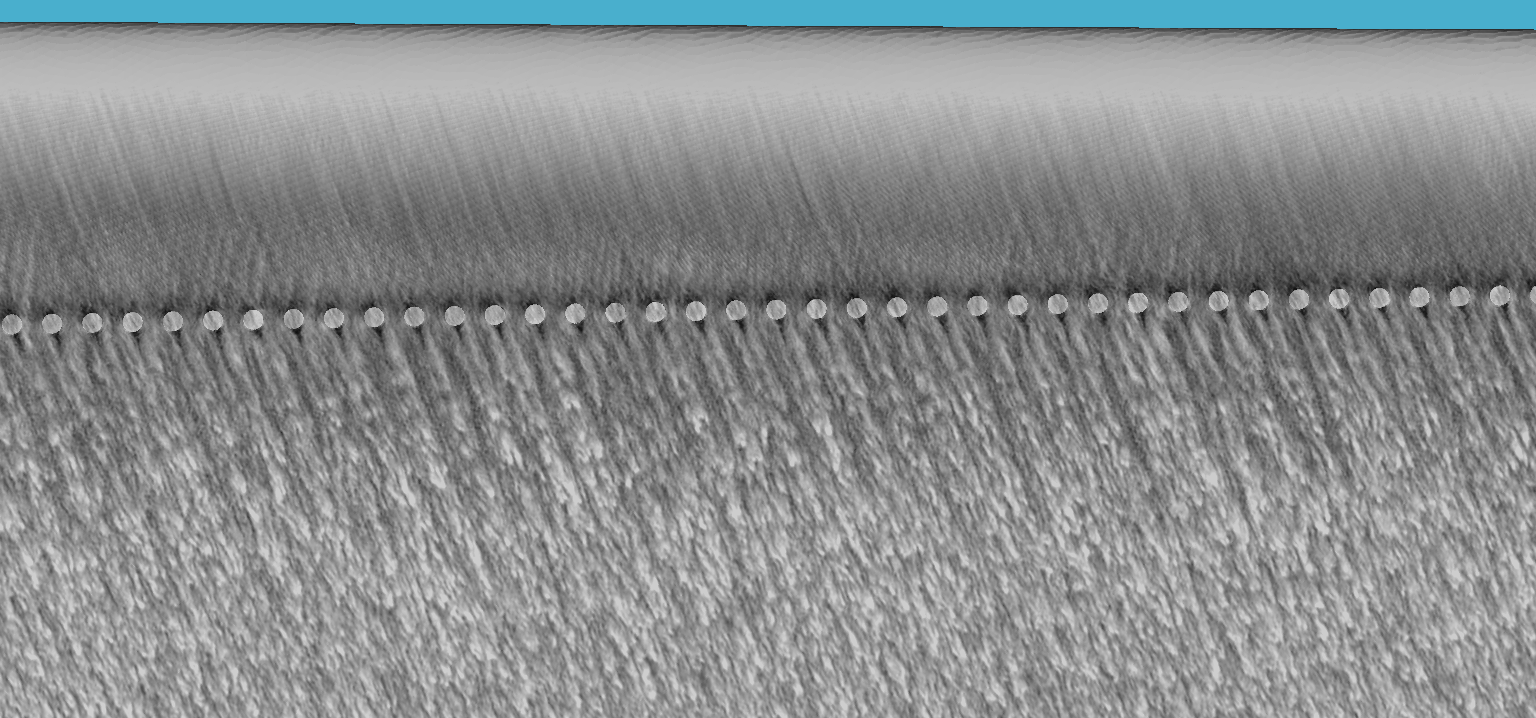}}\\ 
    \centering
    \subfloat[\label{fig:yehudi_hcpL11_num_trip01_zoom}]{\includegraphics[width=0.495\textwidth]{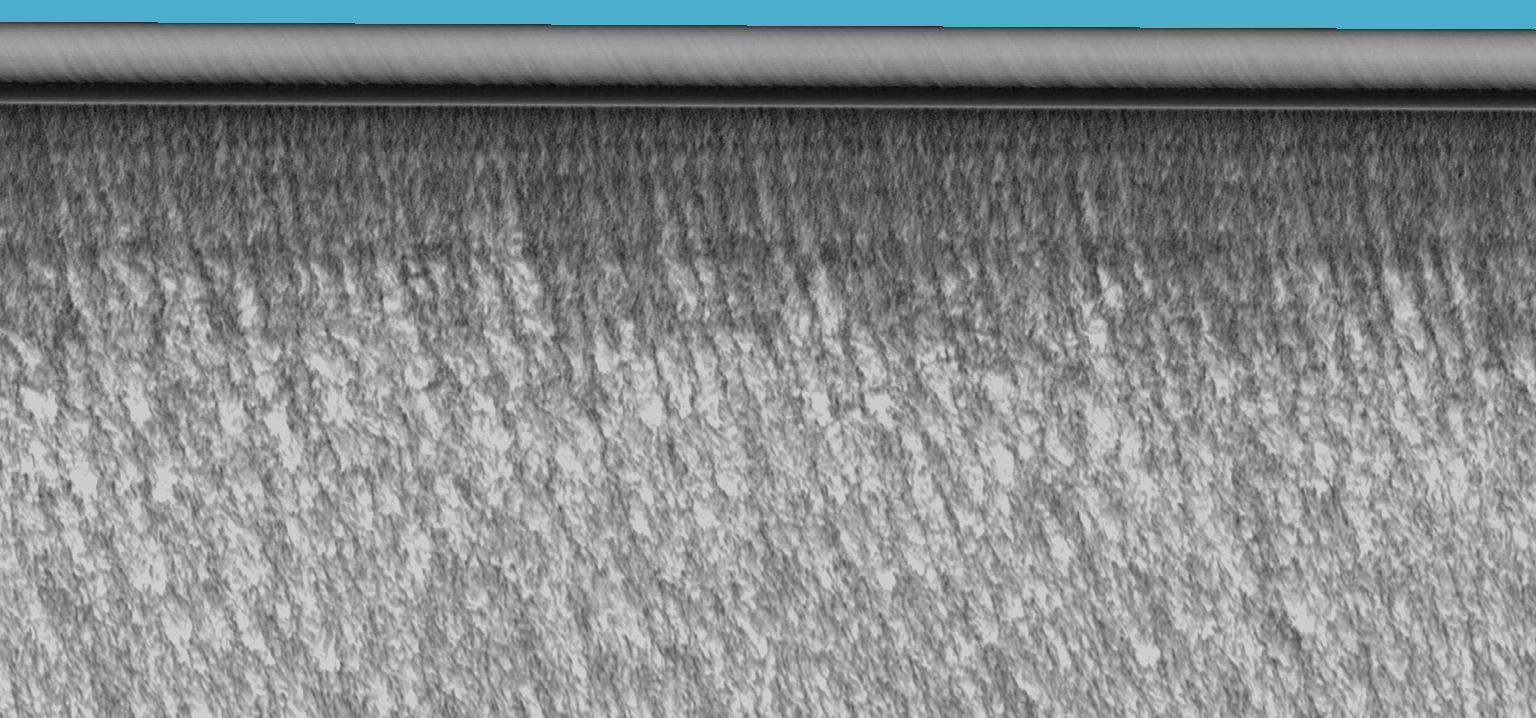}}
    
    \caption{Instantaneous velocity magnitude projection on the suction side of the transonic CRM wing near the Yehudi break showing (a) an untripped  (b) a tripped case where trip dots have been placed along the entire span of the wing at a constant spanline of $10 \%$ and (c) a numerically tripped calculation, all on the HCP-F mesh. Corresponding zoomed in views are shown in subfigures (d), (e), and (f). Flow is from left to right in (a)-(c) and from top to bottom in (d)-(f). \label{fig:crm_inst_trip_dots_yehudi}}
\end{figure}

\begin{figure}[!ht]
    \subfloat[\label{fig:wingtip_hcpL11_untripped}]{\includegraphics[width=0.495\textwidth]{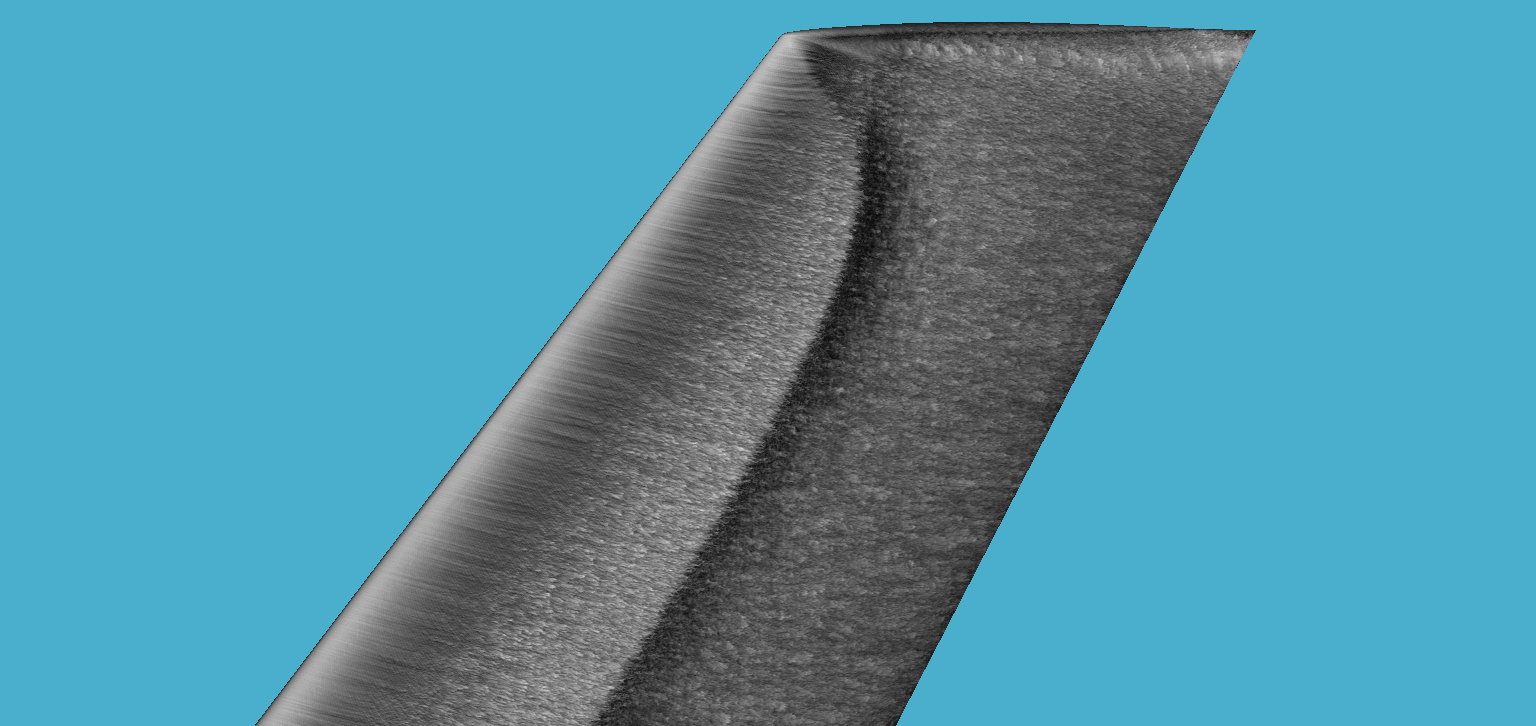}}
    \subfloat[\label{fig:wingtip_hcpL11_tripdot}]{\includegraphics[width=0.495\textwidth]{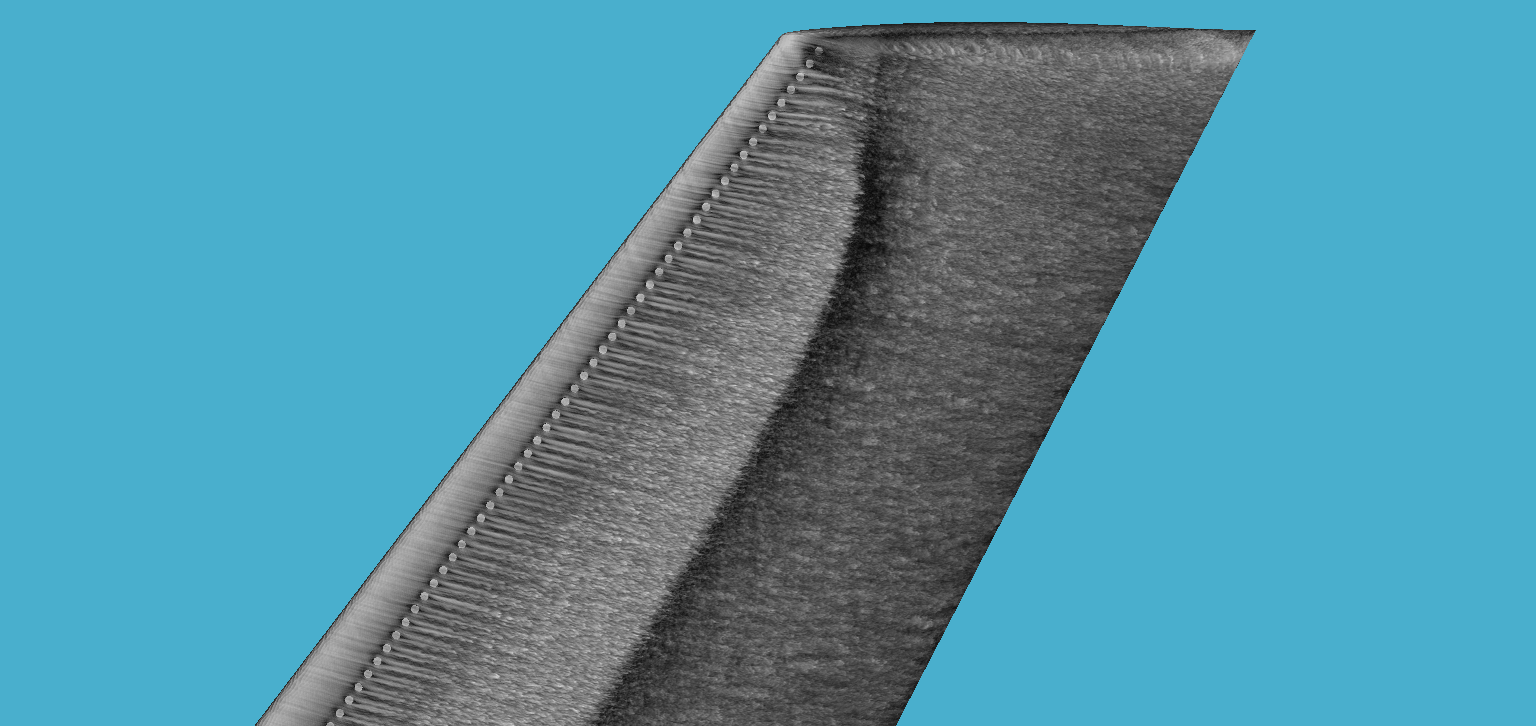}}\\ 
    \centering
    \subfloat[\label{fig:wingtip_hcpL11_num_trip01}]{\includegraphics[width=0.495\textwidth]{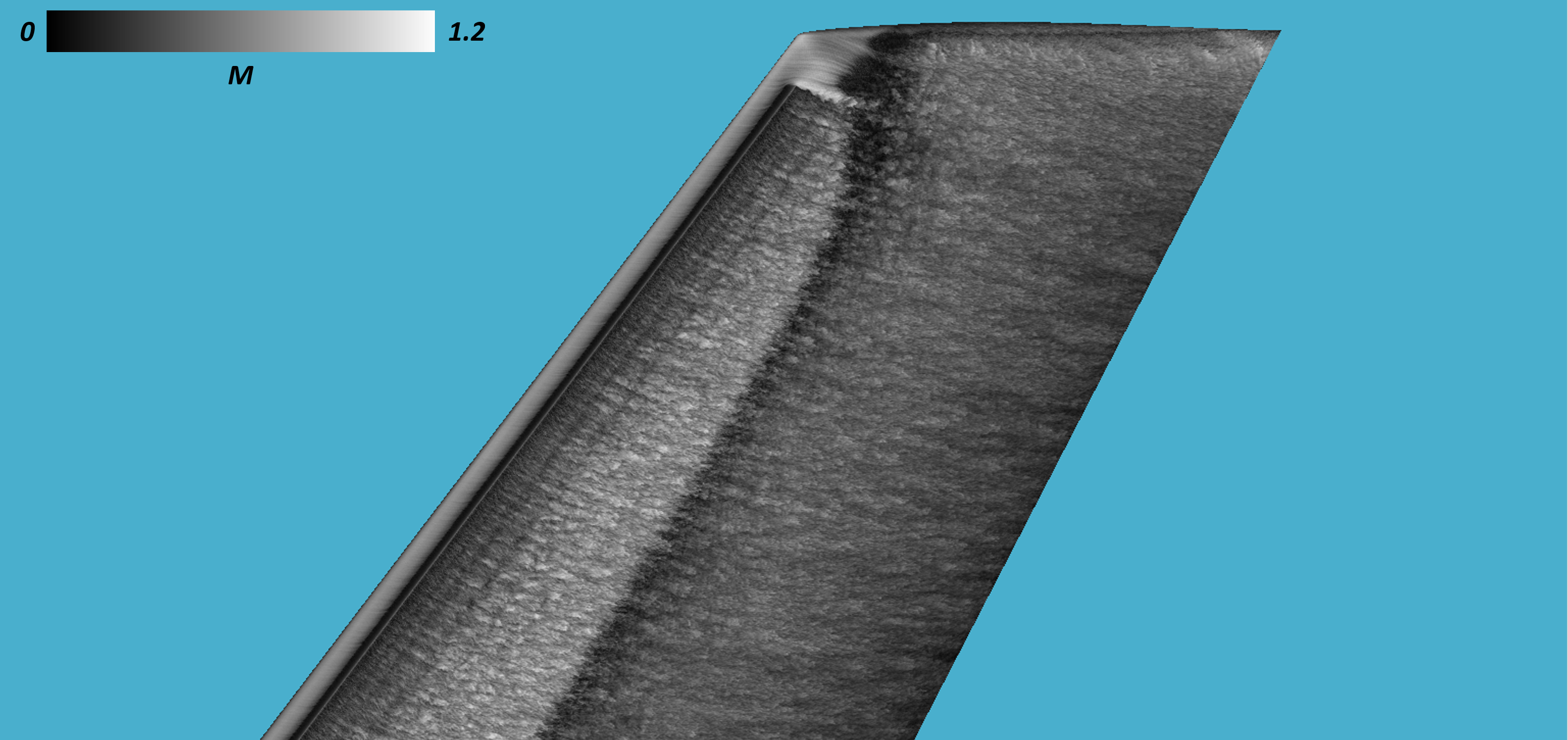}}
     
    \caption{Instantaneous velocity magnitude projection near the wingtip of the transonic CRM showing (a) an untripped, (b) a tripped case where trip dots have been placed along the entire span of the wing at a constant spanline of $10 \%$, and (c) a numerically tripped calculation (tripping at a spanline of $1 \%$), all on the HCP-F mesh. Flow is from left to right. \label{fig:crm_inst_trip_dots_wingtip}}
\end{figure}

Next, a full array of trip dots, matching those used in the experiments \citep{rivers2010experimental,evans2020test}, were incorporated on the transonic CRM aircraft. The trip dot heights reported by the experiments were applied to the CFD geometry, no artificial inflation of the geometric trip dots was employed to make them more easy to resolve on the computational grid. The experiments used cylindrical trip dots placed at $10\%$ of the local chord on the suction side of the wing along the entirety of the span as well as near the fuselage nose. These dots were made of a vinyl adhesive material and measured $0.003'' - 0.0035''$ at model scale ($\approx 0.04\% - 0.045\%$ of a mean aerodynamic chord length). The diameter of these dots was $0.05''$, with a spacing of $0.1''$, resulting in a total of $\approx 315$ trip dots along the span of the wing. Note that resolving the trip dots with 5-6 points (as we concluded to be the ``best practice'' in the single trip dot case) was unachievable due to computational resource constraints (both due to the temporal stability requirement associated with the tiny cells in the trip dot region and the concomitant grid count), and as a consequence, a resolution of only up to 2 points across the trip height was used in the calculations which included the geometric trip dots on the CRM (more in line with the single trip dot ``medium'' mesh simulations). Mesh HCP-XF achieved this resolution of 2 points per trip dot, while mesh HCP-F had a resolution of only 1 point per trip dot and was used as the ``workhorse'' mesh for sensitivity studies (e.g. influence of the sting mounting system). We therefore make no claim that the simulations presented herein are truly trip dot-resolving, but rather that the role of the trip dots is to simply introduce some degree of numerical noise into the leading edge boundary layer, promoting laimar-to-turbulent transition. An alternative approach is to achieve this effect entirely via a numerical blowing/suction-type boundary condition, which will also be explored in this study. Figure \ref{fig:crm_trip_dot_grid} shows crinkle-cut views of the grid from mesh HCP-F in the vicinity of several of the trip dots. The instantaneous visualizations in Fig. \ref{fig:crm_inst_trip_dots_yehudi} and \ref{fig:crm_inst_trip_dots_wingtip} provide evidence that at this resolution level, at least some effect of the trip dots on the flow is captured as there is a clear disturbance in the boundary layer as the flow passes over them. The location at which the projected first cell Mach number (which is the subject of the instantaneous contour plots) abruptly changes from light to dark is the shock location. At this point, the strong adverse pressure gradient of the shock strongly decelerates the flow, and the turbulent structures grow rapidly in size. A clear wake spreading angle is not visible (such as was the case in Section \ref{sec:trip_dots_plate}), but it is unclear whether this is due to the marginal grid resolution or a constraining effect associated with the array of trip dots suppressing this mechanism which was visible in the single trip dot case. We acknowledge this deficiency yet proceed with the subsequent assessment, which can be interpreted as a sensitivity study to the inclusion of trip dots at presently achievable grid resolution levels.

\begin{figure}[!ht]
    \subfloat[\label{fig:crm_cp_trip_dots_a}]{\includegraphics[width=0.495\textwidth]{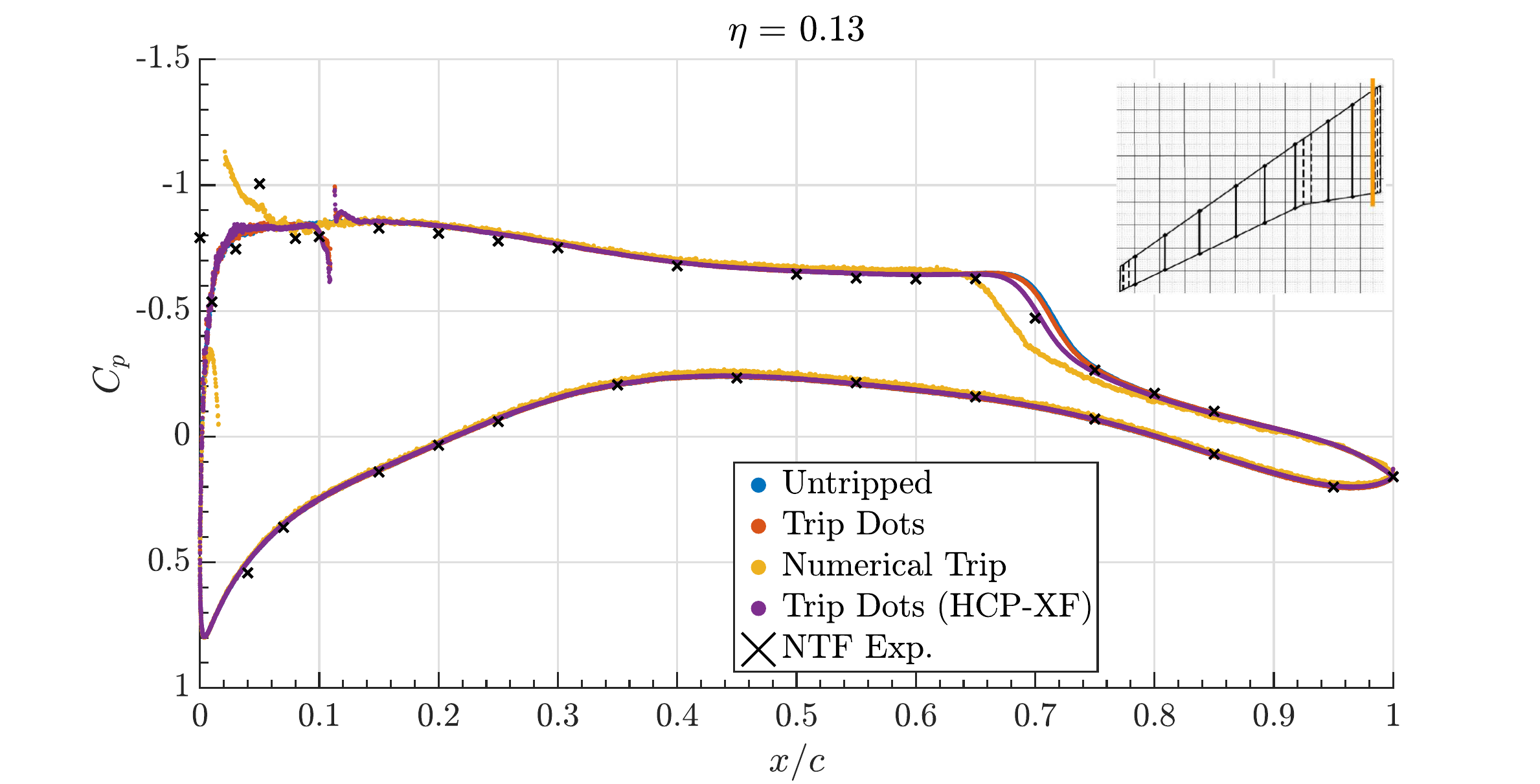}}
    \subfloat[\label{fig:crm_cp_trip_dots_b}]{\includegraphics[width=0.495\textwidth]{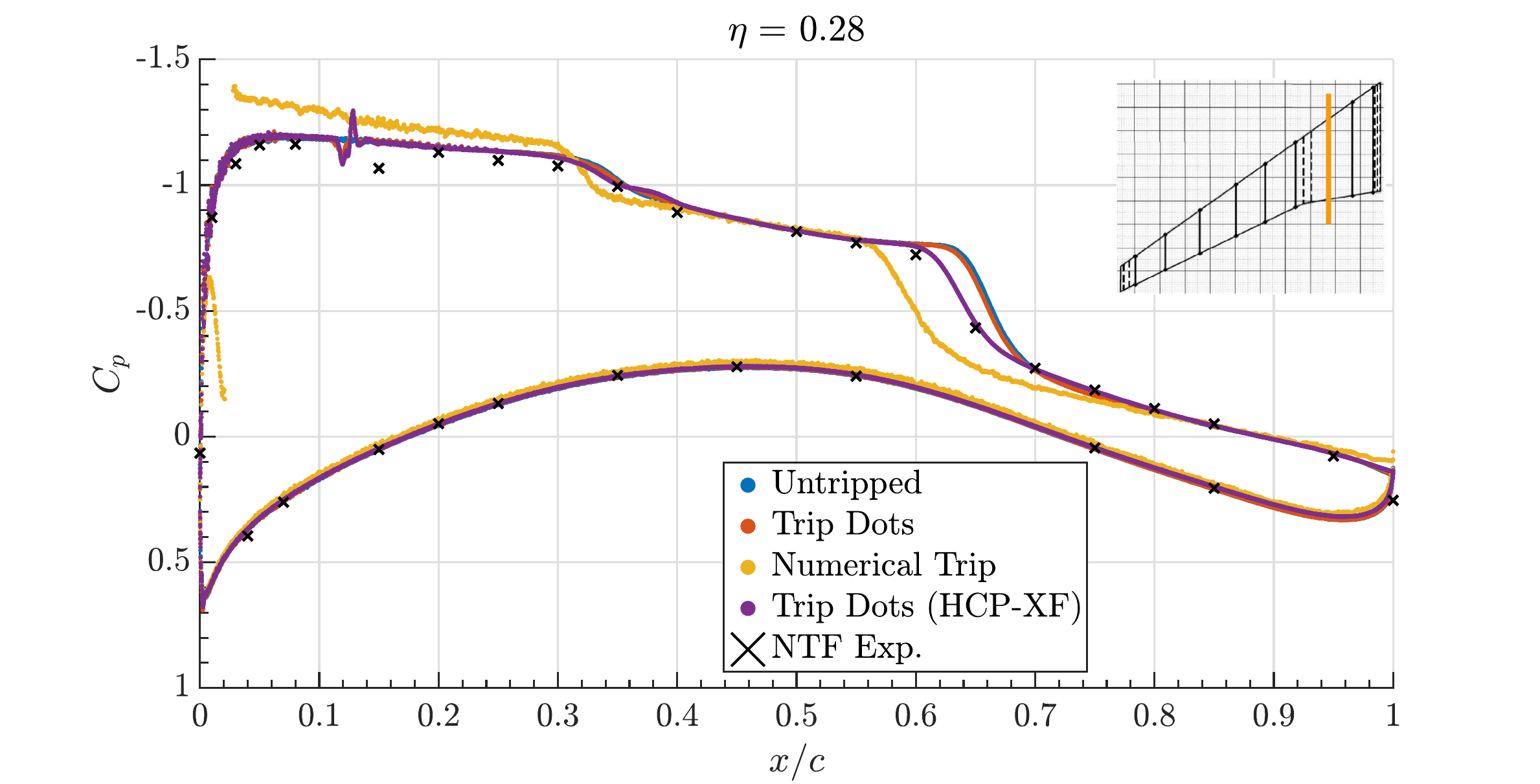}}\\ 
    \centering
    \subfloat[\label{fig:crm_cp_trip_dots_c}]{\includegraphics[width=0.495\textwidth]{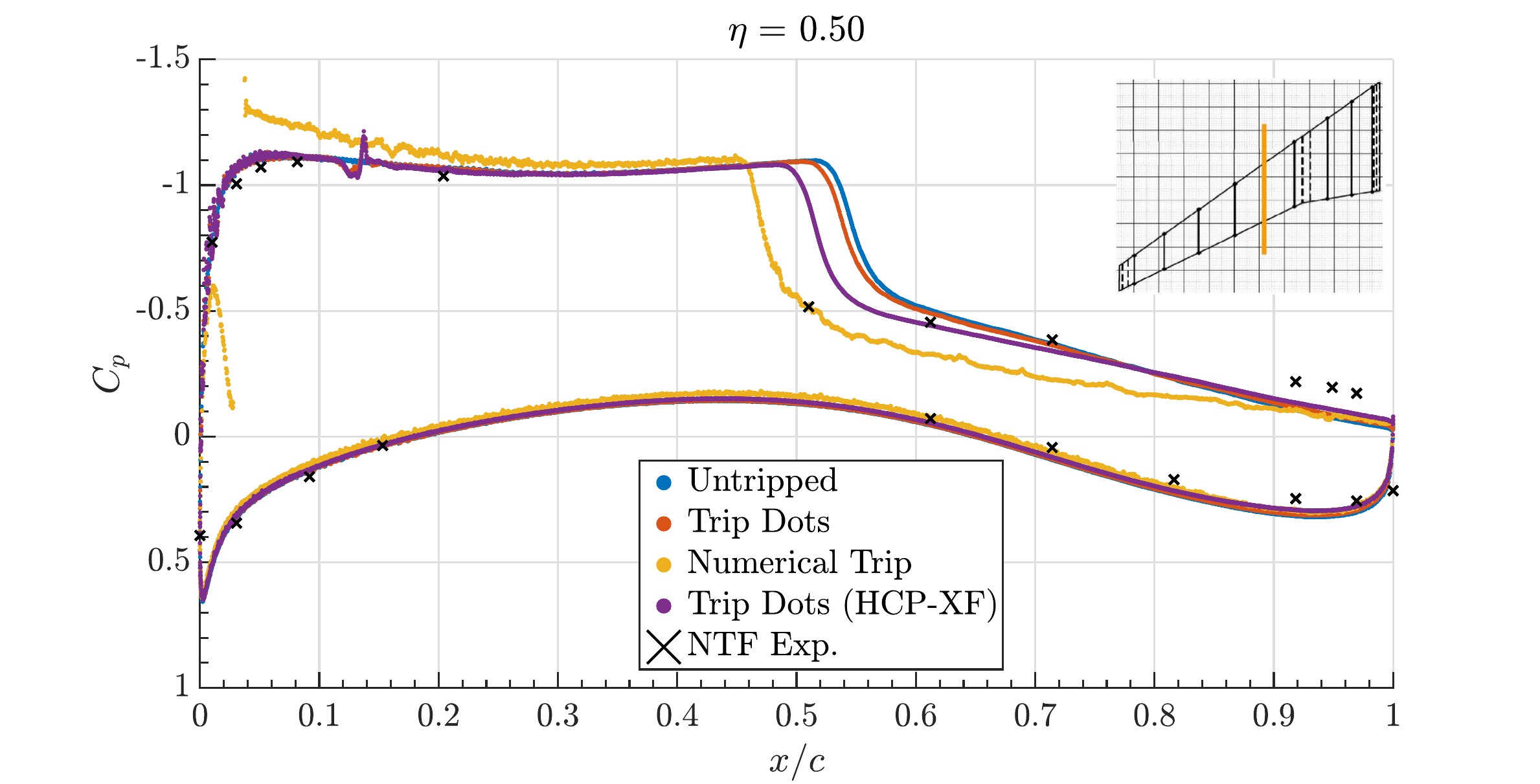}}
    \subfloat[\label{fig:crm_cp_trip_dots_d}]{\includegraphics[width=0.495\textwidth]{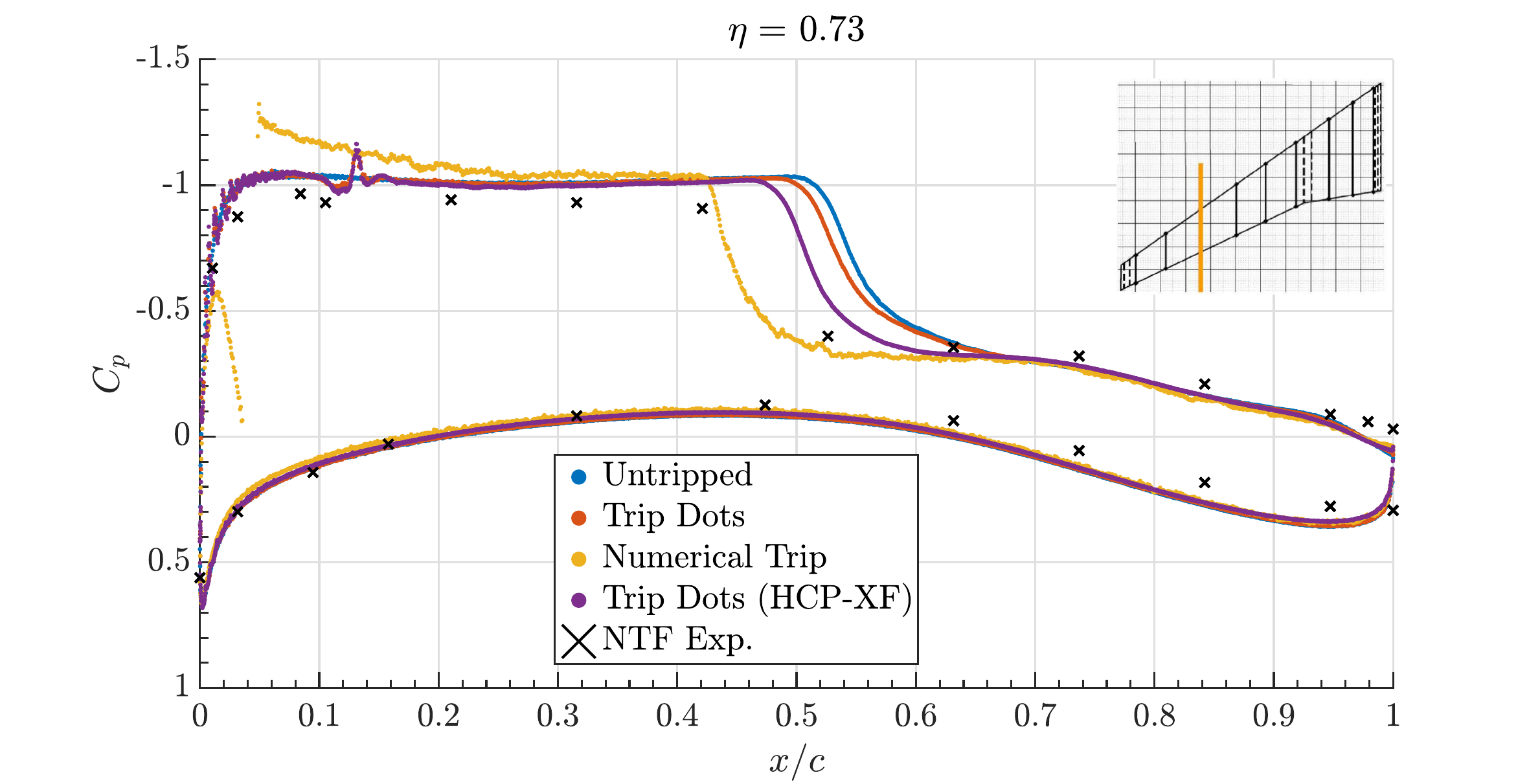}}
     
    \caption{Average pressure coefficient slice through four wing pressure belt stations at $\alpha = 4^{\circ}$ showing the impact of various methods of treating leading edge transition, including untripped, geometric trip dots, and a numerical transition strip, all computed on mesh HCP-F (unless otherwise noted). Pressure belt locations range from the wing root (a), mid-span (b-c), to near the wingtip (d). The vertical orange line in the inset graphic shows the location of the pressure cut along the span of the wing. The NTF experiments were performed by Rivers et al. \cite{rivers2010experimental}}
    \label{fig:crm_cp_trip_dots}
\end{figure}

Fig. \ref{fig:crm_inst_trip_dots_yehudi} and \ref{fig:crm_inst_trip_dots_wingtip} show the qualitative impact of this trip dot array at a location near the Yehudi break (mid-span) and towards the wingtip respectively. These images show that the trip dots force an earlier and more consistent (happening at a uniform chordwise location as a function of span) transition front than the untripped simulation, which is especially the case at the wingtip, where the untripped cases transition later than the location at which the trip dots are installed. Near the Yehudi break, the untripped cases become turbulent at around the same location as the cases that include the geometric trip dots. The impact of the trip dots is therefore attenuated there, which is also visible in the pressure coefficient ($C_p$) plots in Fig. \ref{fig:crm_cp_trip_dots}. These plots show comparisons between the untripped, geometrically tripped (Trip Dots), and numerically tripped (Numerical Trip) simulations, which show a larger sensitivity to the presence of trip dots or numerical transition towards the wingtip than at the wing root, consistent with the notion that lower local chord-based Reynolds numbers exist as the wing tapers off, leading to a flow that is more sensitive to the treatment of laminar-to-turbulent transition in that region. A more detailed discussion on the numerical tripping methodology is included in appendix \ref{sec:appendix_a_tripping}. The result shown in Fig. \ref{fig:crm_cp_trip_dots} is for the Numerical Trip located at the forward trip line position, and the flow is turbulent on nearly the entire suction side of the airfoil as a result. No transition treatment was applied on the lower surface of the airfoil. Parameters for the numerical tripping were chosen based on dimensional arguments from simple boundary layer analysis, with a zero net mass flux sinusoidal forcing applied at the tripping location with amplitude based on the leading edge boundary layer $u_{\tau}$ and frequency constructed from $u_{\tau}$ and the local $\delta_{99}$. The boundary layer parameters used were computed based on the technique described in Appendix \ref{sec:bl_details}.  A thorough parameter study associated with the numerical tripping coefficients (amplitude/frequency) was not undertaken at this time (except for a limited sensitivity to the location of the numerical trip strip discussed in the appendix). All calculations, unless otherwise noted, presented on these plots were conducted on the fine HCP mesh, HCP-F. The meshes only slightly differ in their grid point distribution in the immediate vicinity of the trip dots, otherwise the resolution is identical. 

It was observed that despite the marginal resolution level relative to the trip dot size in the geometrically tripped cases, the earlier transition associated with the inclusion of trip dots on the CRM wing led to slightly improved shock locations relative to the experimental data on the suction side of the wing, with maximum sensitivity to the shock location of only $\approx 1-2\%$ in x/c (on the HCP-F mesh on which the tripped/untripped comparison was made). The perturbation in the $C_p$ plot (see Fig. \ref{fig:crm_cp_trip_dots}) near $10\%$ $x/c$ reflects the location at which the trip dots are installed in the simulations, while the more aggressive perturbation in the numerically tripped cases at the leading edge reflects the location of the numerical trip line. It is clear from the numerical transition cases that transition can be a strong driver of the upper surface pressure coefficient (and, therefore, the global forces/moments) distribution. In fact, up to $\approx 10-15\%$ x/c shift in shock location depending on the pressure belt is observed between the untripped and numerically tripped cases. The effect of tripping was appreciable across the entire span of the wing when considering the numerical transition approach, with larger sensitivities observed towards the wingtip. In this case, the shock location generally moves forward (and in doing so, achieves a more favorable comparison to the experiments) as the transition is forced upon the flow at the wing leading edge. Numerical tripping with the chosen set of forcing parameters lead to a more pronounced effect on the shock location as compared to the geometrically-tripped cases, though it is unclear whether this conclusion would hold up under grid refinement, as there is still significant grid sensitivity demonstrated between the cases which include the trip dot geometries on the HCP-F and HCP-XF meshes. For the numerically tripped cases in contrast, the transition mechanism is likely too strong, as it leads to a substantial change to the pressure rooftop $C_p$ and sometimes a different pressure recovery behavior after the shock (as is the case at station $\eta = 0.5$). 

Overall, the simulations described in this section highlight the strong sensitivity of transonic aircraft flows to the treatment of laminar-to-turbulent transition, as variations in shock location of $\approx 10-15\%$ x/c are routinely possible depending on the transition approach used. They also motivate a more thorough exercise in which a range of numerical tripping approaches are studied on this configuration. A desirable outcome would be a numerical tripping approach that robustly triggers a turbulent state with a minimal ``numerical signature'' of its own. It remains unclear whether such an approach can work reliably on a complex aircraft configuration and whether this approach is preferable to directly resolving the experimental trip dots in the simulations. Unfortunately, grid-converging the trip dot-resolving simulations in order to answer this question is presently untenable, even on leadership-class computing resources such as OLCF Frontier on which the 2.2 Bcv HCP-XF calculation was run. Note that the geometric trip dots were used on the baseline geometry in all alpha sweeps which used an HCP mesh (Mesh HCP-C/HCP-M/HCP-F/HCP-XF), though they were only ever even marginally resolved on meshes HCP-F or HCP-XF.

\section{Influence of the Sting Mounting System}
\label{sec:sting}

\begin{figure}[!ht]
\centering
    \subfloat[\label{fig:crm_sting_on}]{\includegraphics[width=0.495\textwidth]{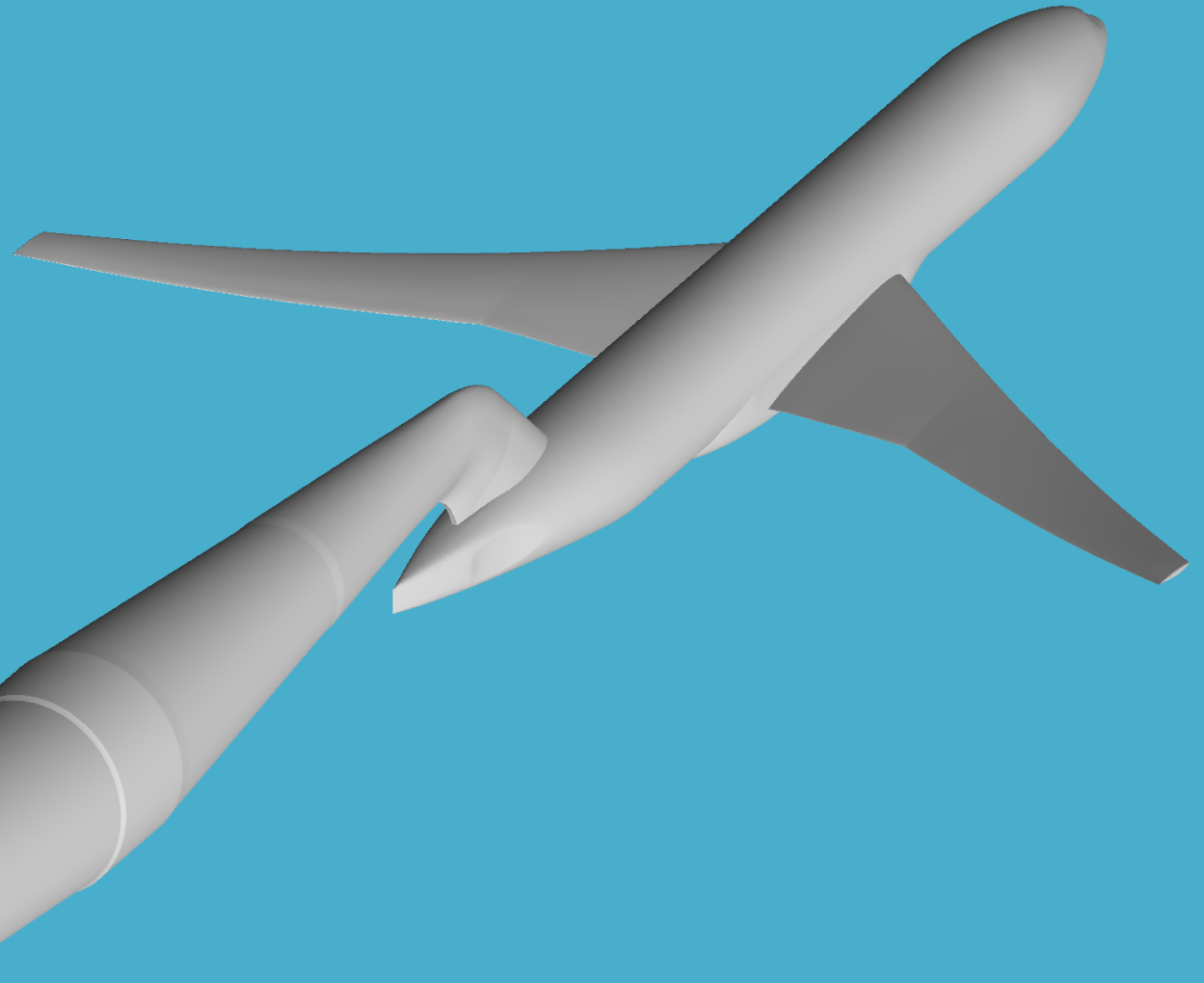}}
    \centering
    \subfloat[\label{fig:crm_sting_off}]{\includegraphics[width=0.488\textwidth]{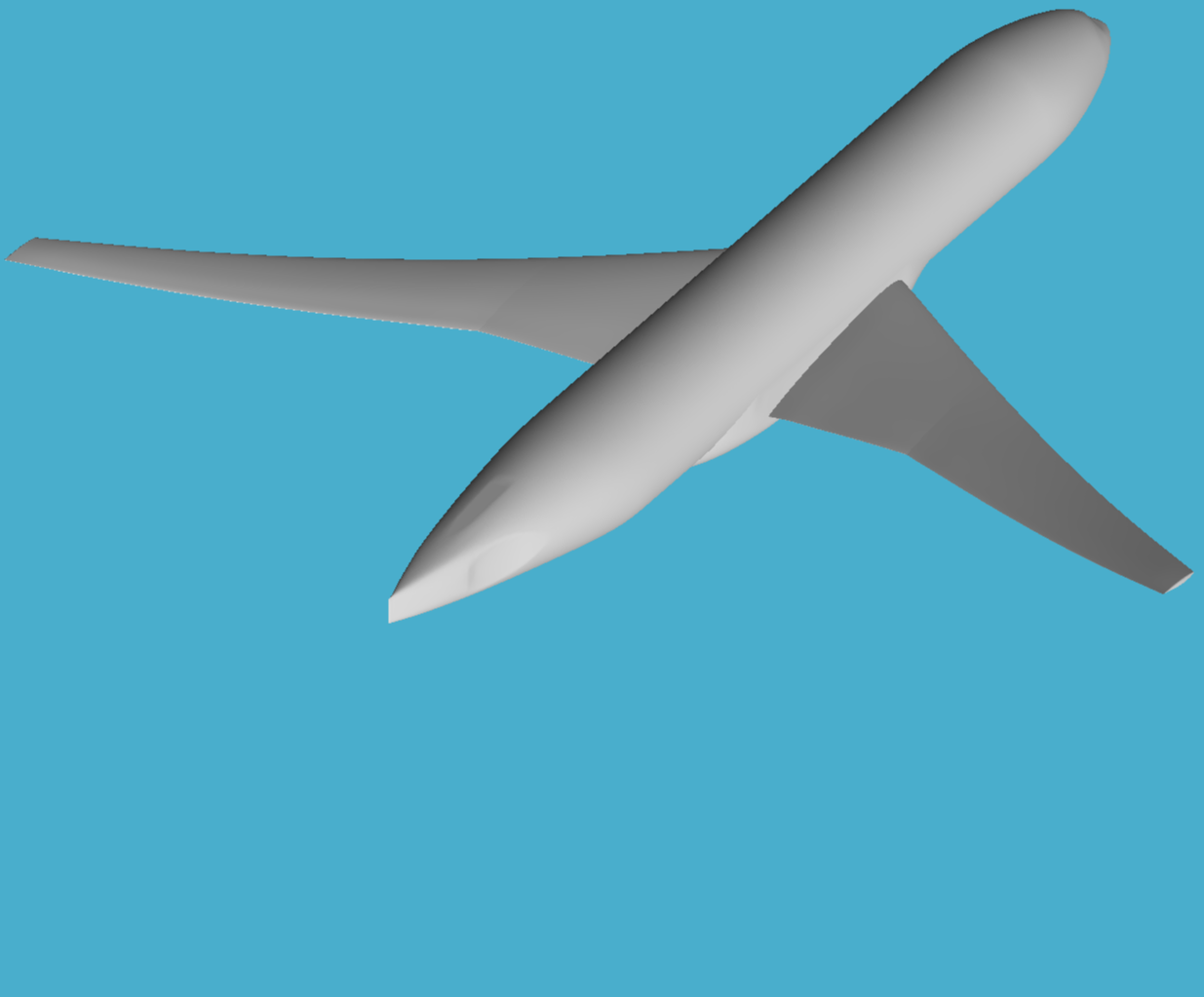}}
    \caption{Image of (a) the simulated geometry including the wind tunnel sting mounting system and (b) the same geometry excluding the sting (both mirrored about the symmetry plane for ease of viewing).  \label{fig:crm_sting_on_v_off_geom}}
\end{figure}

As previously mentioned, the transonic CRM has been a recurring focus of the DPW series \citep{levy2003data,laflin2005data,vassberg2008abridged,vassberg2010summary,levy2013summary,tinoco2018summary,tinoco2023summary}, but to the best of our knowledge, a study of the influence of the wind tunnel mounting system has not been carried out for this configuration. Figure \ref{fig:crm_sting_on_v_off_geom} shows views of the CRM both with the aft sting mount installed and removed. The typical practice for this configuration is to simulate the geometry depicted in subfigure (b) of Fig. \ref{fig:crm_sting_on_v_off_geom}. The influence of this mounting system was assessed using the HCP-F mesh for angles of attack between $2.50^{\circ}-4.00^{\circ}$. The sting mount, in this case, is not connected to any boundary and is an extension of the ``floating'' airplane. This calculation can be interpreted as giving insight into partial wind tunnel installation effects since the test section is excluded from the simulations.

\begin{figure}[!ht]
    \subfloat[\label{fig:crm_cp_sting_on_v_off_a}]{\includegraphics[width=0.495\textwidth]{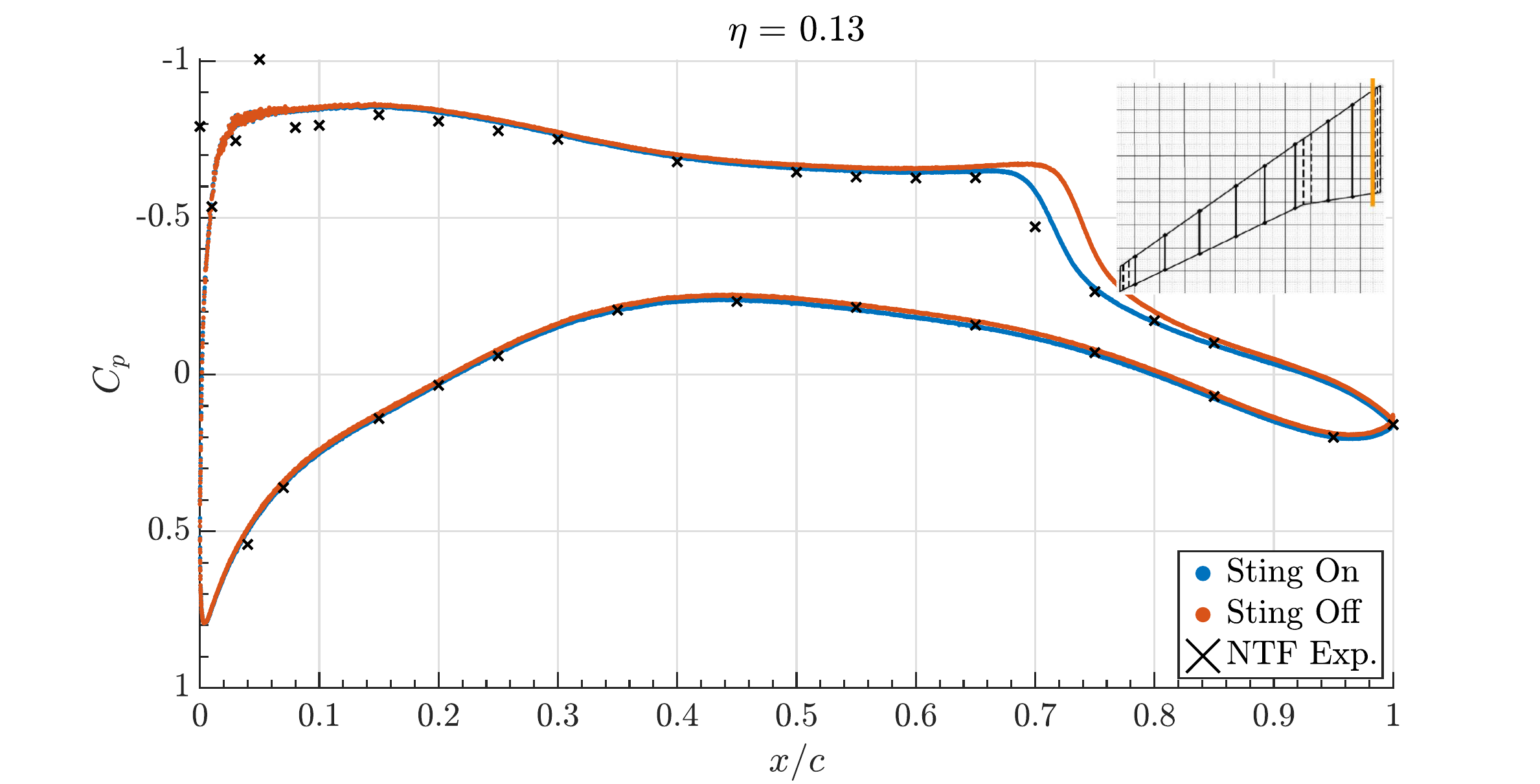}}
    \subfloat[\label{fig:crm_cp_sting_on_v_off_b}]{\includegraphics[width=0.495\textwidth]{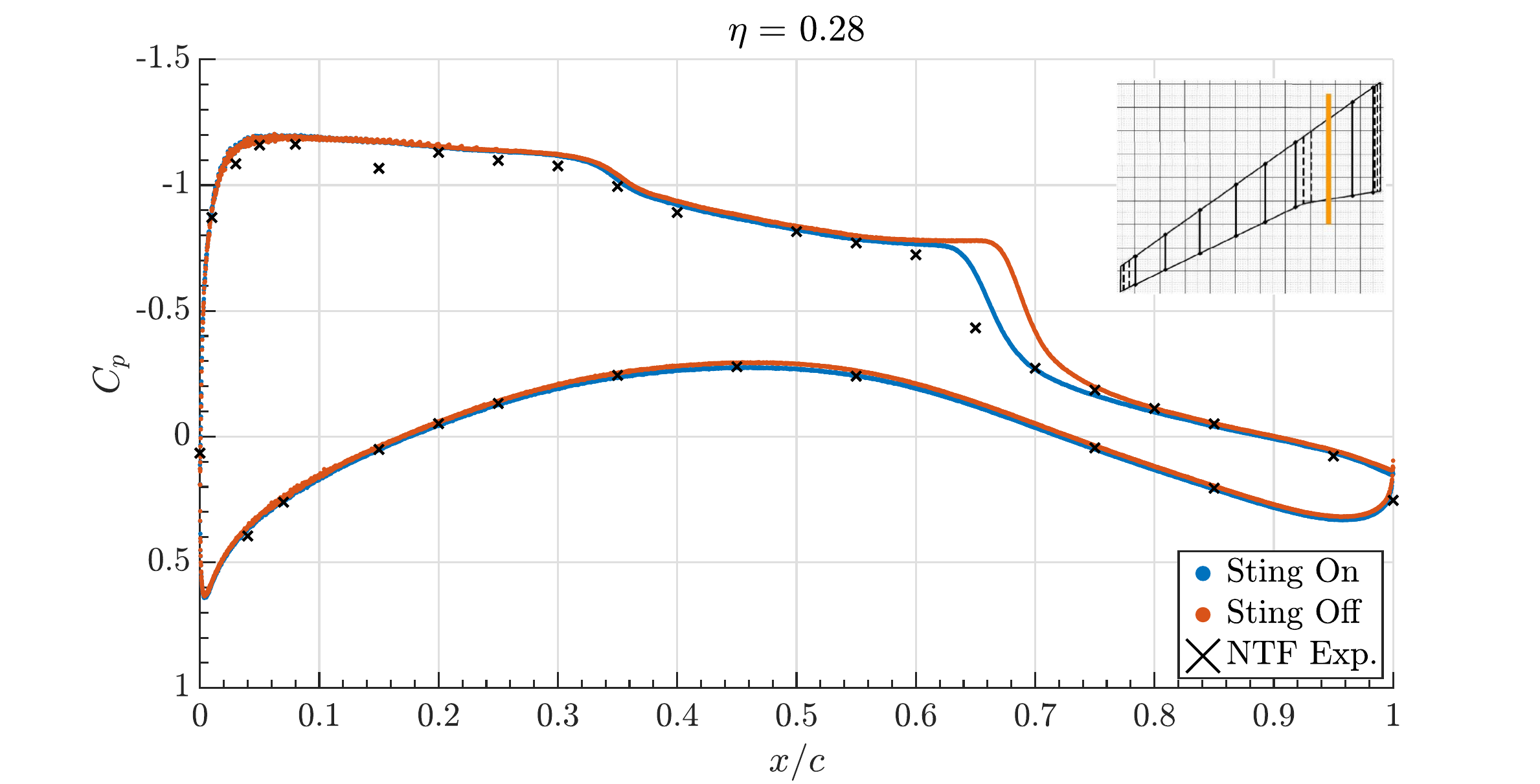}}\\ 
    \centering
    \subfloat[\label{fig:crm_cp_sting_on_v_off_c}]{\includegraphics[width=0.495\textwidth]{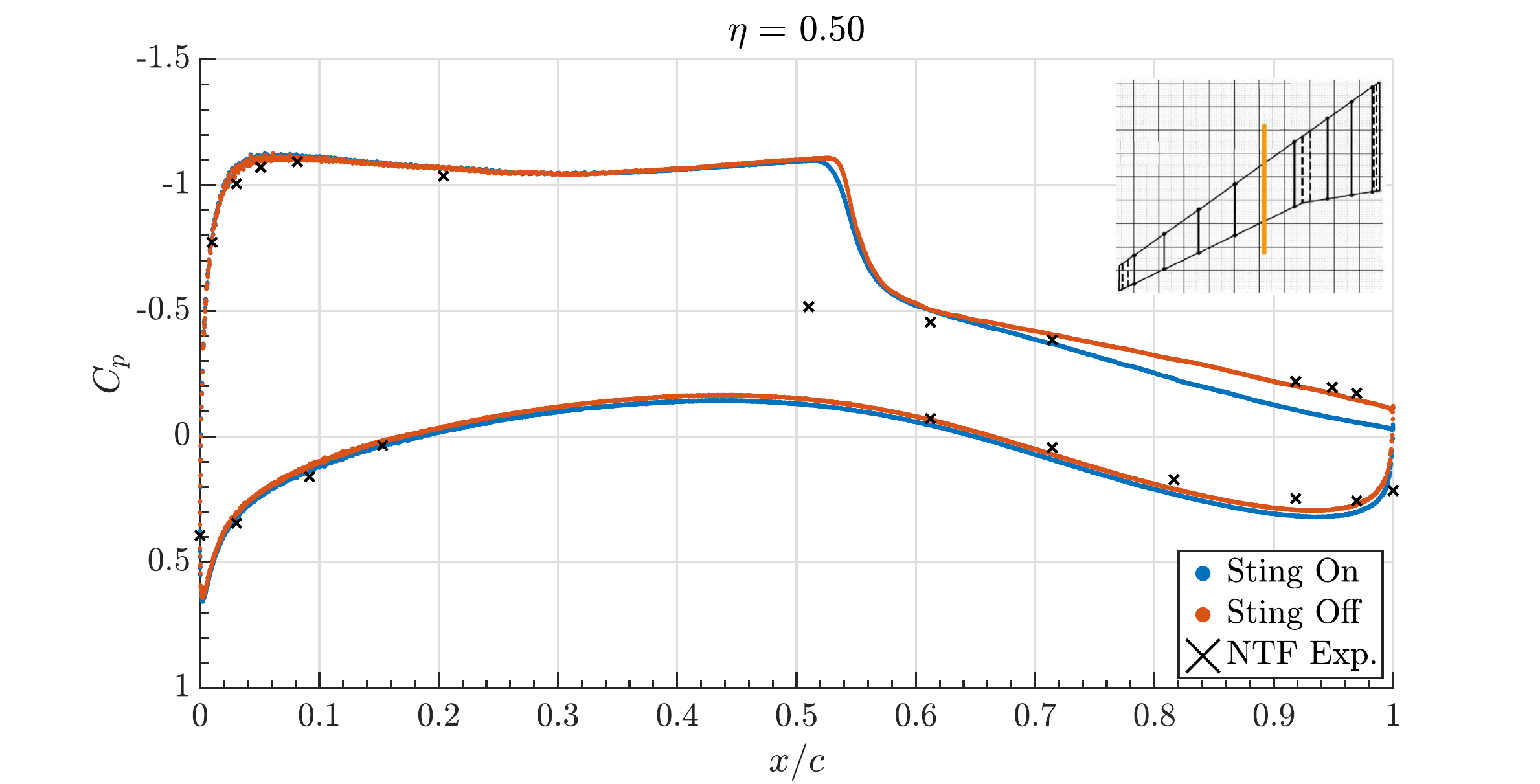}}
    \subfloat[\label{fig:crm_cp_sting_on_v_off_d}]{\includegraphics[width=0.495\textwidth]{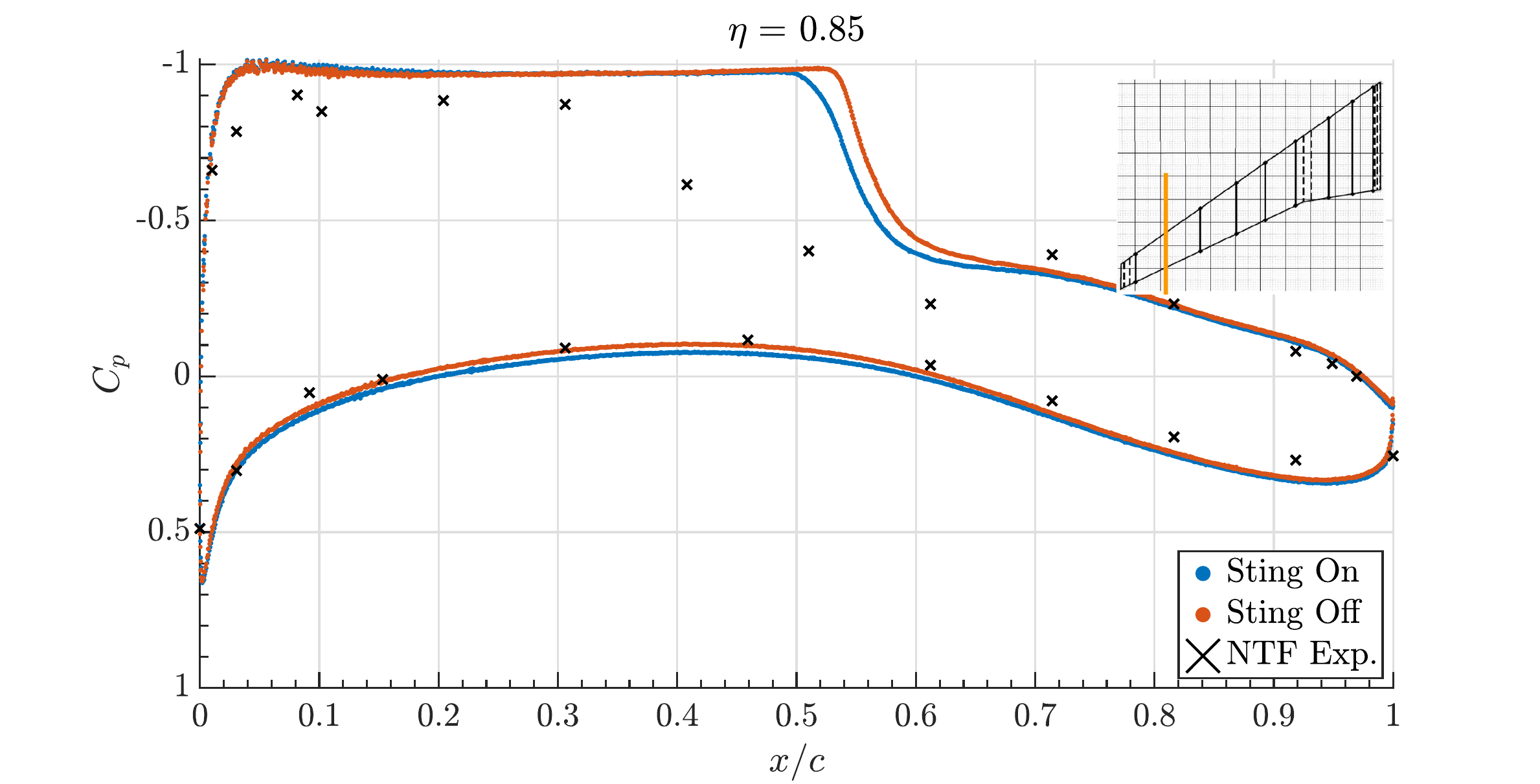}}
     
    \caption{Average pressure coefficient slice through four wing pressure belt stations at $\alpha = 4^{\circ}$ showing the impact of the sting mounting system, ranging from the wing root (a-b), mid-span (c), to near the wingtip (d) on Mesh HCP-F (without trip dots). The vertical orange line in the inset graphic shows the location of the pressure cut along the span of the wing. The NTF experiments were performed by Rivers et al. \cite{rivers2010experimental}}
    \label{fig:crm_cp_sting_on_v_off}
\end{figure}

\begin{figure}[!ht]
    \subfloat[\label{fig:crm_forces_sting_a}]{\includegraphics[width=0.495\textwidth]{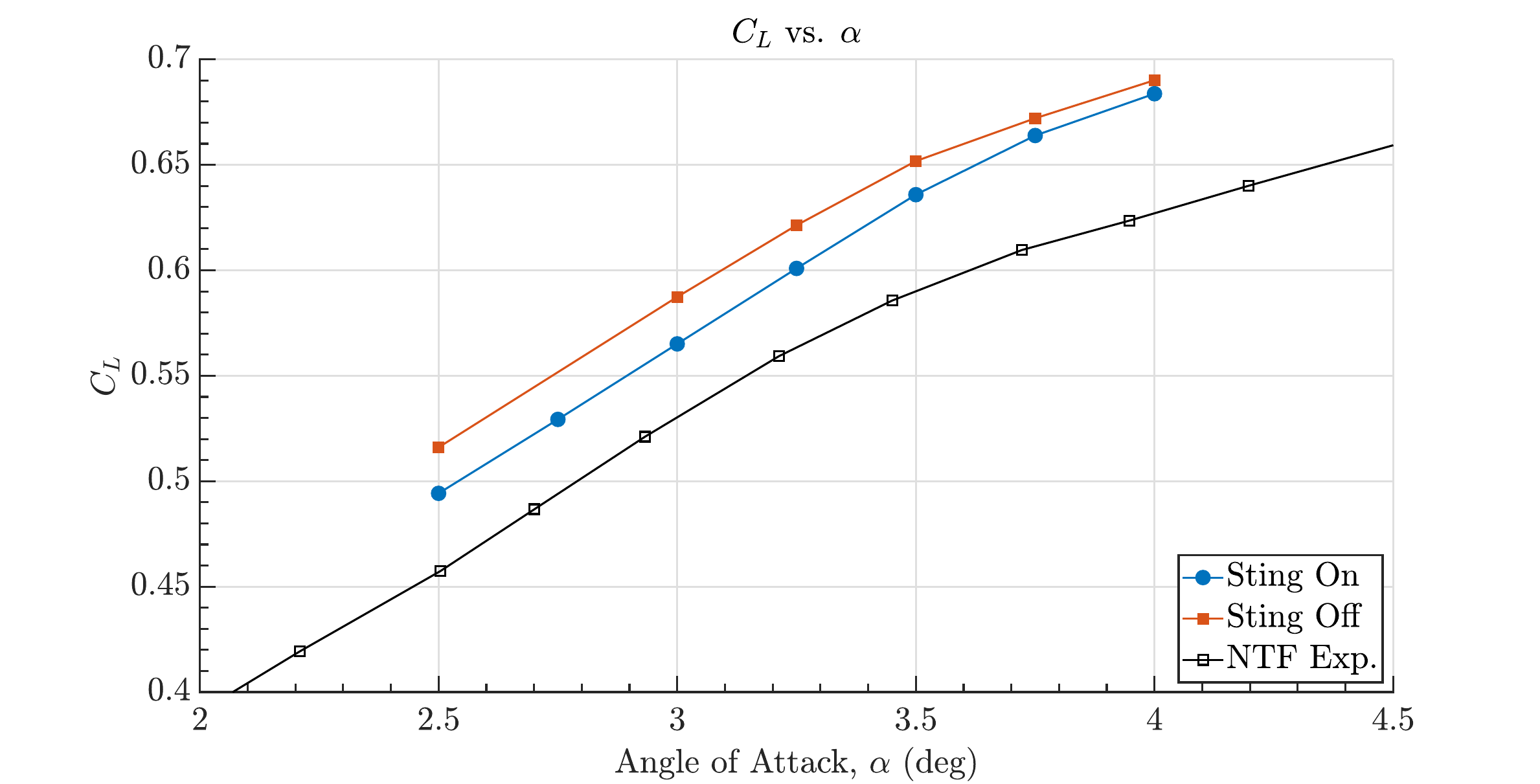}}
    \subfloat[\label{fig:crm_forces_sting_b}]{\includegraphics[width=0.495\textwidth]{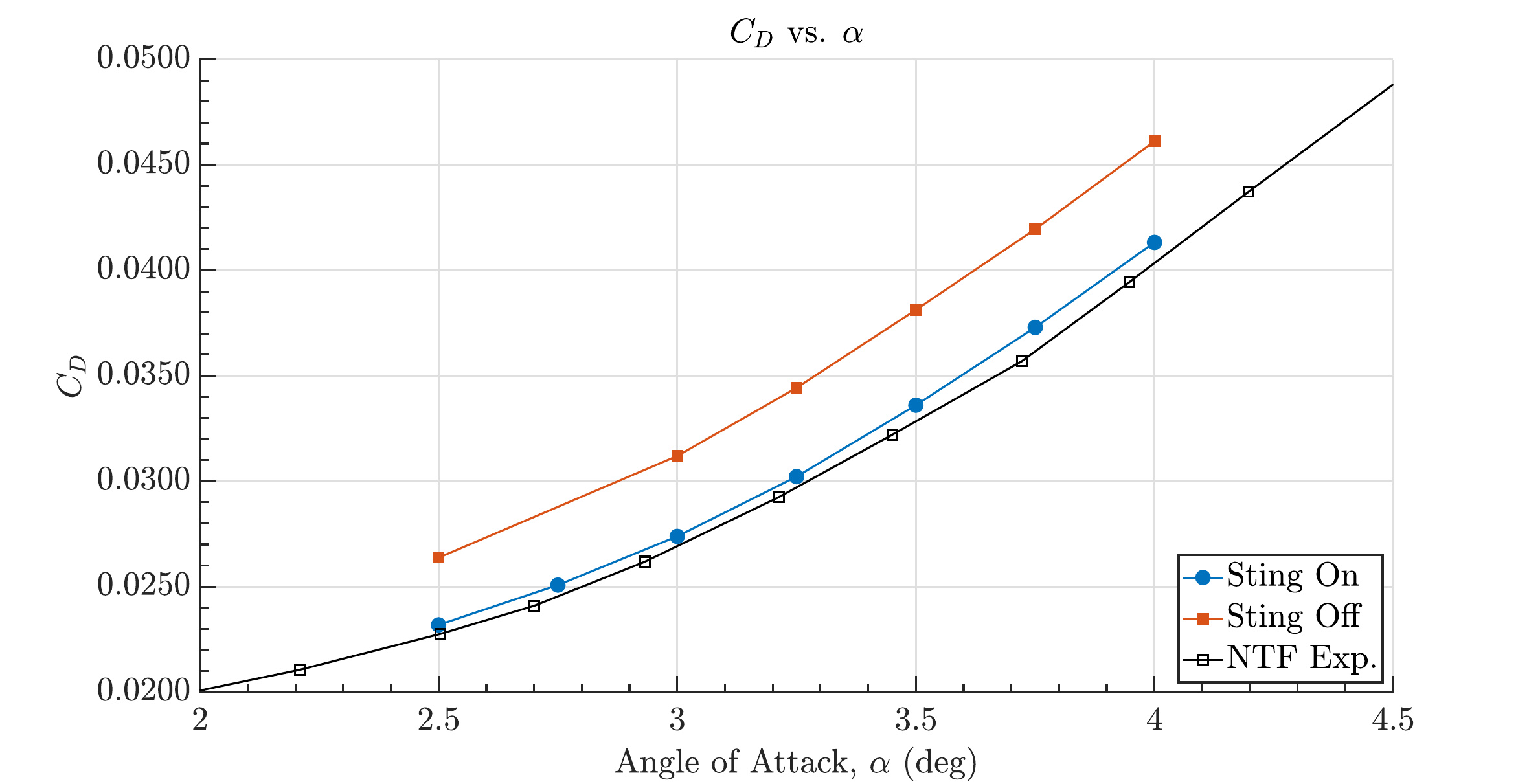}}\\ 
    \centering
    \subfloat[\label{fig:crm_forces_sting_c}]{\includegraphics[width=0.495\textwidth]{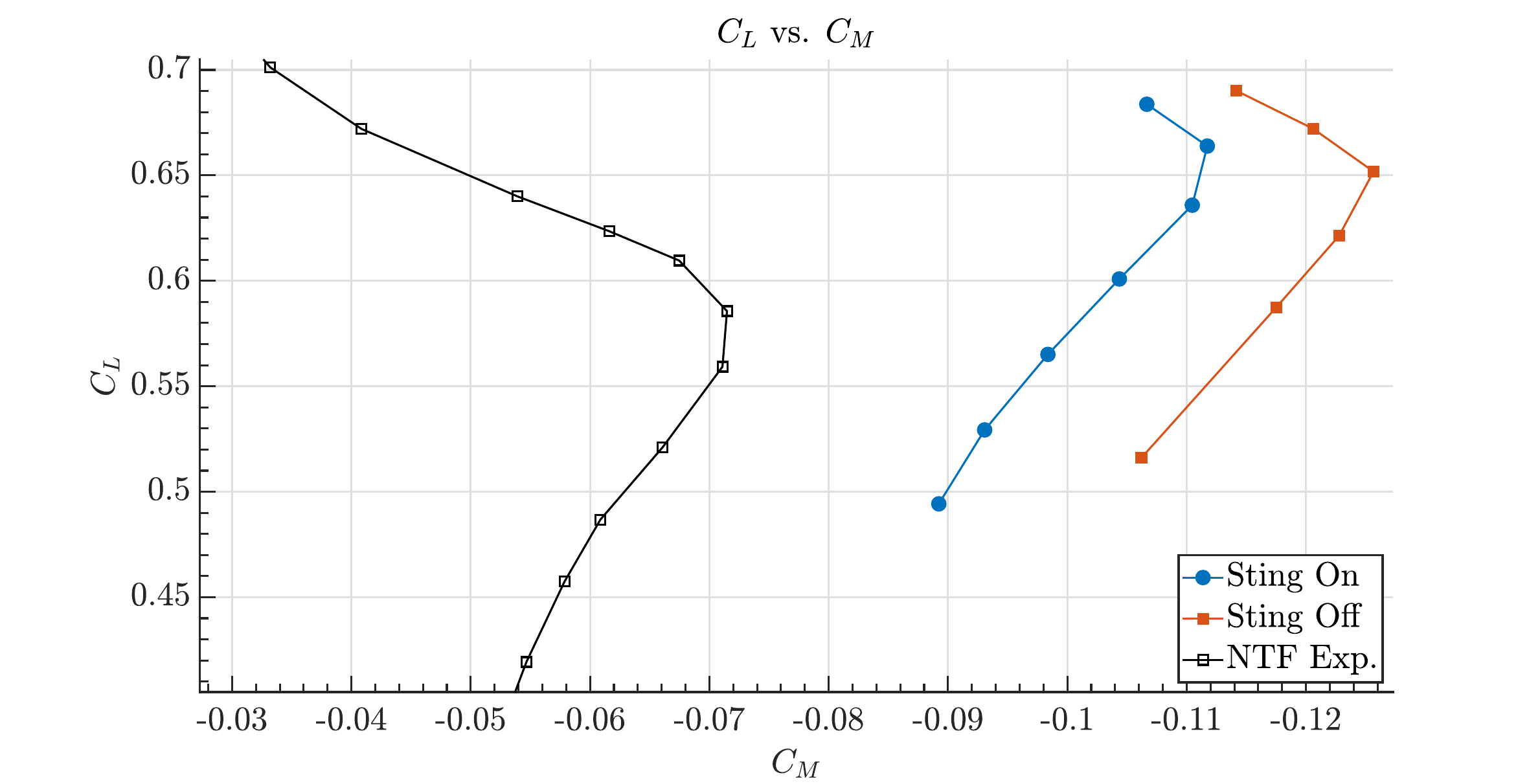}}
     
    \caption{Time-averaged forces/moments from Mesh HCP-F (without trip dots) solutions showing the impact of the sting mounting system, including (a) the lift, (b) the drag, and (c) the pitching moment. \label{fig:crm_forces_sting}}
\end{figure}

The surface pressure measurements for the sting on vs. off comparison at $\alpha=4.00^{\circ}$ are shown in Figure \ref{fig:crm_cp_sting_on_v_off}, and they show that the inclusion of the aft sting mount generally improves the prediction of the shock location and subsequent pressure recovery, except the pressure recovery behavior at the mid-span pressure belt. It is clear from this experiment that although the mounting system is at a significant distance from the wing, its influence is felt across the entire span of the wing and tends to move the shock location forward by $2-3\%$ x/c, a trend that is consistent with the experimental data. One possible explanation for this is the additional inviscid blockage imparted on the upstream flow due to the presence of the aft sting mount, slightly increasing the upstream Mach number and leading to an earlier shock location. Additionally, a full alpha sweep was conducted for the sting on vs. off comparison, and its results are shown in Fig. \ref{fig:crm_forces_sting}. Clearly, the inclusion of the sting mounting system makes a substantial impact on the prediction of lift, drag, and pitching moment at all angles of attack. At every angle, the solution that includes the sting is closer to the experimental data than its sting-off counterpart, even if, in an absolute sense, the data are not close to the experimental measurements. This section should, therefore, be viewed as a sensitivity study which, because it was not conducted on the best-performing mesh (accuracy improvements due to stranded boundary layer meshes will be discussed in subsequent sections of the manuscript), offers useful guidance toward which pieces of the experimental apparatus need to be reproduced in simulations in order to reproduce certain trends observed in the experimental data. Since these numerical tests were performed before studying the effects of grid topology, the results maintain a lingering bias away from the experimental data. Note that the forces on the sting itself are excluded from the force/moment calculation, as was the case in the experiments, which did not measure the force on the sting itself. Based on this result, it was deemed as best practice to include the sting mounting system in all subsequent calculations described in this manuscript. This result motivates a future, more comprehensive validation study of the influence of wind tunnel installation effects on the CRM, potentially including simulation of the NTF test section, diffuser/contraction sections, and plenum.

\section{Sensitivity to Grid Topology}
\label{sec:topology}

The shock location in transonic flows is sensitive to the details of the boundary layer because it is set by the inviscid acceleration, which, in turn, is dictated by the effective body shape (the combined shape of the airplane surface plus viscous regions). For this reason, it was of interest to explore alternate grid topologies in the near-wall region to assess their impact on the prediction of the boundary layer turbulence and engineering quantities of interest ($C_L$,$C_D$,$C_M$,$C_p$,$\tau_w$). Since the near-wall region of the mesh interacts closely with the wall model and is also prone to the most numerical errors, the mesh topology used in this region is believed to be important for determining the accuracy of the solution. Two alternate strategies were employed (relative to the baseline HCP meshing approach for which results have already been shown) both of which fall within the umbrella of Voronoi \citep{fortune2017voronoi,du2006convergence} meshing strategies. 

\begin{figure}[!ht]
    \subfloat[\label{fig:crm_grid_topology_a}]{\includegraphics[width=0.495\textwidth]{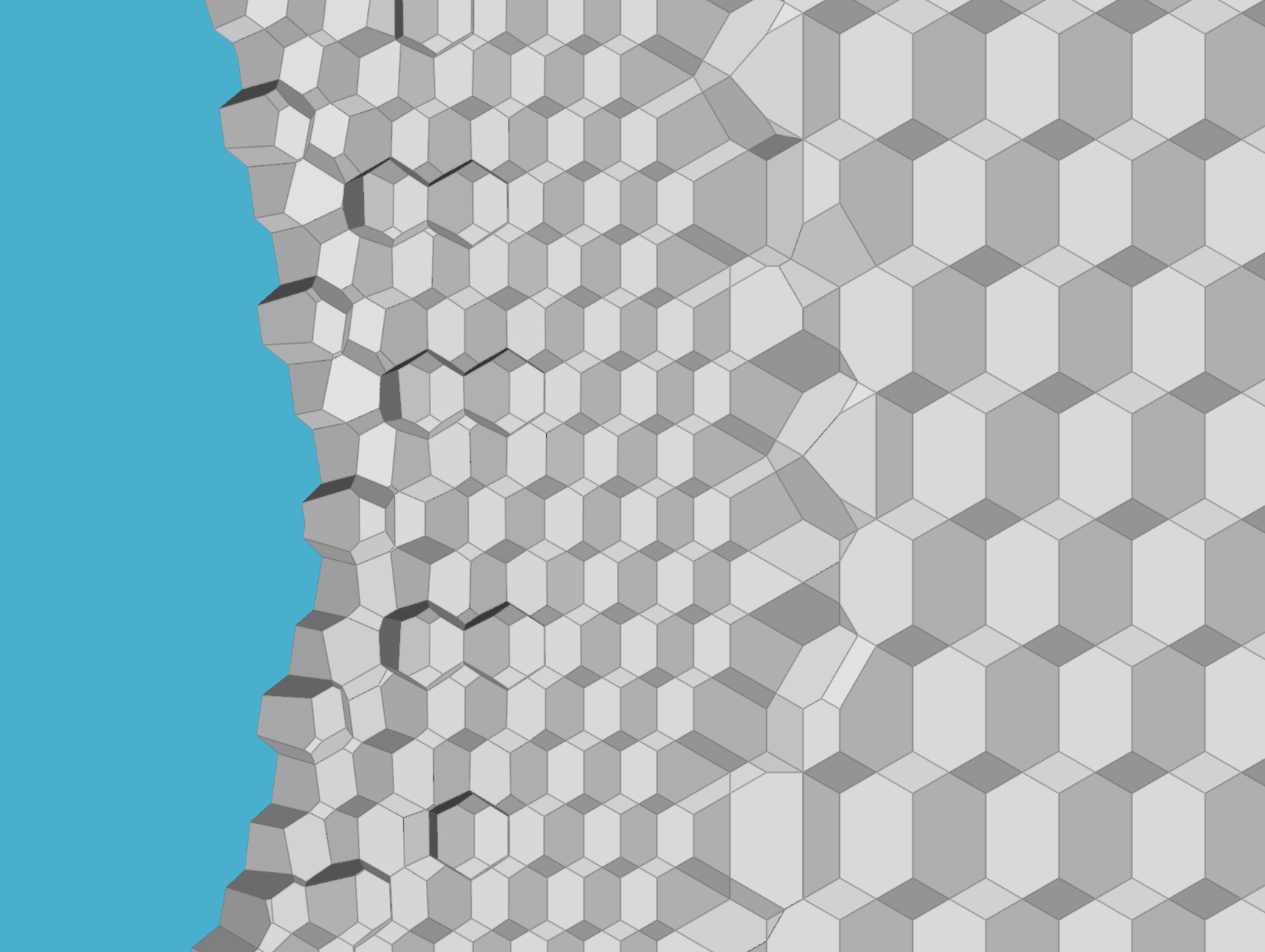}}
    \subfloat[\label{fig:crm_grid_topology_b}]{\includegraphics[width=0.495\textwidth]{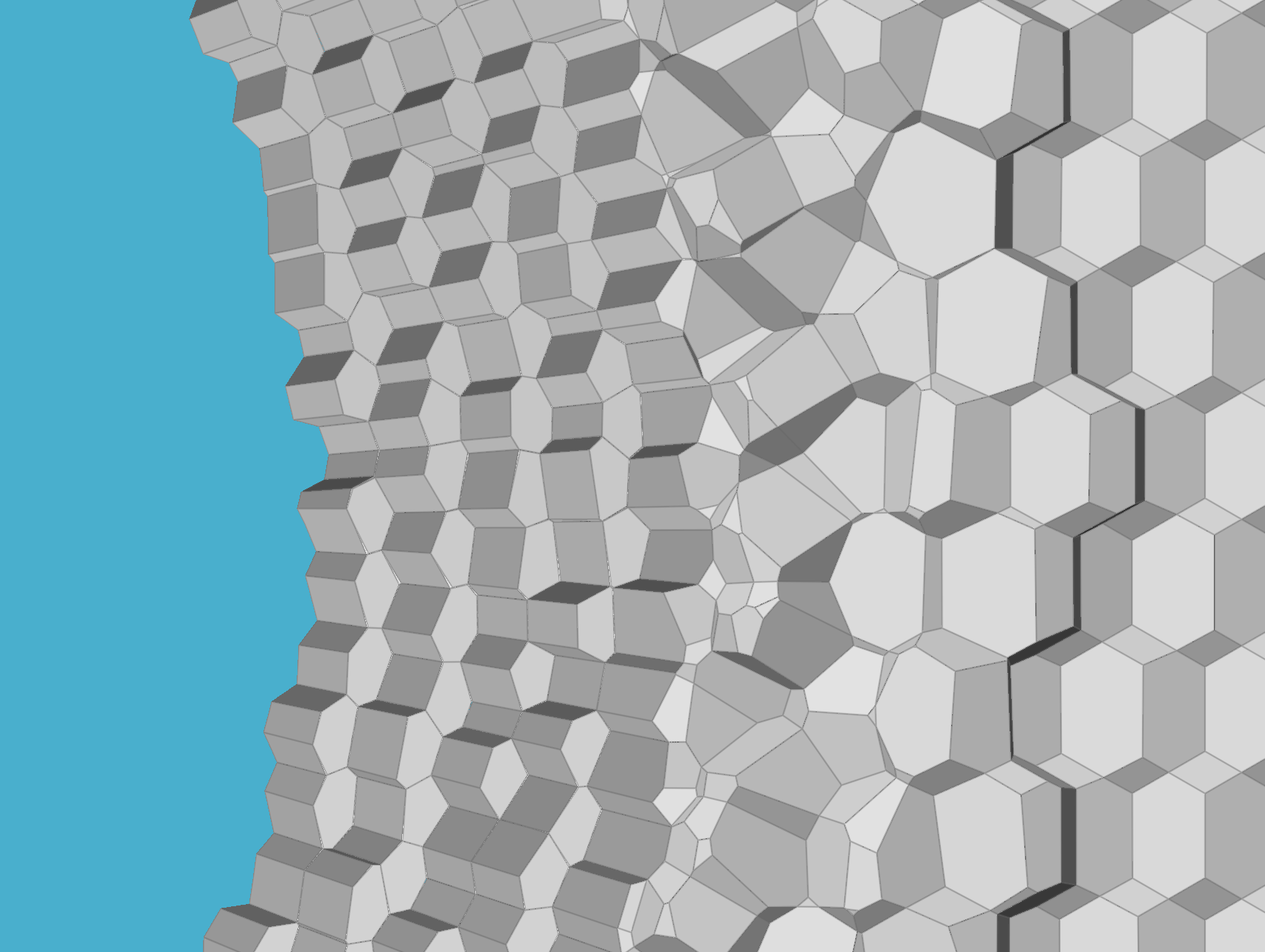}} \\ 
    \subfloat[\label{fig:crm_grid_topology_c}]{\includegraphics[width=0.495\textwidth]{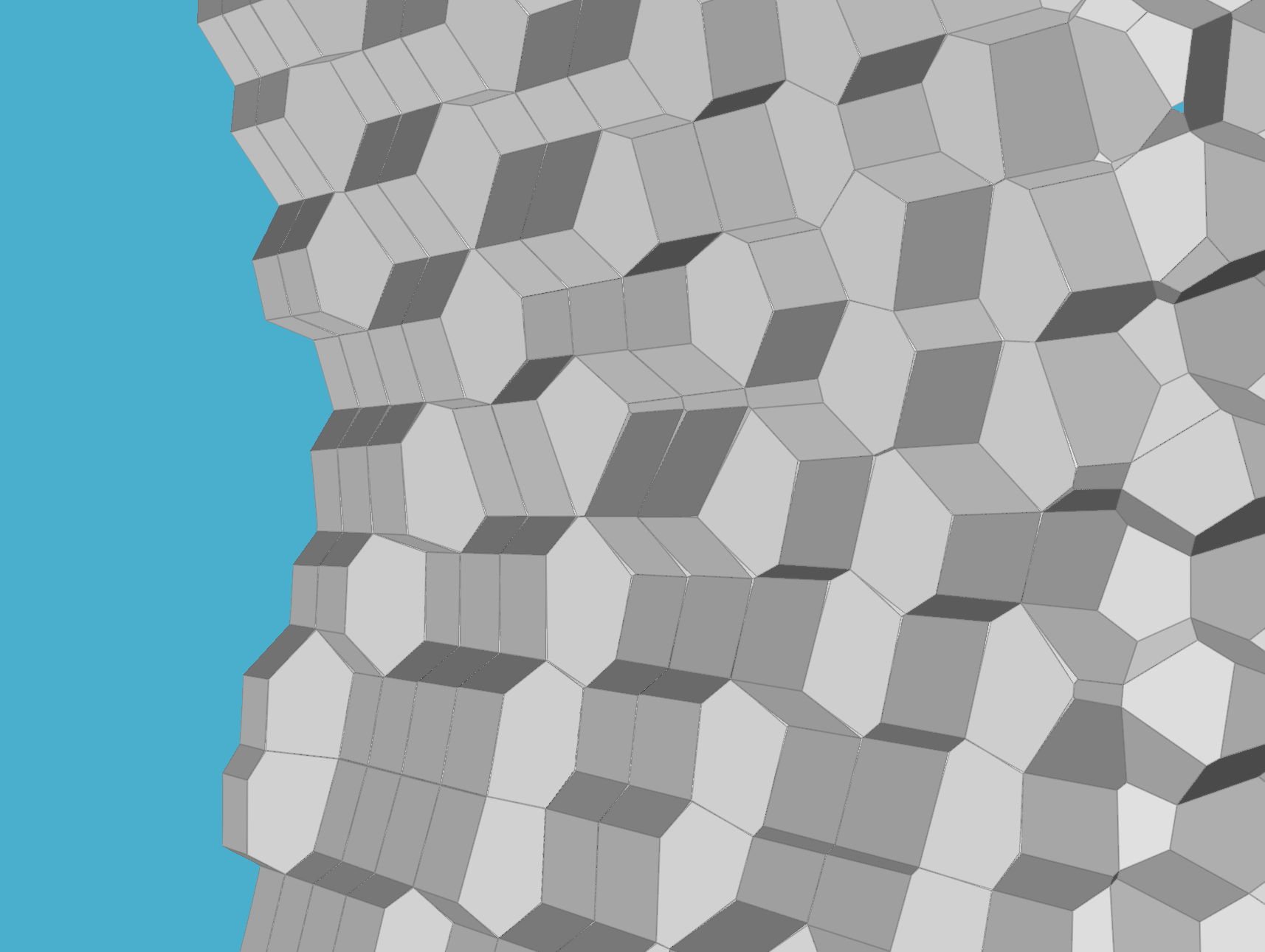}}
    \subfloat[\label{fig:crm_grid_topology_d}]{\includegraphics[width=0.495\textwidth]{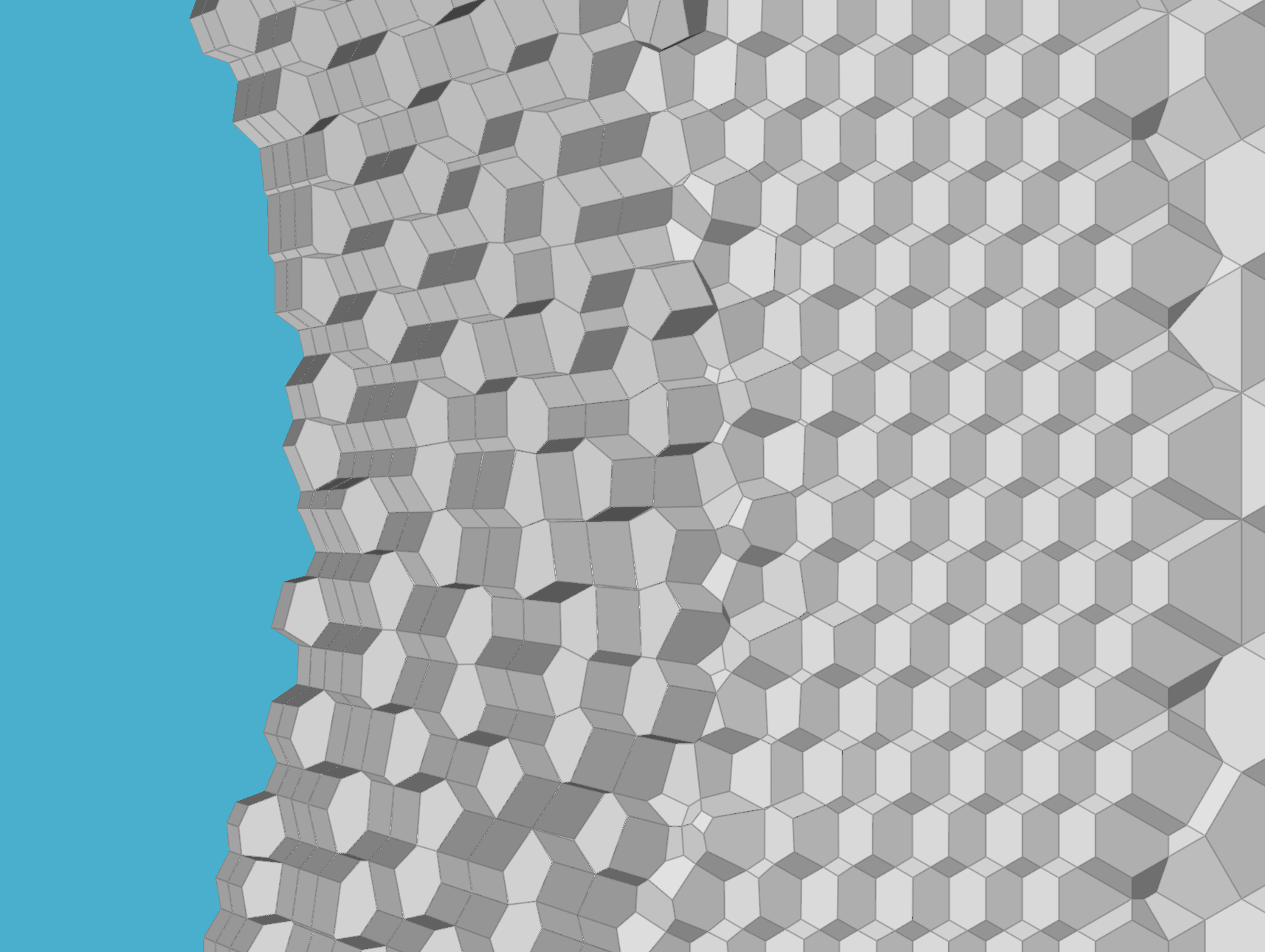}}
     
    \caption{Slice through the leading edge of the CRM wing showing crinkle-cut mesh views from (a) mesh HCP-F, (b) mesh S-IF, (c) mesh S-AM, and (d) mesh S-AF. \label{fig:crm_grid_topology}}
\end{figure}

The alternate topological structures are presented in Fig. \ref{fig:crm_grid_topology} as crinkle-cut views (and were previously shown schematically in Fig. \ref{fig:grid_schematics}); subfigure (a) shows the HCP grid in which a staggered Voronoi seeding approach results in HCP cells that have many neighbors (14 on average), while subfigure (b) shows a stranded gridding approach in which an ordered Voronoi diagram seeding is used in the near-wall region, resulting in a more regularly ordered boundary layer mesh of hexagonal prisms. To isolate the effect of grid topology, isotropic stranded meshes were first used (the stranded meshes need not be uniformly refined in all coordinate directions, unlike the HCP cells) with identical near-wall spacing to their HCP counterparts. For this purpose, Mesh HCP-F is analogous to the Isotropic Strand Mesh S-IF and so on. Tables \ref{tab:hcp_meshes} and \ref{tab:iso_strand_meshes} show that the wall-normal and stream/span-wise resolutions of these meshes on the wings are identical. Differences in how the grid transitions between the strand layer and the off-body HCP region account for the discrepancies in the aggregate grid counts in these cases (stranded meshes have slightly more points for a given refinement level than their HCP counterpart due to how this strand-to-HCP region is handled). This transition region can be seen in Fig. \ref{fig:crm_grid_topology} (b) as being more refined than in (a). Also shown in subfigures (c) and (d) are crinkle-cut slices from meshes S-AM and S-AF, respectively, which highlight the ability of the stranded meshing paradigm to preferentially refine the grids in the wall-normal direction.

All the stranded cases presented in this work employ a second cell exchange of the velocity between the outer LES and the equilibrium wall model, unlike the hexagonal meshes that used the first cell exchange. In the authors' experience, the stranded meshes exhibit the log-layer mismatch \cite{kawai2012wall,kawai2013dynamic} with first point matching, whereas the hexagonal close-packed meshes do not exhibit any significant differences in solutions when using either of the two exchange locations. Additionally, because of the difficulty associated with creating a high-quality prismatic mesh around the leading-edge trip dots, the geometric trip dots were not included in the stranded mesh calculations and the stranded cases have been allowed to transition freely.

\subsection{Isotropic Strand}
\label{sec:isotropic_strand}

\begin{figure}[!ht]
    \subfloat[\label{fig:crm_forces_isostrand_a}]{\includegraphics[width=0.495\textwidth]{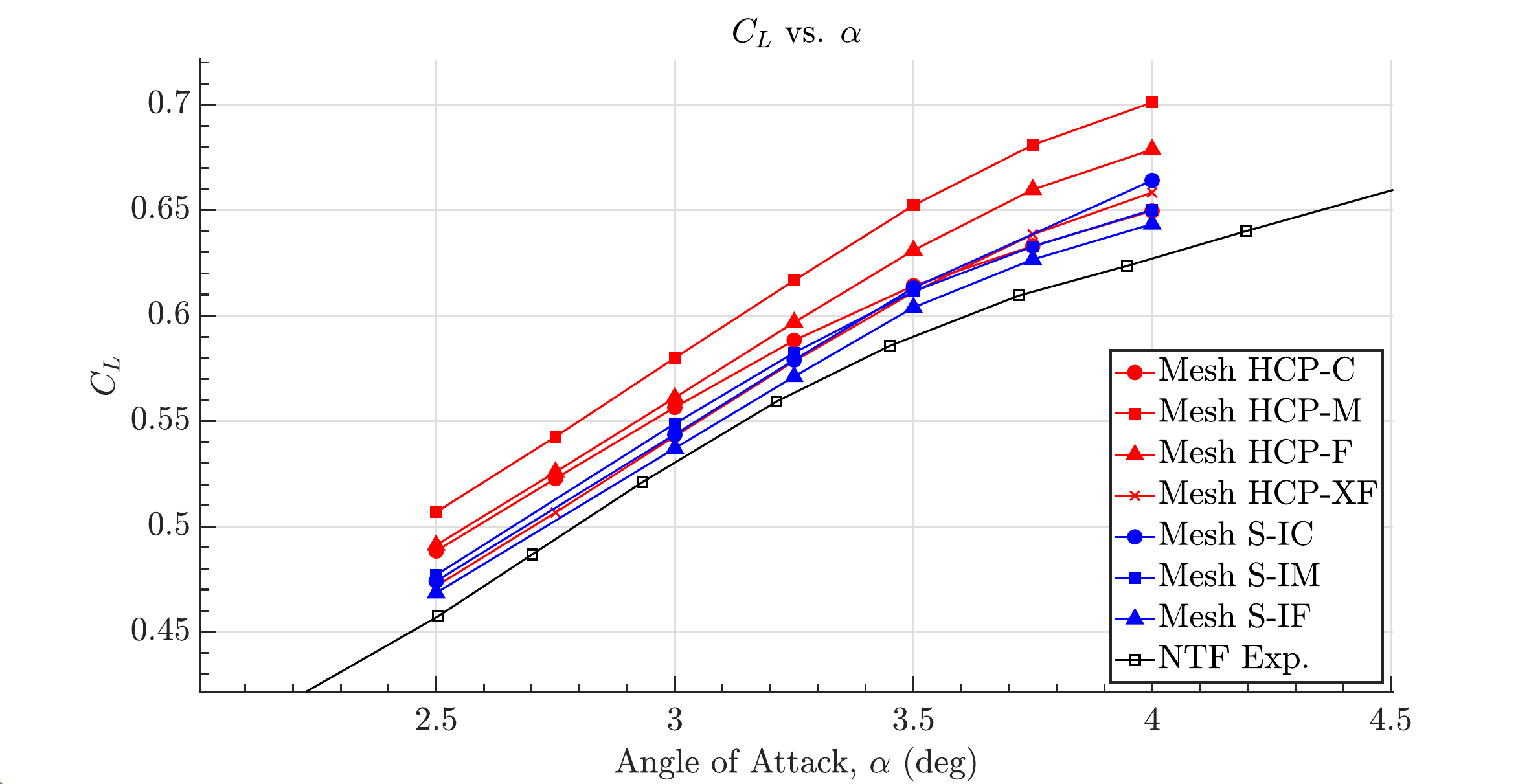}}
    \subfloat[\label{fig:crm_forces_isostrand_b}]{\includegraphics[width=0.495\textwidth]{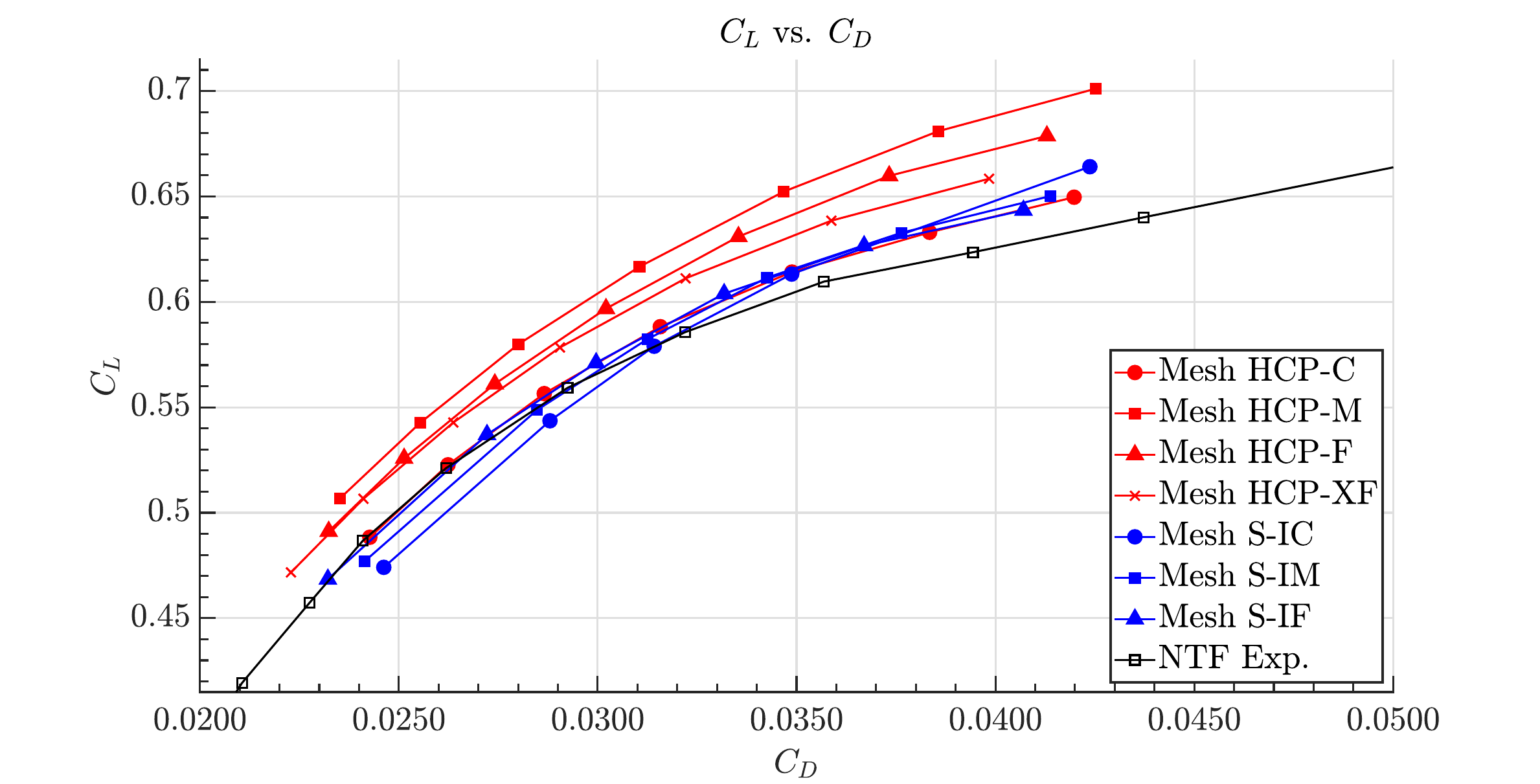}}\\ 
    \centering
    \subfloat[\label{fig:crm_forces_isostrand_c}]{\includegraphics[width=0.495\textwidth]{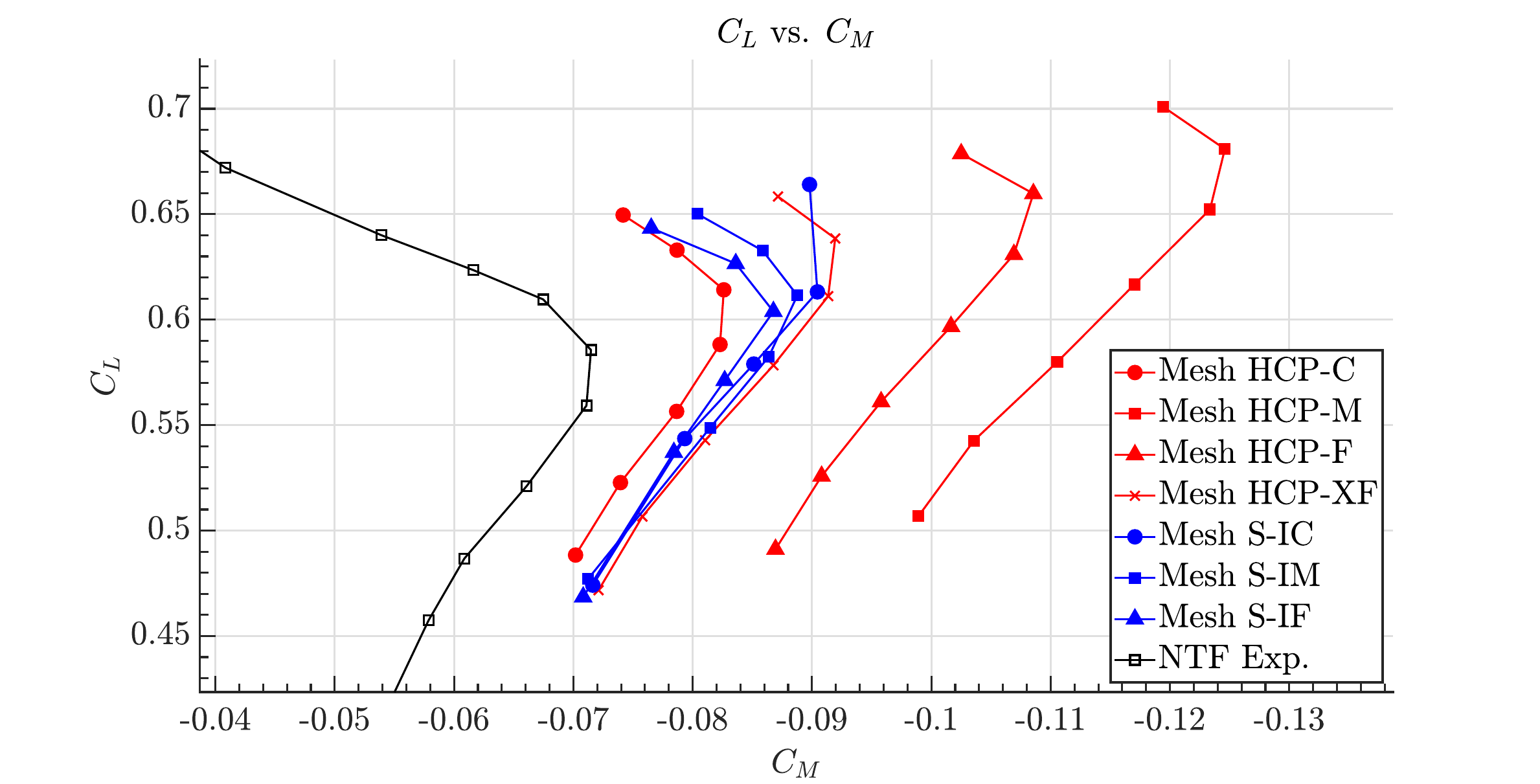}}
     
    \caption{Time-averaged forces/moments from the Isotropic Strand mesh solutions (S-IC through S-IF) and Isotropic HCP mesh solutions (HCP-C through HCP-XF) on coarse, medium, fine, and xfine grids for the transonic CRM, including (a) the lift, (b) the drag polar and (c) the pitching moment. \label{fig:crm_forces_isostrand}}
\end{figure}

\begin{figure}[!ht]
    \subfloat[\label{fig:crm_cp_strand_a}]{\includegraphics[width=0.495\textwidth]{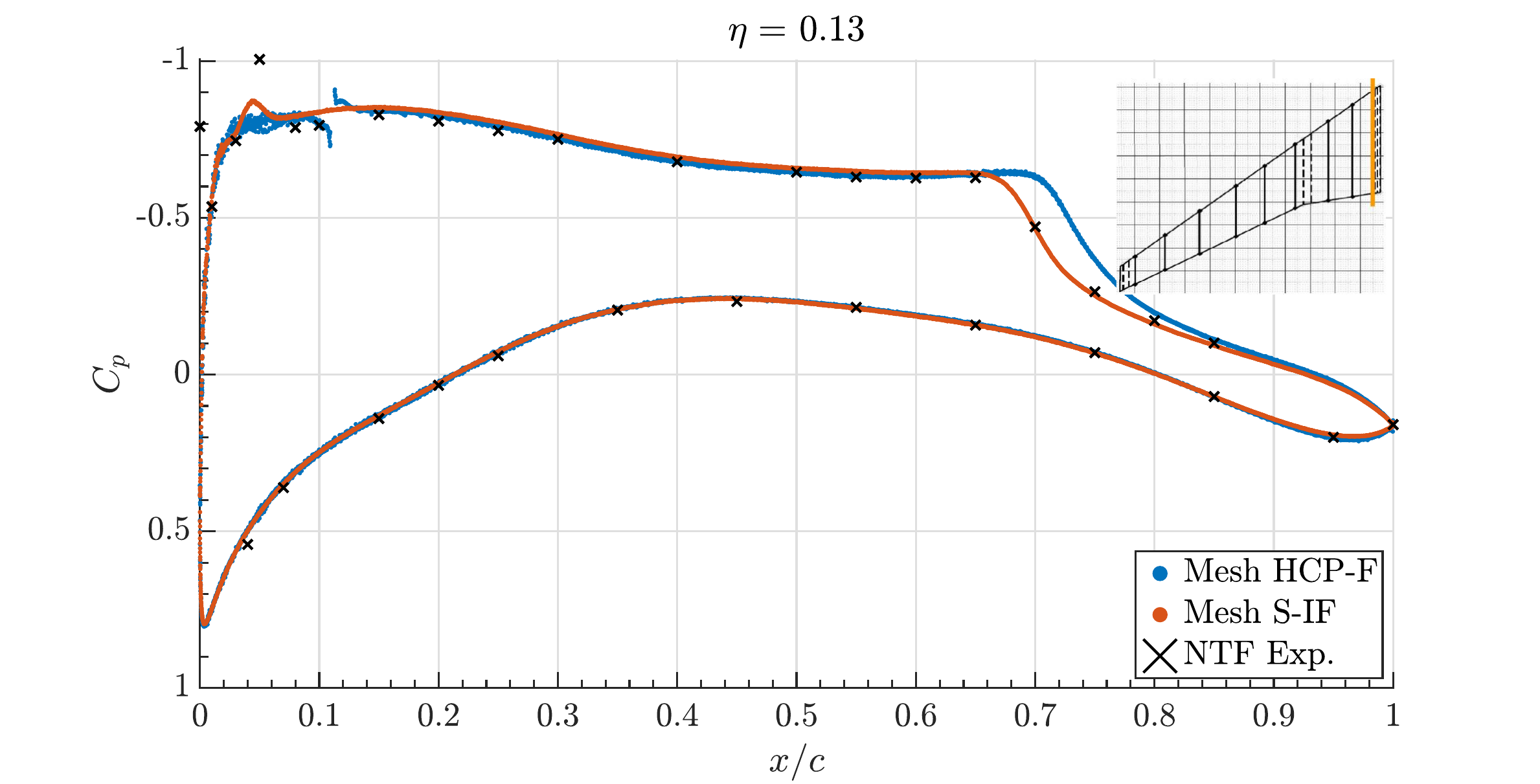}}
    \subfloat[\label{fig:crm_cp_strand_b}]{\includegraphics[width=0.495\textwidth]{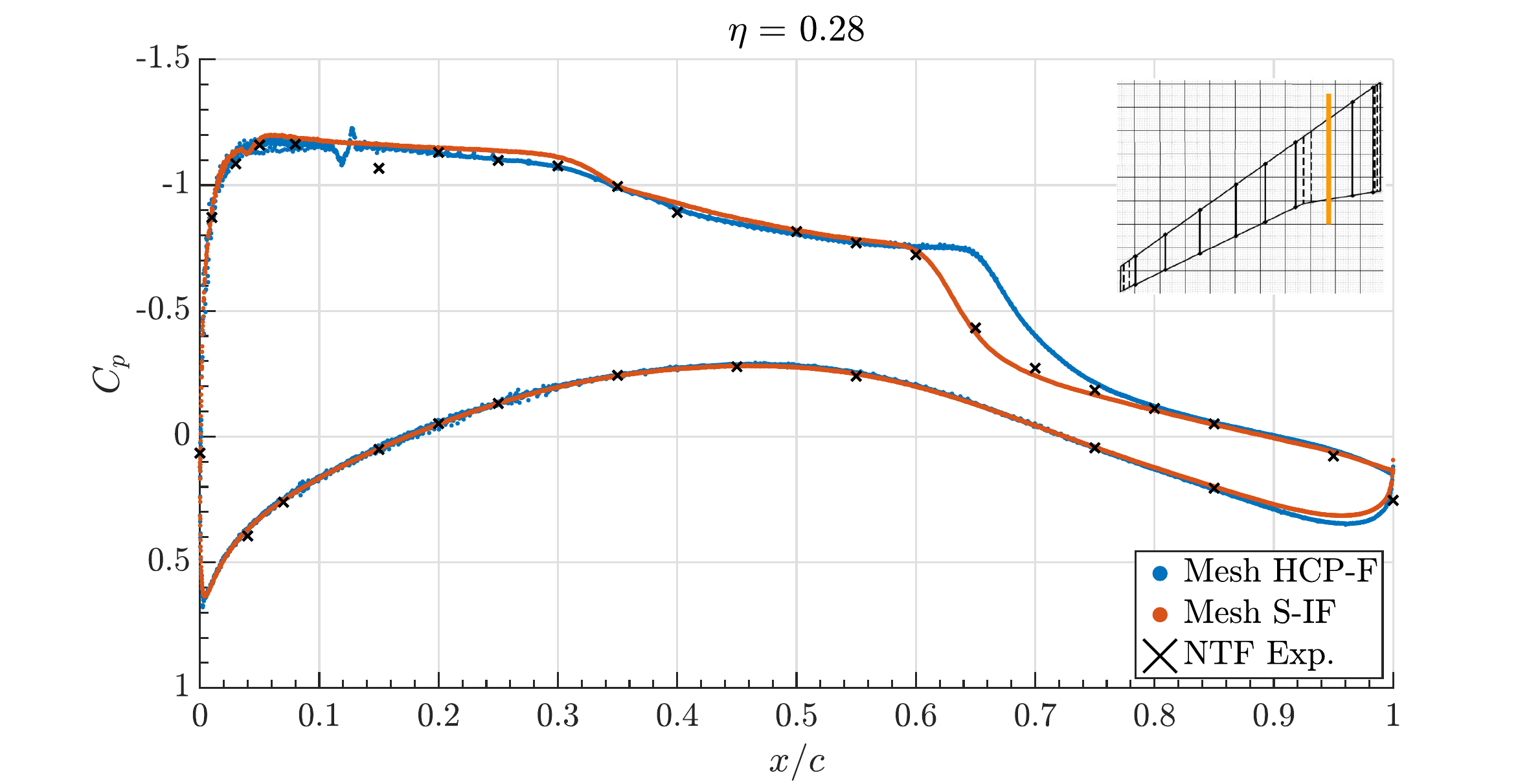}}\\ 
    \centering
    \subfloat[\label{fig:crm_cp_strand_c}]{\includegraphics[width=0.495\textwidth]{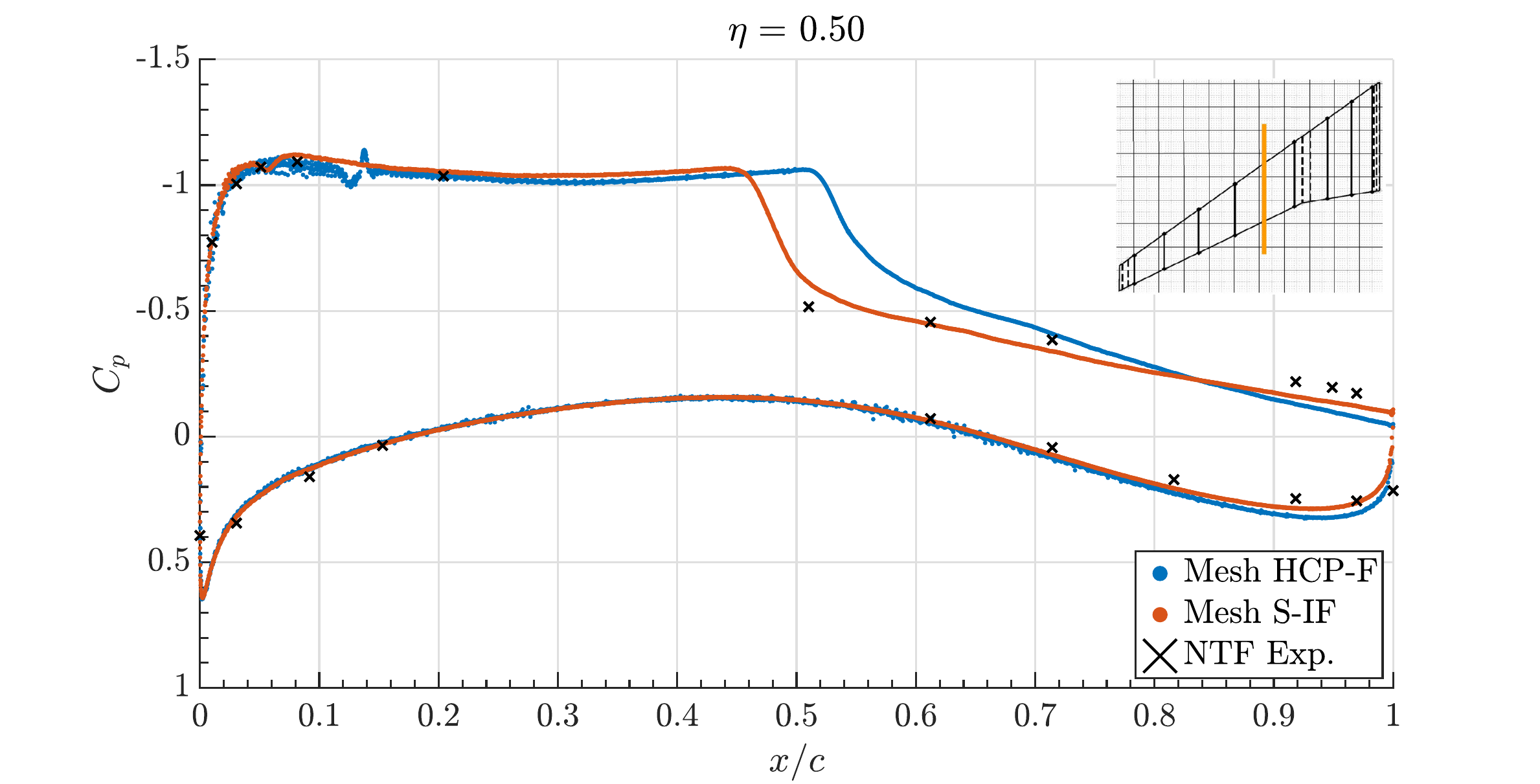}}
    \subfloat[\label{fig:crm_cp_strand_d}]{\includegraphics[width=0.495\textwidth]{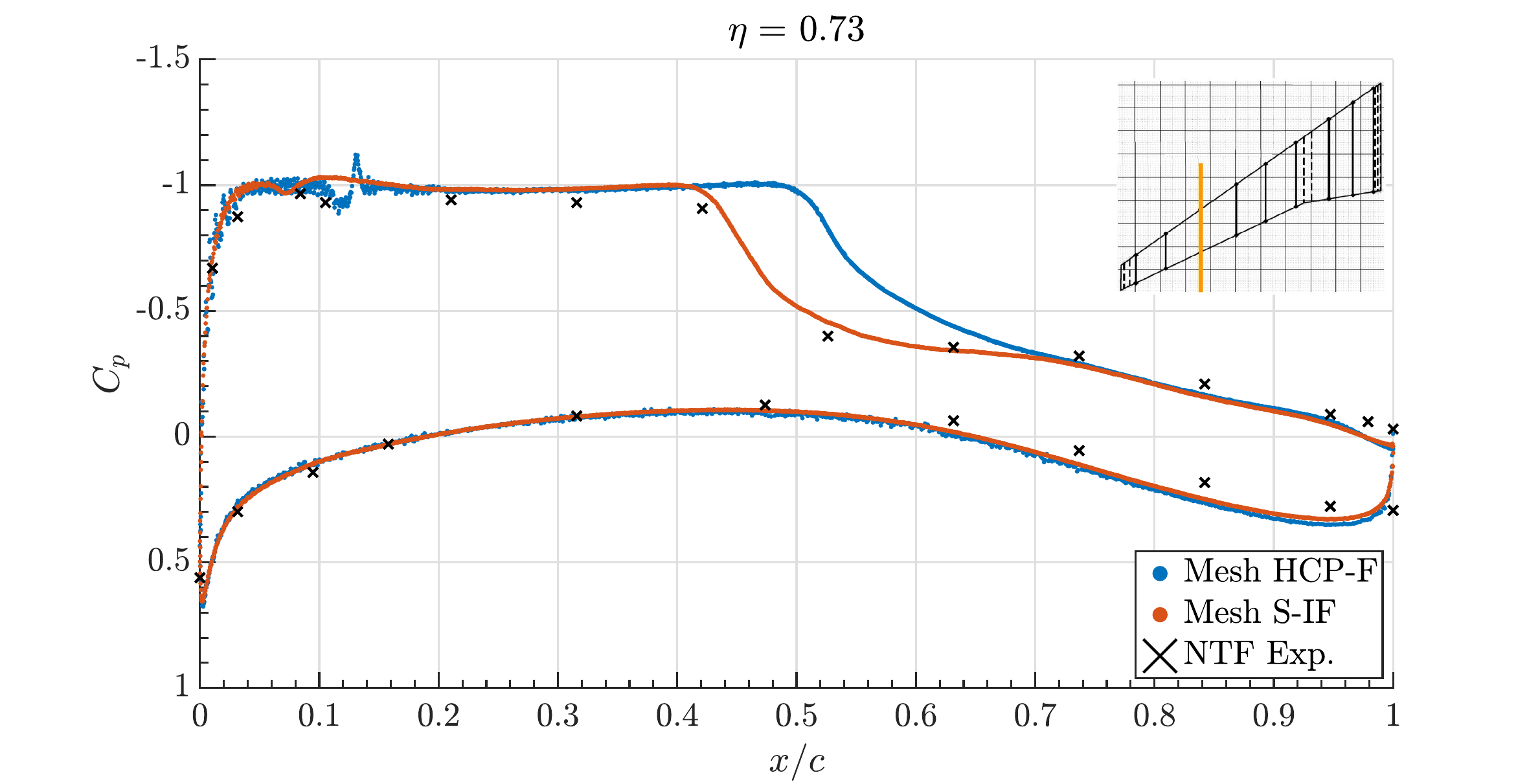}}
    
    \caption{Average static pressure measurements for the transonic CRM at an angle of attack of $\alpha = 4.0^{\circ}$ at four different stations along the span of the wing, ranging from (a) inboard to (b-c) mid-span to (d) outboard. The comparison includes results computed on a fine mesh of isotropic HCP grid elements ($\approx 570$ Mcv, Mesh HCP-F) and an isotropic strand mesh (S-IF) solution at the same near-wall resolution as the corresponding HCP mesh. The glitch in the HCP results is due to the presence of geometric trip dots in these calculations which are absent from the stranded cases.  \label{fig:crm_cp_strand}}
\end{figure}

The results of the isotropic stranded grid study are presented in Fig. \ref{fig:crm_forces_isostrand} and \ref{fig:crm_cp_strand}. The HCP solutions exhibit a non-monotonic response upon grid refinement (mesh HCP-M solution is worse in all force/moment metrics than the mesh HCP-C solution and then improves on mesh HPC-F before further improving on mesh HCP-XF); this has also been observed in the context of other smooth-body separation flows using the same gridding approach and solver \citep{agrawal2022non, whitmorebump}. On the contrary, for the stranded solutions, a monotonic improvement is observed. The solutions predicted using mesh S-IM offer improvements over mesh S-IC solutions in all metrics and S-IF improves further upon S-IM. A stark contrast can be observed in the surface pressure plots, where a simple change to the near-wall grid topology at fixed grid size (mesh S-IF vs. HCP-F) offered even up to $10\%$ improvement in the shock location prediction towards the wingtip. The accuracy of the prediction on the stranded mesh was improved relative to the HCP mesh at the same grid resolution across the span of the wing. The improved shock locations corroborate the improved accuracy observed in the prediction of integrated quantities ($C_L$, $C_D$, $C_M$) on isotropic strand meshes. 

While the stranding method provides tangible benefits over HCP meshes in monotonic approaches toward the solution, further predictive accuracy may be achievable via preferential refinement in the wall-normal direction. To explore the requirements for solution accuracy while maintaining monotonic grid convergence, an anisotropic stranding approach was explored next. 

\subsection{Anisotropic Strand}
\label{sec:anisotropic_strand}

Meshes S-AXC through S-AF were utilized in this subsection to study the effect of mesh anisotropy (by preferentially increasing wall normal refinement) on solution accuracy in comparison to isotropic stranded grids. Although the previous subsection compared isotropic hexagonally close-packed and stranded cells, it is possible within the strand mesh paradigm to selectively increase the refinement in wall-normal directions compared to the wall-parallel directions. Physically, this is consistent with the length scales of wall-bounded turbulence shrinking at a faster rate in the wall-normal direction than in the transverse directions in the near-wall region \citep{marusic2019attached}. 

\begin{figure}[!ht]
    \subfloat[\label{fig:crm_forces_moment_a}]{\includegraphics[width=0.495\textwidth]{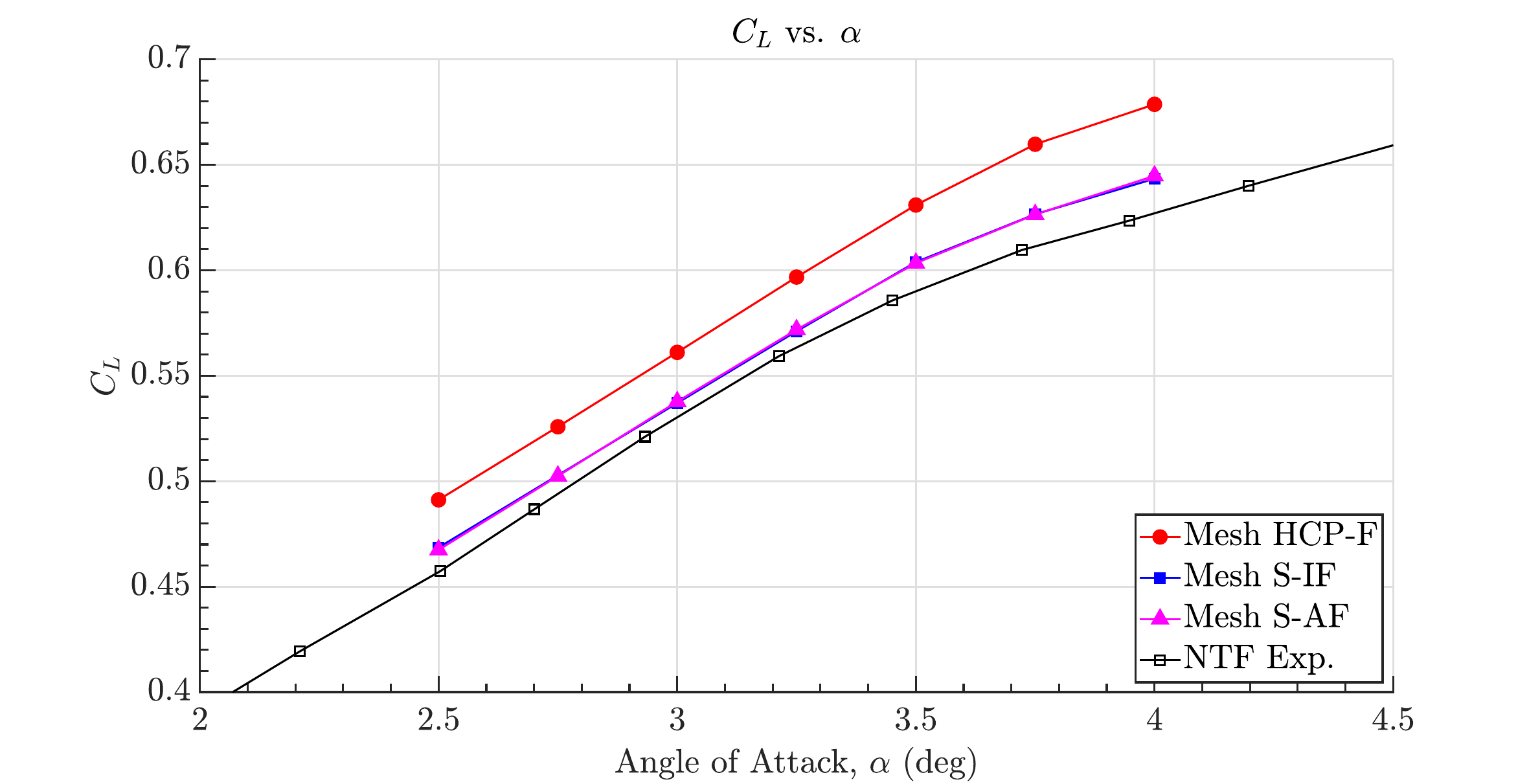}}
    \subfloat[\label{fig:crm_forces_moment_b}]{\includegraphics[width=0.495\textwidth]{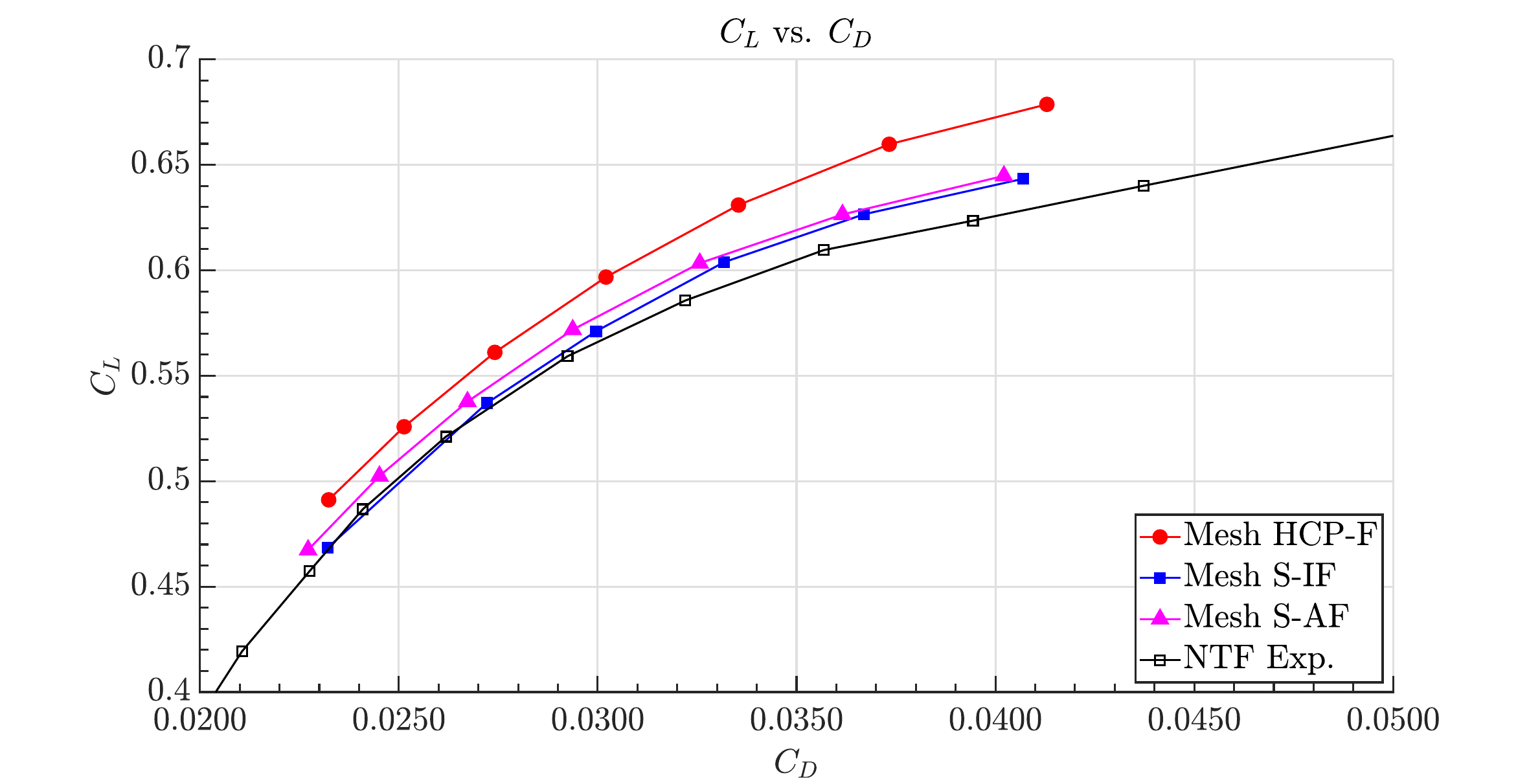}}\\ 
    \centering
    \subfloat[\label{fig:crm_forces_moment_c}]{\includegraphics[width=0.495\textwidth]{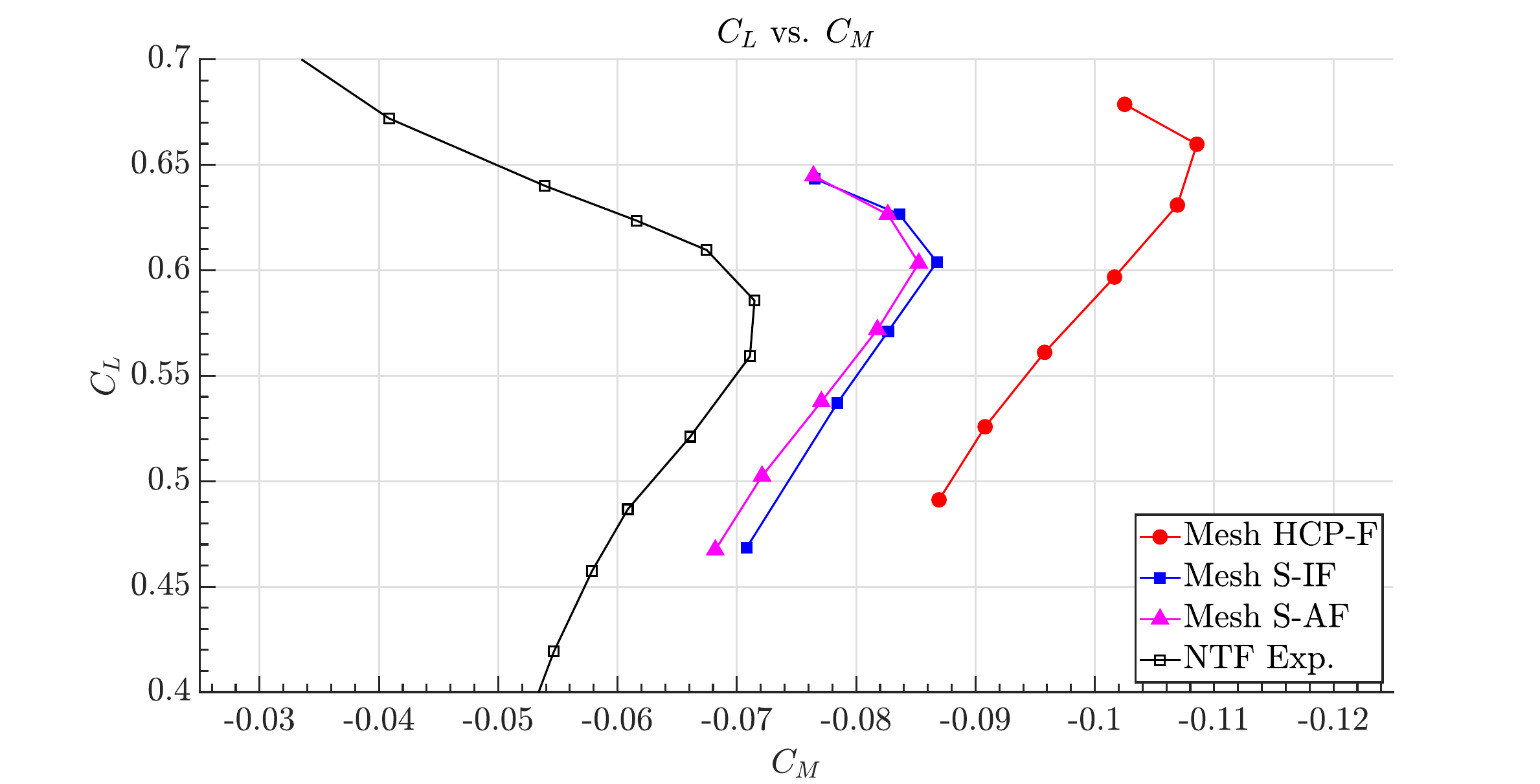}}
     
    \caption{Time-averaged forces/moments from the finest HCP (HCP-F), isotropic strand (S-IF), and anisotropic strand (S-AF) mesh LES simulations of the transonic CRM, including (a) the lift, (b) the drag polar and (c) the pitching moment.\label{fig:crm_forces_moment}}
\end{figure}

\begin{figure}[!ht]
    \subfloat[\label{fig:crm_iso_aniso_strand_a}]{\includegraphics[width=0.495\textwidth]{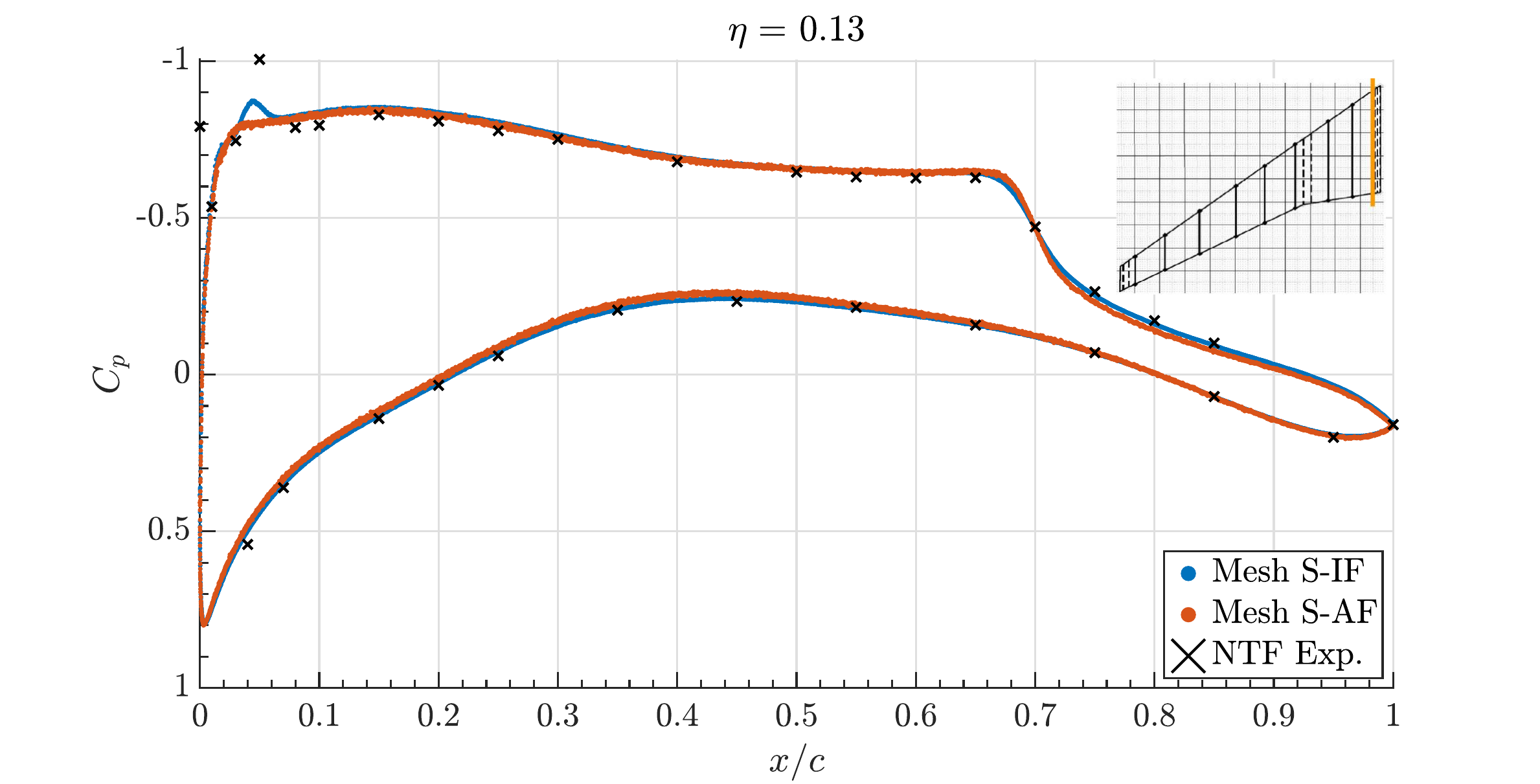}}
    \subfloat[\label{fig:crm_iso_aniso_strand_b}]{\includegraphics[width=0.495\textwidth]{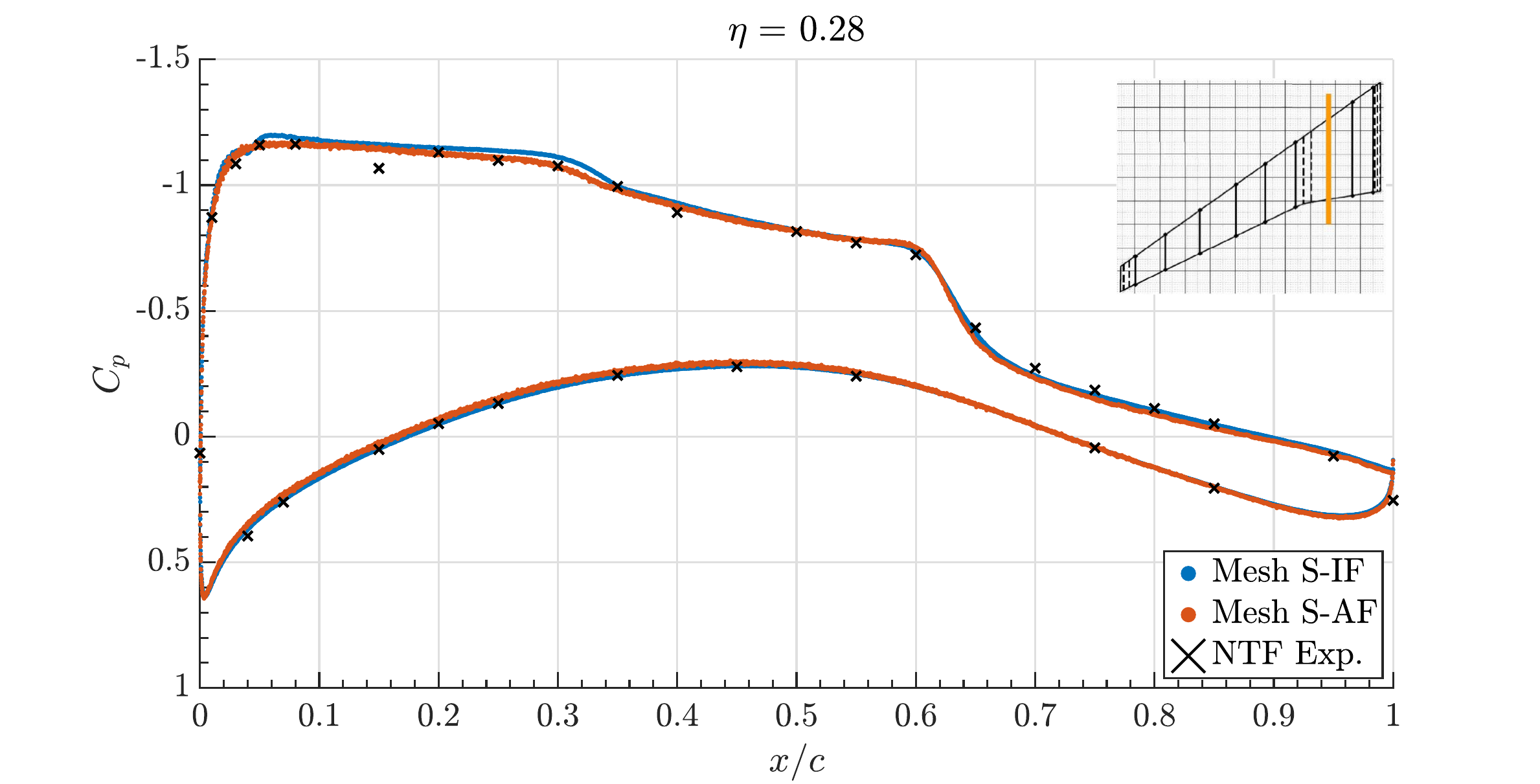}}\\ 
    \centering
    \subfloat[\label{fig:crm_iso_aniso_strand_c}]{\includegraphics[width=0.495\textwidth]{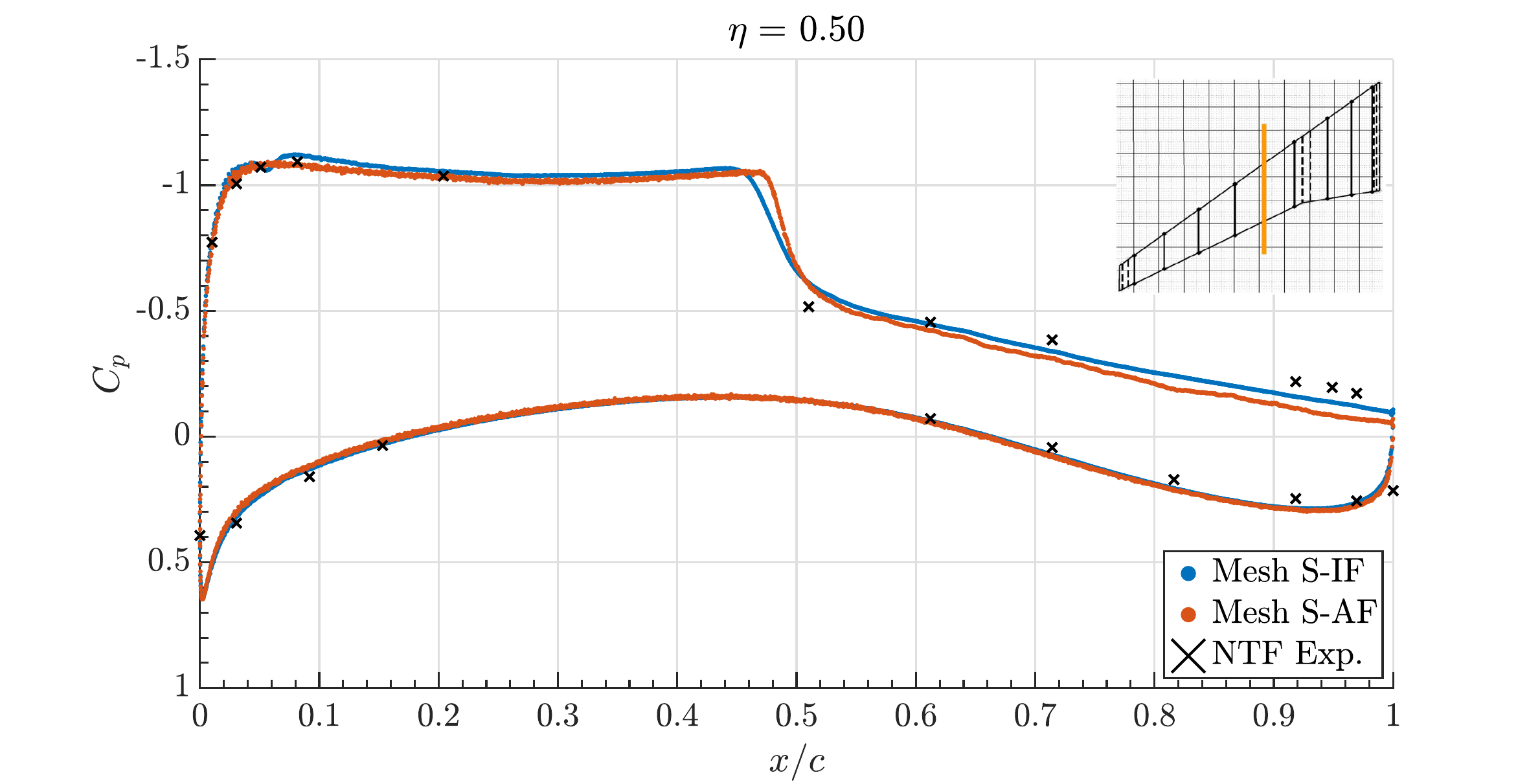}}
    \subfloat[\label{fig:crm_iso_aniso_strand_d}]{\includegraphics[width=0.495\textwidth]{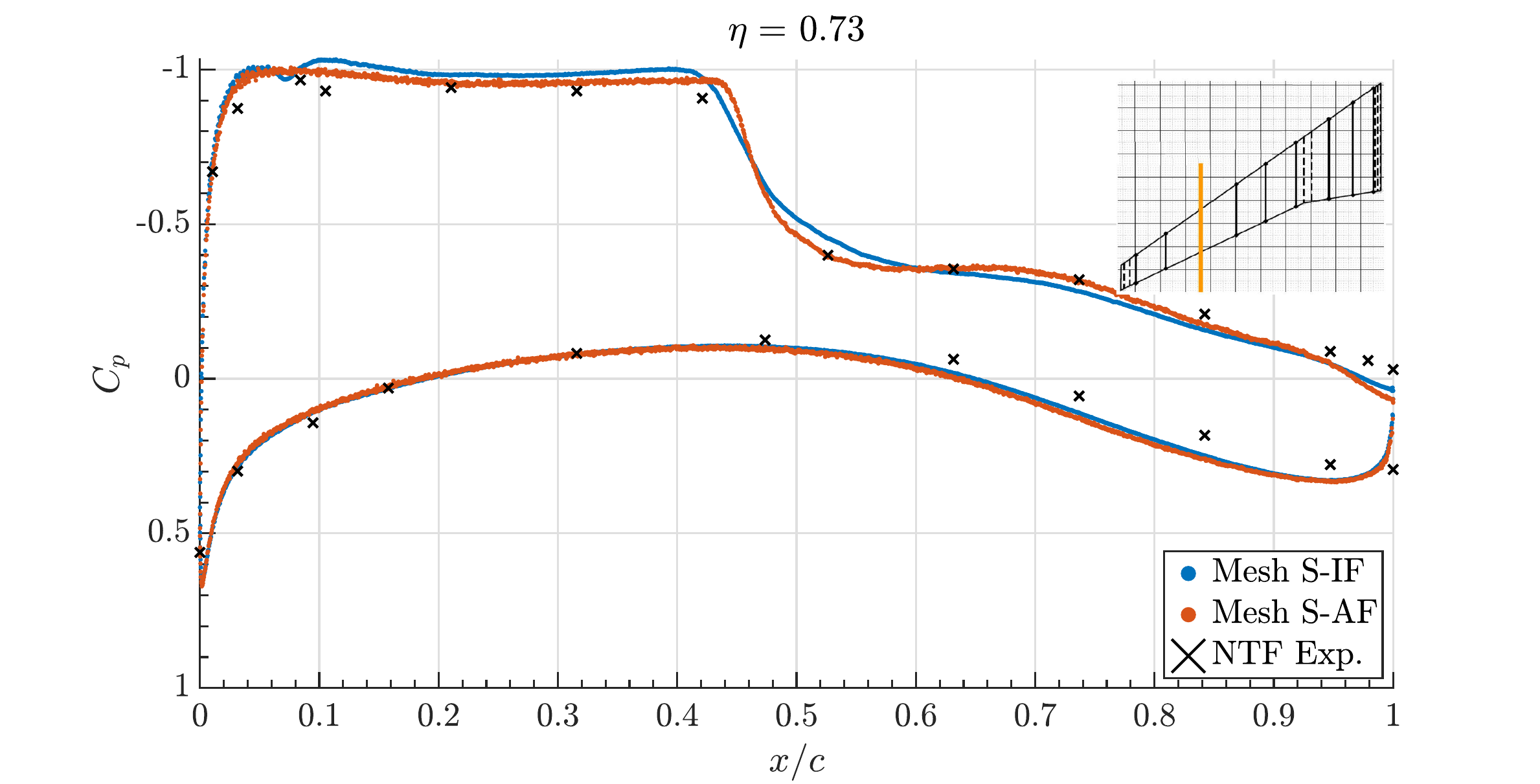}}
    
    \caption{Average static pressure measurements for the transonic CRM at an angle of attack of $\alpha = 4.0^{\circ}$ at four different stations along the span of the wing, ranging from (a) inboard to (b-c) mid-span to (d) outboard. The comparison includes results computed on a mesh of isotropic strand grid elements (Mesh S-IF) and an anisotropic strand mesh solution (Mesh S-AF). \label{fig:crm_iso_vs_aniso_strand_cp}}
\end{figure}

The number of cells in the stranded region had to increase from 5 in the isotropic strand case to 10 in the anisotropic strand case to maintain a limited wall-normal stretching ratio ($1.15$), and therefore no effort was made to make grid counts consistent between corresponding isotropic and anisotropic meshes. Instead, a given anisotropic mesh at coarse/medium/fine resolution can be interpreted as having identical wall-parallel resolution to the corresponding isotropic strand mesh (coarse/medium/fine) while having $4\times$ finer wall-normal resolution. Beyond an anisotropy of 4$:$1, a degradation in the accuracy of the solutions was observed, which is expected, as once the grids are sufficiently refined, the near-wall turbulent structures express anisotropy that is capped above certain limits, especially at increasingly high Reynolds numbers \citep{toosi2017anisotropic}. The stretching ratio of $1.15$ was chosen as the suitable upper limit for these simulations based on unpublished NACA 0012 simulations in the same flow regime. It is the case that additional accuracy may be achievable in the anisotropic stranded results from further limiting the maximum wall-normal stretching ratio, but this would require introducing more cells in the stranded region and, therefore, drive additional costs. These internal studies on a simple transonic NACA 0012 configuration showed that the present choice of anisotropic strand grid settings presented a reasonable compromise between the accuracy and affordability on anisotropic grids. Figures \ref{fig:crm_forces_moment} and \ref{fig:crm_iso_vs_aniso_strand_cp} show the results of this sensitivity exploration to grid topology. The anisotropic stranded meshing offers advantages over HCP, particularly as the angle of attack increases and a significant shock-induced separation is observed. However, the results from the best-practice strand meshes still do exhibit a slight delay in the pitching moment break (where the $C_M$ curve abruptly becomes less negative) in Fig. \ref{fig:crm_forces_moment_c} that is associated with a delay in the onset of shock-induced separation. This delay is, however, significantly less than on the HCP-F mesh.

\begin{figure}[!ht]
    \subfloat[\label{fig:crm_cf_hcp}]{\includegraphics[width=0.495\textwidth]{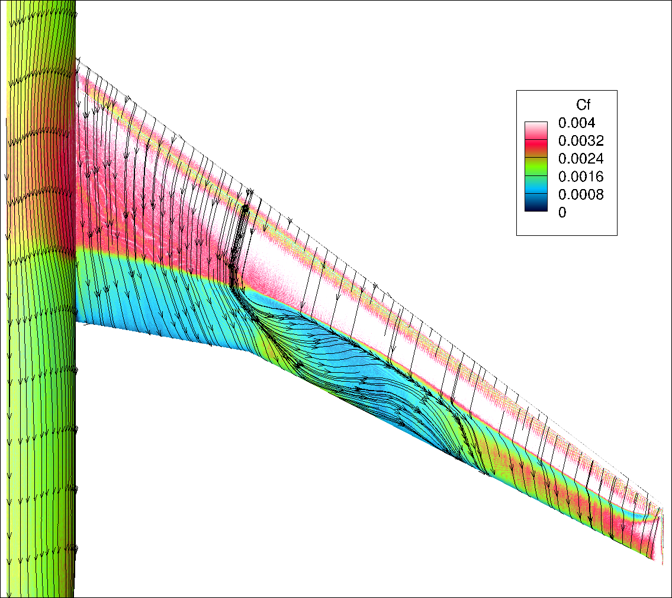}}
    \subfloat[\label{fig:crm_cf_strand}]{\includegraphics[width=0.495\textwidth]{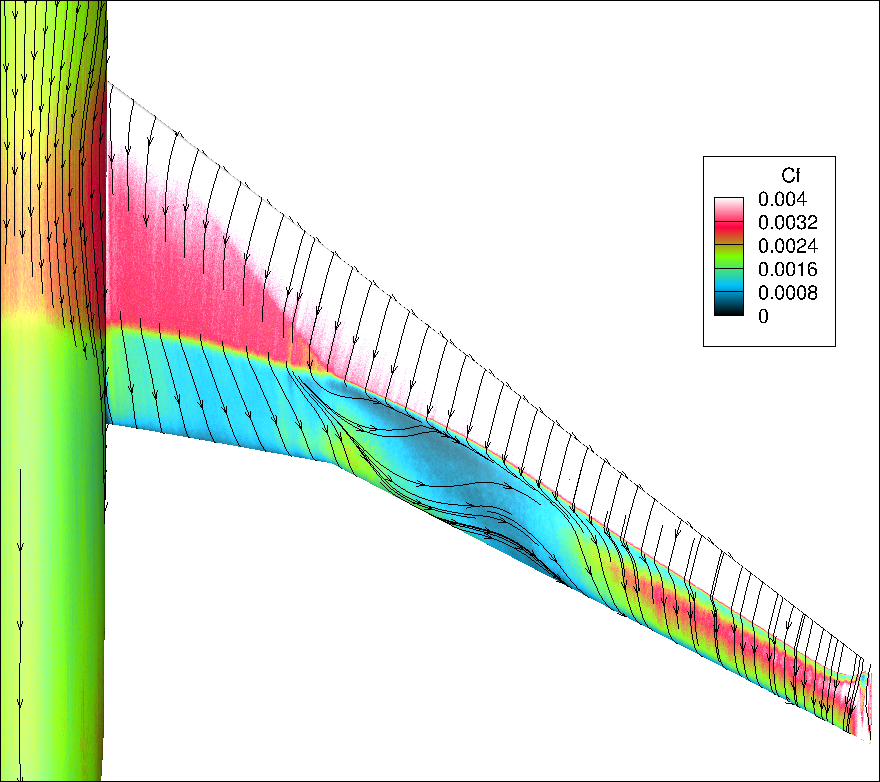}}\\ 
    \centering
    \subfloat[\label{fig:crm_cf_rans}]{\includegraphics[width=0.495\textwidth]{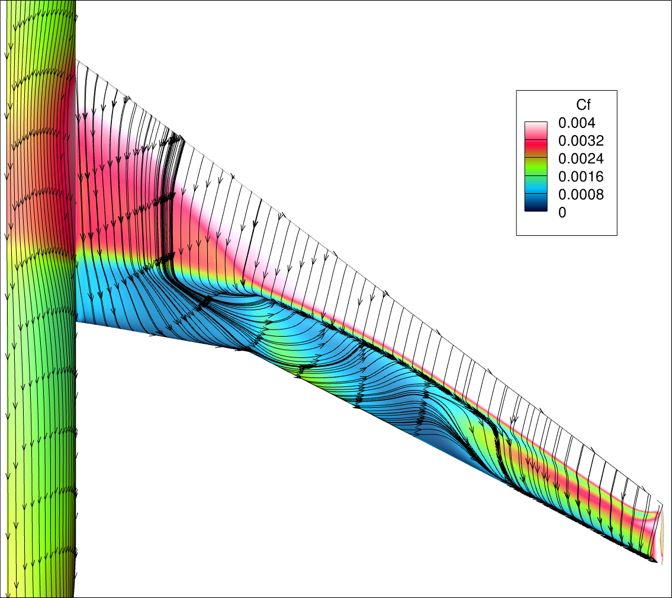}}

    \caption{Average surface skin friction magnitude and surface streamlines from three simulations, including LES simulations with solutions on (a) mesh HCP-F, (b) mesh S-AM, and from (c) a RANS SA-QCR simulation using the OVERFLOW solver \citep{sclafani2014drag}. \label{fig:crm_cf_streamlines}}
\end{figure}

The surface flow patterns predicted by each mesh topology are shown in Fig. \ref{fig:crm_cf_streamlines}. In the absence of experimental reference data for the surface flow, an OVERFLOW RANS solution is compared to the LES solutions. The RANS data represents a best practice result from the contribution of the Boeing Advanced Concepts Group to the Sixth Drag Prediction Workshop (DPW6) \citep{sclafani2014drag}. The RANS simulation was performed with the Spalart-Allmaras turbulence model with Quadratic Constitutive Relation (SA-QCR) without any trip dots but run in a fully turbulent mode. It is noteworthy that the SA-QCR model has been known to perform well in this flow, particularly in the wing juncture region \citep{tinoco2018summary}. The shock that develops in this flow induces a pocket of flow separation in the mid-span region, as visible in the surface streamline patterns (by the regions in which the streamlines become parallel to the trailing-edge sweep). Both the HCP and the anisotropic stranded mesh solutions predict slightly smaller regions of separation than the RANS solution.

While it is clear from the results shown in figures \ref{fig:crm_forces_moment}-\ref{fig:crm_iso_vs_aniso_strand_cp} that the solutions obtained on fine anisotropic stranded grids systematically outperform those computed on fine HCP grids, it is not as clear that the preferential wall-normal refinement of anisotropic stranded meshes offer significant additional benefit to the accuracy of the calculations relative to isotropic stranded meshes, especially when computational cost is taken into account. For instance, the finest isotropic strand mesh (mesh S-IF) was refined by a factor of four in the wall-normal direction while keeping the stream/span-wise resolution unaltered, resulting in mesh S-AF. As can be seen in Table \ref{tab:iso_strand_meshes} and \ref{tab:aniso_strand_meshes}, this leads to a relatively appreciable growth in the total cell count while also resulting in a significant drop to the time step required for stability, making the mesh S-AF calculations $\approx 4.5 \times$ more expensive than their mesh S-IF counterparts. The cost measure includes both grid count and time step cost implications as measured in GPU hours and is shown in Table \ref{tab:cost}. In contrast, the fine isotropic strand calculation incurs only a slight cost penalty relative to its fine HCP counterpart due to the slightly increased grid count and has a comparable timestep due to having the same near-wall mesh length scale (an approx. $10-15\%$ penalty in time step is incurred due to stability robustness considerations on the stranded mesh), but offers significant accuracy benefits.

\begin{table}[ht!]
\begin{center}
    \caption{Computational cost summary for various grids using the charLES solver for the CRM configuration at angle of attack of $4.00^\circ$. ``Wall Time to Solution'' is 100 flow passes based on MAC. NVIDIA V100 GPUs on the Oak-Ridge Summit cluster and on internal Boeing clusters were used in all simulations.}
    \begin{tabular}{cccccl}
        \cline{1-5}
        \cline{1-5}
        \textbf{Grid} & \textbf{\begin{tabular}[c]{@{}c@{}}Grid Size \\ (Mcv)\end{tabular}} & \textbf{\begin{tabular}[c]{@{}c@{}}Wall Time to \\ Solution\end{tabular}} & \textbf{\begin{tabular}[c]{@{}c@{}}Number of \\ GPU's\end{tabular}} & \textbf{\begin{tabular}[c]{@{}c@{}}GPU \\ Hours\end{tabular}}\\ \cline{1-5}
        HCP-F & 570 & 3 days 8 hr & 40 & 3,200 \\ 
        S-IF & 670 & 1 day 11 hr & 128 & 4,480 \\ 
        S-AM & 310 & 1 day 1 hr & 128 & 3,200 \\ 
        S-AF & 1100 & 5 days 21 hr & 128 & 18,050 \\ \cline{1-5}\cline{1-5}
    \end{tabular}
    \label{tab:cost}
\end{center}
\end{table}

\begin{figure}[!ht]
    \subfloat[\label{fig:crm_aniso_strand_a}]{\includegraphics[width=0.495\textwidth]{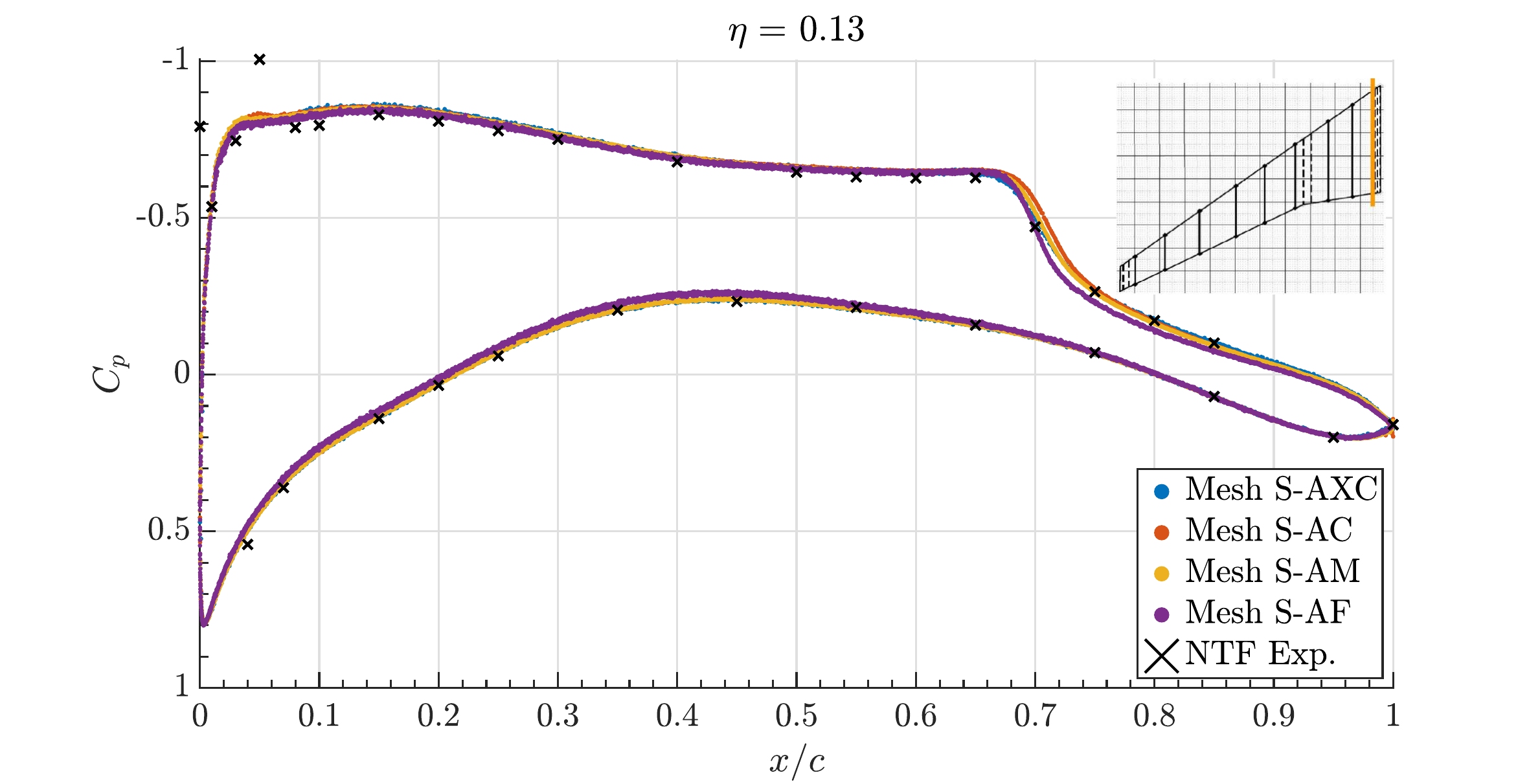}}
    \subfloat[\label{fig:crm_aniso_strand_b}]{\includegraphics[width=0.495\textwidth]{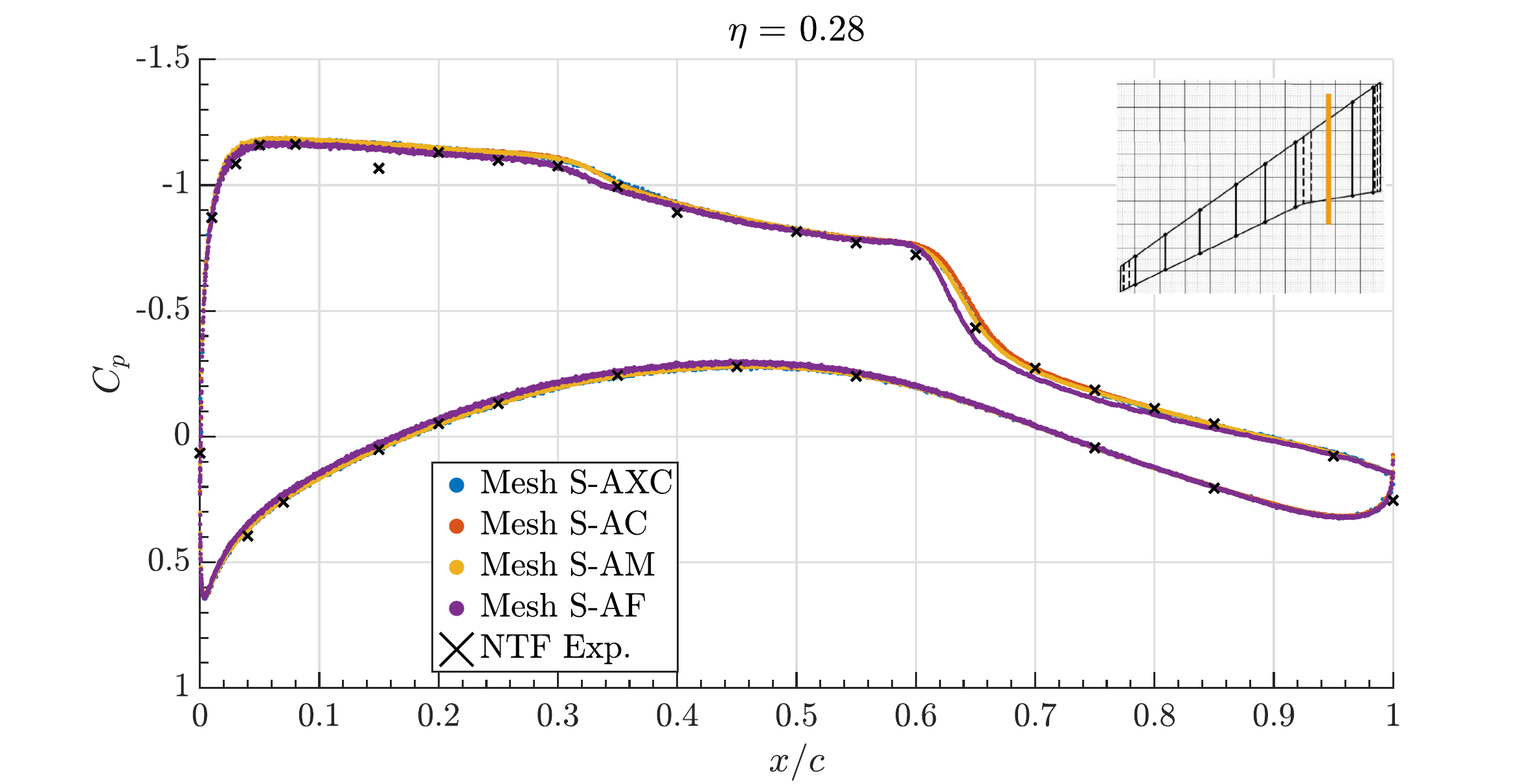}}\\ 
    \centering
    \subfloat[\label{fig:crm_aniso_strand_c}]{\includegraphics[width=0.495\textwidth]{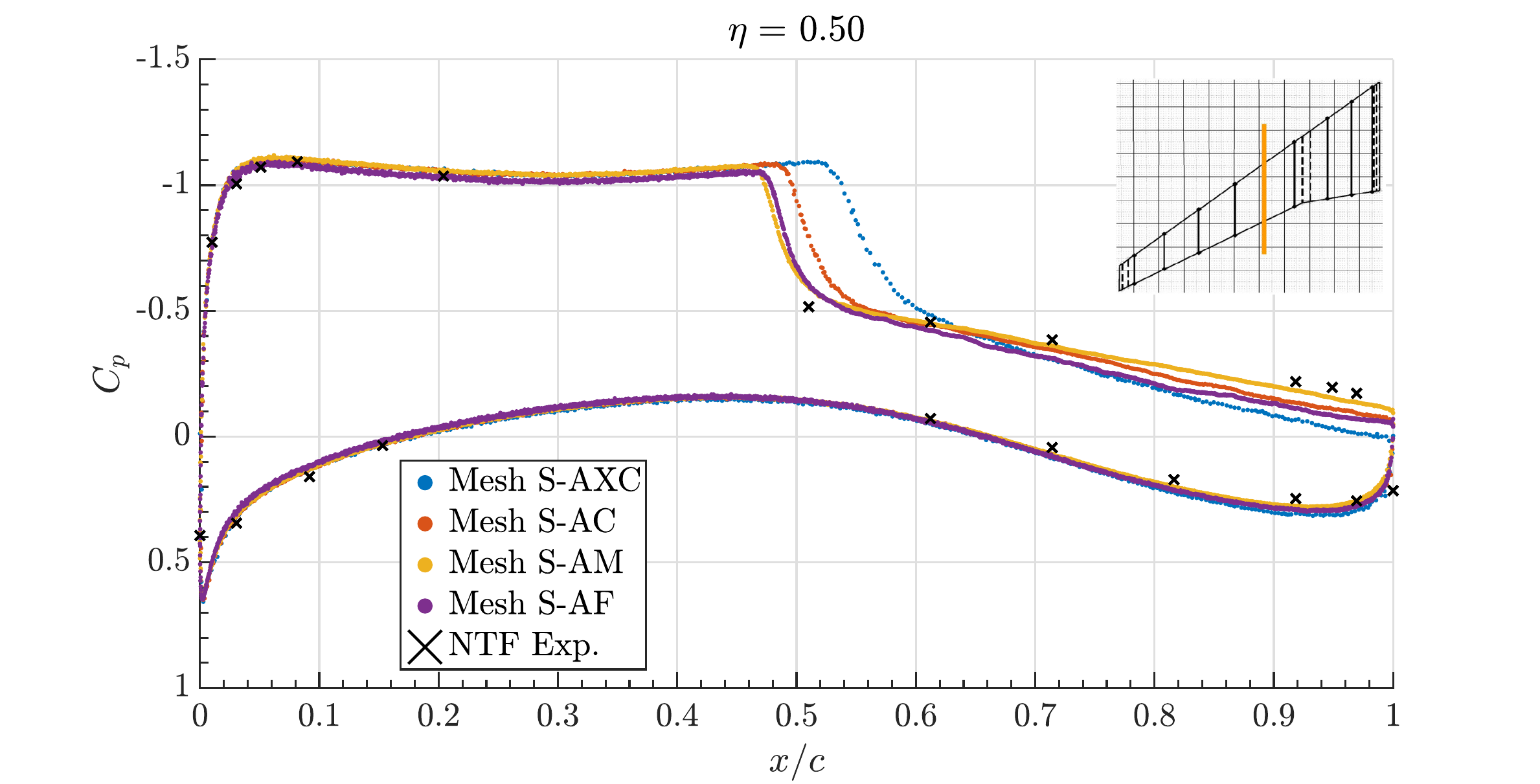}}
    \subfloat[\label{fig:crm_aniso_strand_d}]{\includegraphics[width=0.495\textwidth]{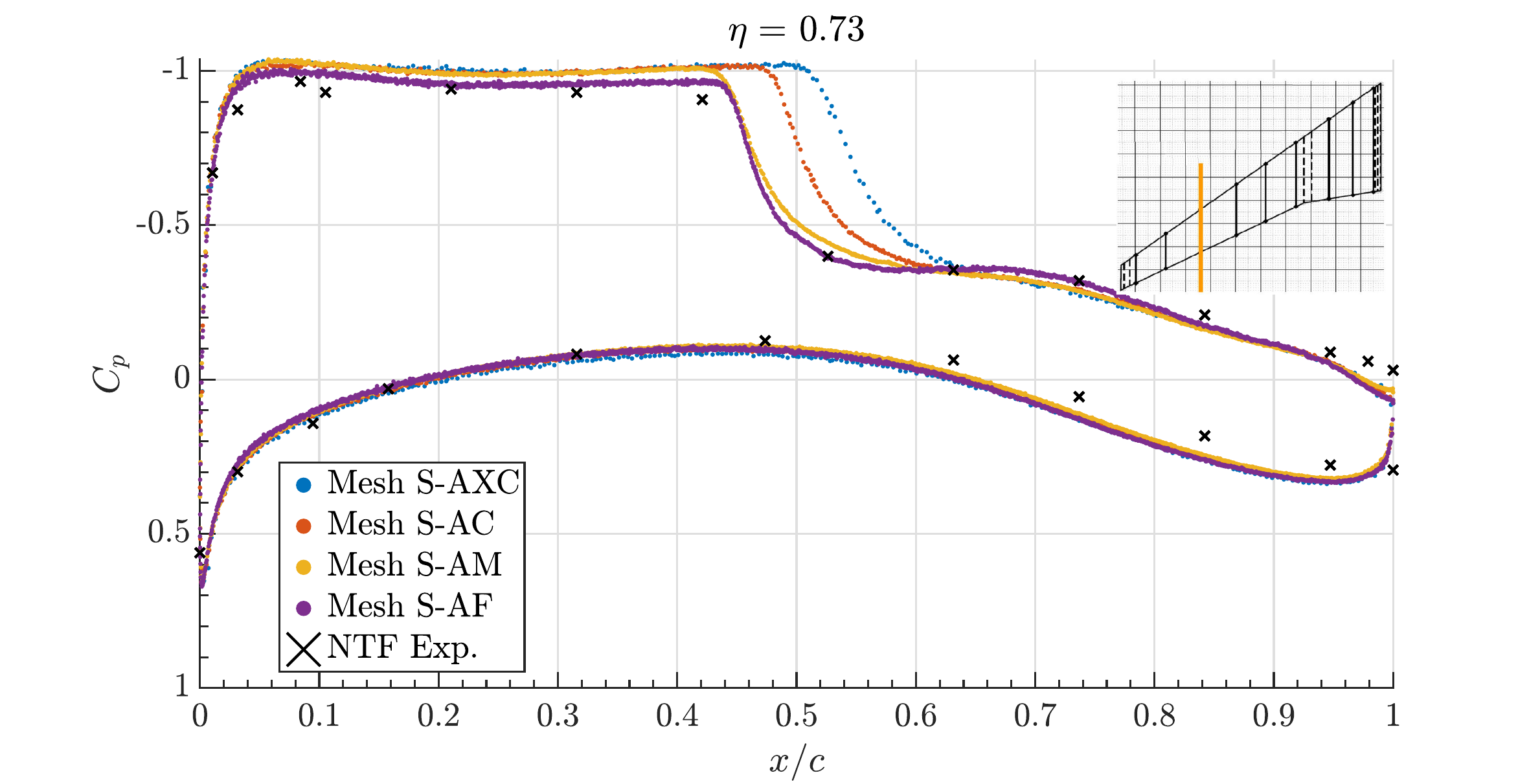}}
    
    \caption{Average static pressure measurements for the transonic CRM at an angle of attack of $\alpha = 4.0^{\circ}$ at four different stations along the span of the wing, ranging from (a) inboard to (b-c) mid-span to (d) outboard. The comparison includes results computed on three meshes containing anisotropic stranded boundary layer elements sequentially refined by a factor of 2 uniformly in all three coordinate directions ($\approx 30-1100$ Mcv, Meshes S-AXC through S-AF). \label{fig:crm_aniso_strand_cp}}
\end{figure}

Inspired by the observations made in Figs. \ref{fig:crm_forces_isostrand} and \ref{fig:crm_iso_vs_aniso_strand_cp}, an additional set of numerical experiments were performed to study the grid convergence behavior of anisotropic stranded meshes. For this purpose, a sequence of four grids was simulated, which were each refined by a factor of 2 uniformly in all three coordinate directions (meshes S-AXC through S-AF). Fig. \ref{fig:crm_aniso_strand_cp} shows the sectional pressure predictions from these four meshes at the 4-degree angle of attack condition. The grid sensitivities are preferentially concentrated towards the wingtip (potentially owing to the finer geometric length scales present there associated with the wing taper). Importantly, at the outboard stations (Fig. \ref{fig:crm_aniso_strand_c} and \ref{fig:crm_aniso_strand_d}), it is observed that the solutions converge monotonically towards the experimental predictions. Furthermore, near grid-insensitivity of the solution across two consecutive grids has been achieved between mesh S-AM and S-AF. This result demonstrates that LES calculations that use anisotropic stranded boundary layer elements have the potential to provide accurate, grid-converged solutions on fine meshes that monotonically approach the test data, all of which are important criteria for the more widespread adoption of wall-modeled LES technology in the aerospace industry. 

An alternative perspective regarding the comparison of anisotropic stranded meshes against their isotropic counterparts is arrived at when considering the question of cost at fixed accuracy. For this purpose, the S-AM mesh is included in Table \ref{tab:cost} and compared against mesh S-IF. Mesh S-AM is $2\times$ coarser in the wall-parallel directions, but $2\times$ finer in the wall-normal direction than mesh S-IF, but achieves nearly the same accuracy (as can be deduced from Fig. \ref{fig:crm_forces_moment} - \ref{fig:crm_aniso_strand_cp}) at a discount in cost of $\approx 25\%$. Though we do not directly overlay the S-AM solution along with S-IF, we see that the S-AM solution is essentially converged to the S-AF solution in the $C_p$ plot in Fig. \ref{fig:crm_aniso_strand_cp}, which itself yielded a solution that was very comparable to that obtained on the S-IF mesh. These observations lead to two key conclusions regarding the use of LES grids that contain wall-normal stretching: the ability to preferentially refine the grid in the wall-normal direction provides a viable route to 1) achieving accurate solutions at reduced cost compared to isotropic meshes (because wall-parallel resolution can be traded for wall-normal resolution in a way that lowers overall solution cost) and 2) achieving ``grid-converged'' LES solutions (as can be seen between solutions on S-AM and S-AF which are nearly identical in their prediction of $C_p$). \\

Overall, a significant benefit is derived from the use of isotropic strand meshes over their HCP counterparts in the prediction of lift, drag, pitching moment, and sectional pressures for this flow, while the main advantage of anisotropic stranded meshes (over their isotropic counterparts) is achieved not as much in terms of accuracy but in terms of cost and grid convergence considerations. One exception to note regarding this point is that the anisotropic strand calculation (most pronounced in Fig. \ref{fig:crm_iso_aniso_strand_a}) does not feature the leading edge ``blip'' in $C_p$ associated with a small spurious ``quasi-laminar'' separation that occurs on isotropic stranded meshes. This is likely due to the improved characterization of the thin leading edge boundary layer on the anisotropic stranded mesh afforded by the increased resolution in the wall-normal coordinate direction, though this claim was not investigated further. These results confirm that the stranded meshing paradigm offers significant advantages over HCP meshes in both its grid refinement behavior and superior accuracy. The computing capacity (as detailed in Table \ref{tab:cost}) required for these grid resolutions exists within the aerospace industry today, and hence LES is, based on our experience, poised to make an impact as a design tool in this flow regime. \\ \\

\section{Initial Buffet Prediction}

\label{sec:buffet}

Transonic buffet is a corner of the flight envelope flow condition in which, either by increasing the angle of attack or Mach number, unsteady flow associated with large shock oscillations develops over the aircraft wing. The phenomenon has been studied extensively using Unsteady RANS (URANS) and analytical techniques \cite{crouch2019global} in the context of a canonical NACA 0012 flow. However, limited data concerning buffet onset is available for a full-scale aircraft. The onset of buffet in aircraft design and subsequent certification is typically identified as the point at which an accelerometer placed in the cockpit experiences a minimum of $10\%$ oscillations in lift about the mean in the vertical direction. The point at which a $10\%$ oscillation about the mean lift value computed by the LES is observed is used as the criterion for identification of buffet onset in this work. This approach neglects the vibration contributions to buffet in realistic aircraft flows but is the best that can be achieved from rigid CFD calculations such as those pursued in this paper.

\begin{figure}[!ht]
\begin{center}
\includegraphics[width=1\textwidth]{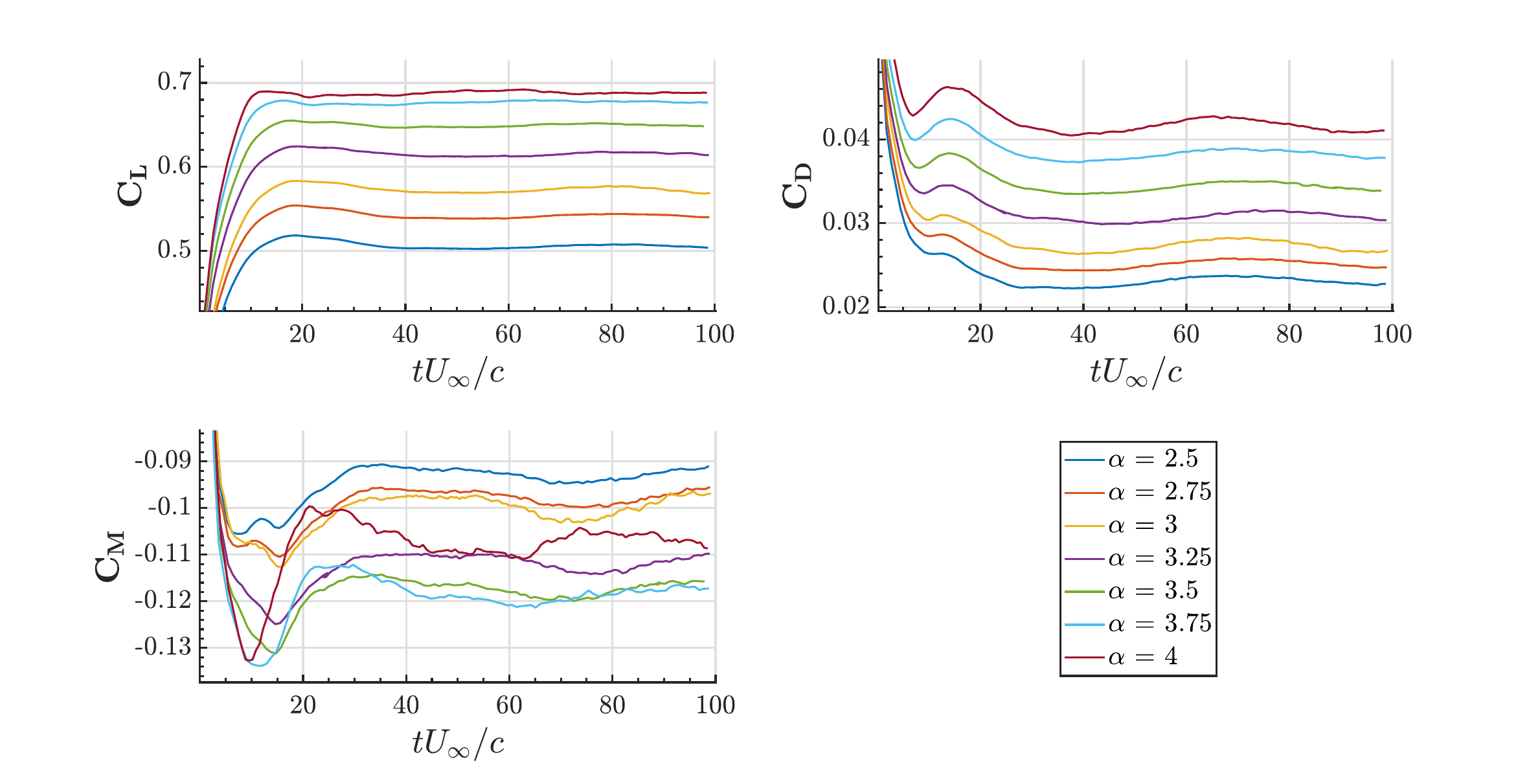}
\caption{Sample time history of the forces/moments from an HCP mesh LES simulation (Mesh HCP-F) of the Transonic CRM, including the lift, drag, and pitching moment. The calculations presented in this report are run for 100 flow pass times based on the mean aerodynamic chord and averaging takes place over the final 50 flow passes. \label{fig:crm_force_time_history}}
\end{center}
\end{figure}

Figure \ref{fig:crm_force_time_history} shows the instantaneous force and moment time histories from the HCP-F mesh calculation discussed in the previous sections. Clearly, over the range of angles of attack considered, the intensity of force oscillations required to trigger buffer onset was not observed (since significant oscillations about the mean are not observed). The pitching moment $C_M$ shows more sensitivity to flow separation than the integrated lift, $C_L$. For $C_M$, some meaningful oscillations can be observed at an alpha of $4.00^{\circ}$, which can be interpreted as a precursor to buffet. This data suggests that while buffet onset may be within the predictive scope of LES, a more controlled test case is needed to explore this phenomenon further, as data availability for the CRM is insufficient to adequately characterize this phenomenon. 

\begin{figure}[!ht]
\begin{center}
\includegraphics[width=0.60\textwidth]{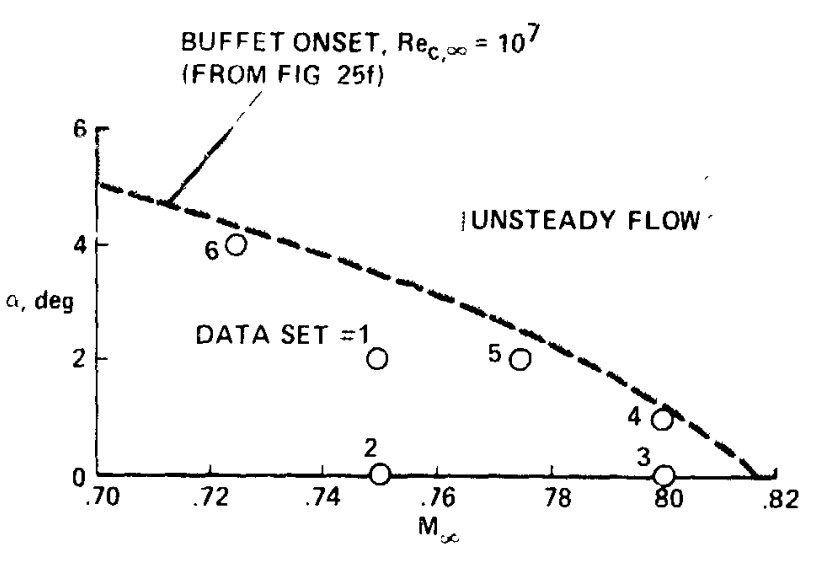}
\caption{Buffet boundary diagram reproduced from Ref. \cite{mcdevitt1985static}. Buffet can be achieved either by increasing the angle of attack or by increasing the Mach number. \label{fig:mcdevitt_test_matrix}}
\end{center}
\end{figure}

\begin{figure}[!ht]
\begin{center}
\includegraphics[width=0.85\textwidth]{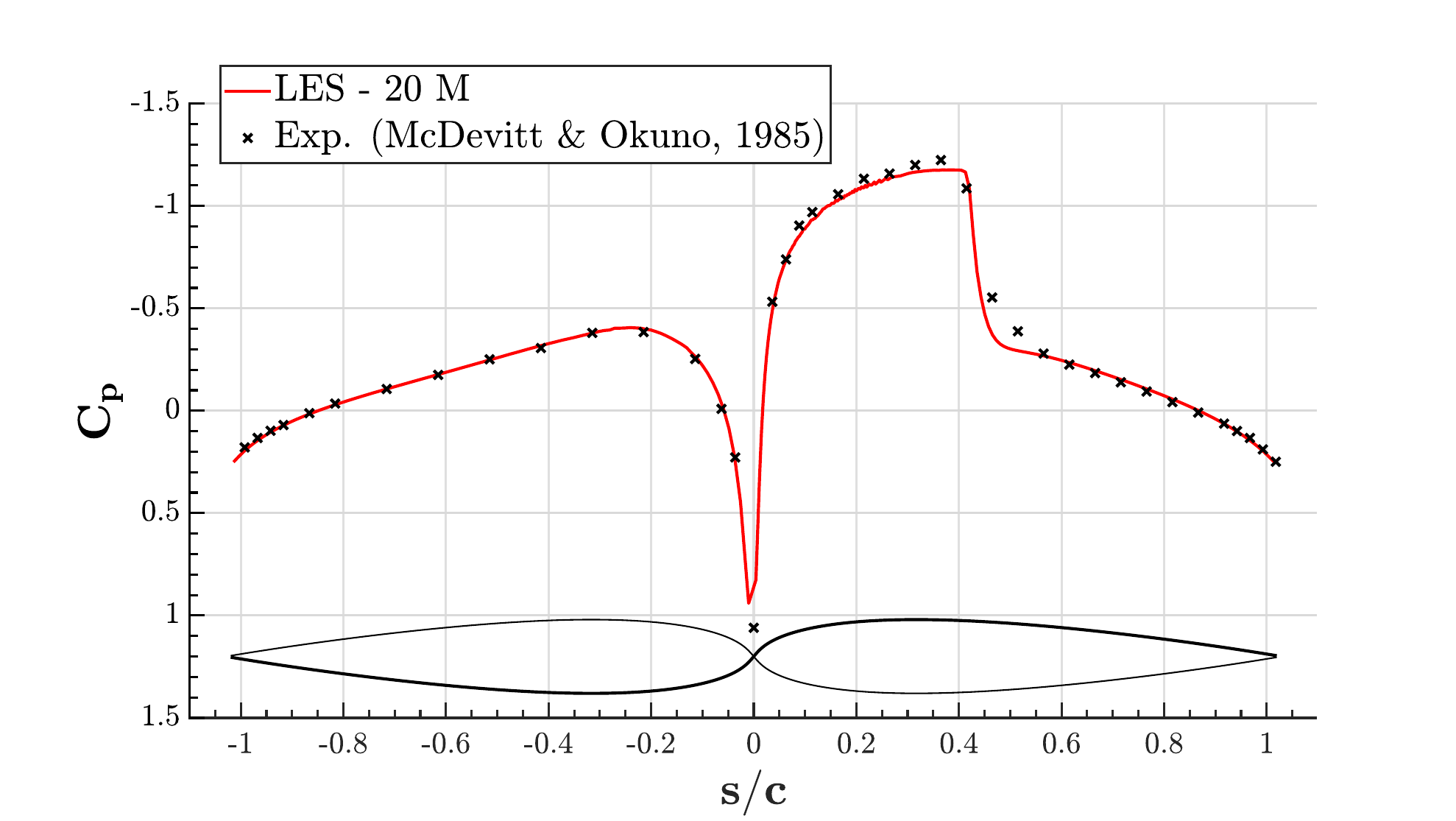}
\caption{Centerline pressure prediction for the NACA 0012 flow by LES at M = 0.75, Re = 10M, and $\alpha = 2^{\circ}$, before the onset of buffet.  \label{fig:mcdevitt_static_alpha}}
\end{center}
\end{figure}

For these reasons, the buffeting flow over the transonic NACA 0012 is considered. The experimental observations of McDevitt and Okuno \cite{mcdevitt1985static} serve as a baseline for this assessment. A key contribution of this work was to experimentally establish a buffet boundary envelope for the NACA 0012 configuration at a range of Mach numbers $Ma = 0.70-0.82$. This result is reproduced in Fig. \ref{fig:mcdevitt_test_matrix}, and this diagram shows that buffet can be achieved by either increasing the angle of attack at a fixed Mach number or by increasing the Mach number at a fixed angle of attack or some combination of the two. Also provided in the report were data at several flow conditions before the onset of buffet where the flowfield was ``static''. This static data can provide a useful benchmark for the calibration of flow solvers and for this manuscript. The static $\alpha = 2^{\circ}$ case at $Ma = 0.75$ is used to establish confidence in the LES predictions before the onset of buffet. Fig. \ref{fig:mcdevitt_static_alpha} shows that excellent agreement between the LES and the experimental results in the predictions of the shock location and strength were achieved for this case on a mesh of 20 Mcv, which is consistent in its refinement approach to mesh HCP-F on the transonic CRM. Having achieved excellent prediction of a static alpha condition, further increasing the angle of attack at a Mach number of $0.75$ was pursued to push closer to the buffet boundary.

\begin{figure}[!ht]
\begin{center}
\includegraphics[width=0.85\textwidth]{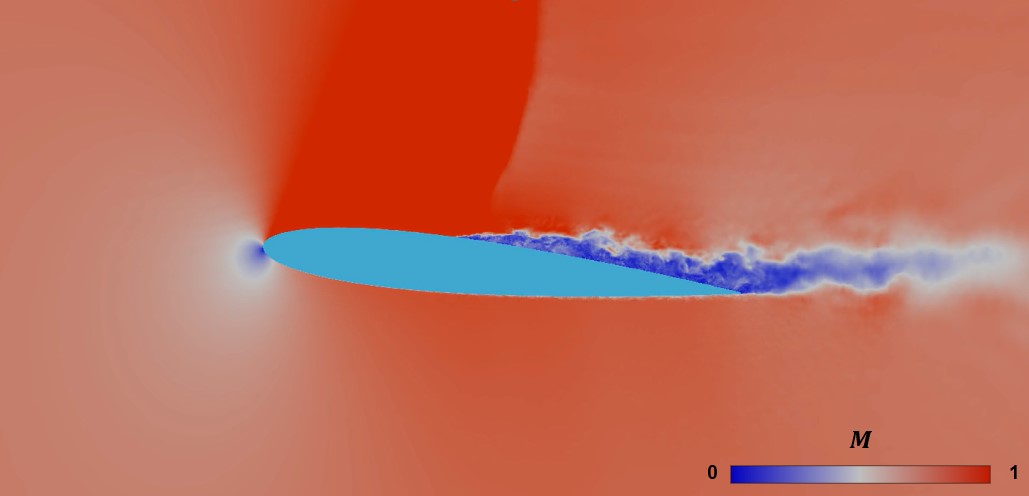}
\caption{Center plane Mach number contours of the rotating mesh NACA 0012 simulation. The airfoil was dynamically rotated between $1^{\circ}-5^{\circ}$ with time to identify the angle at which significant force oscillations were observed. Depicted here is an intermediate angle between $1^{\circ}-5^{\circ}$ showing the rotating airfoil. \label{fig:0012_rotating_mach}}
\end{center}
\end{figure}

\begin{figure}[!ht]
\begin{center}
\includegraphics[width=0.85\textwidth]{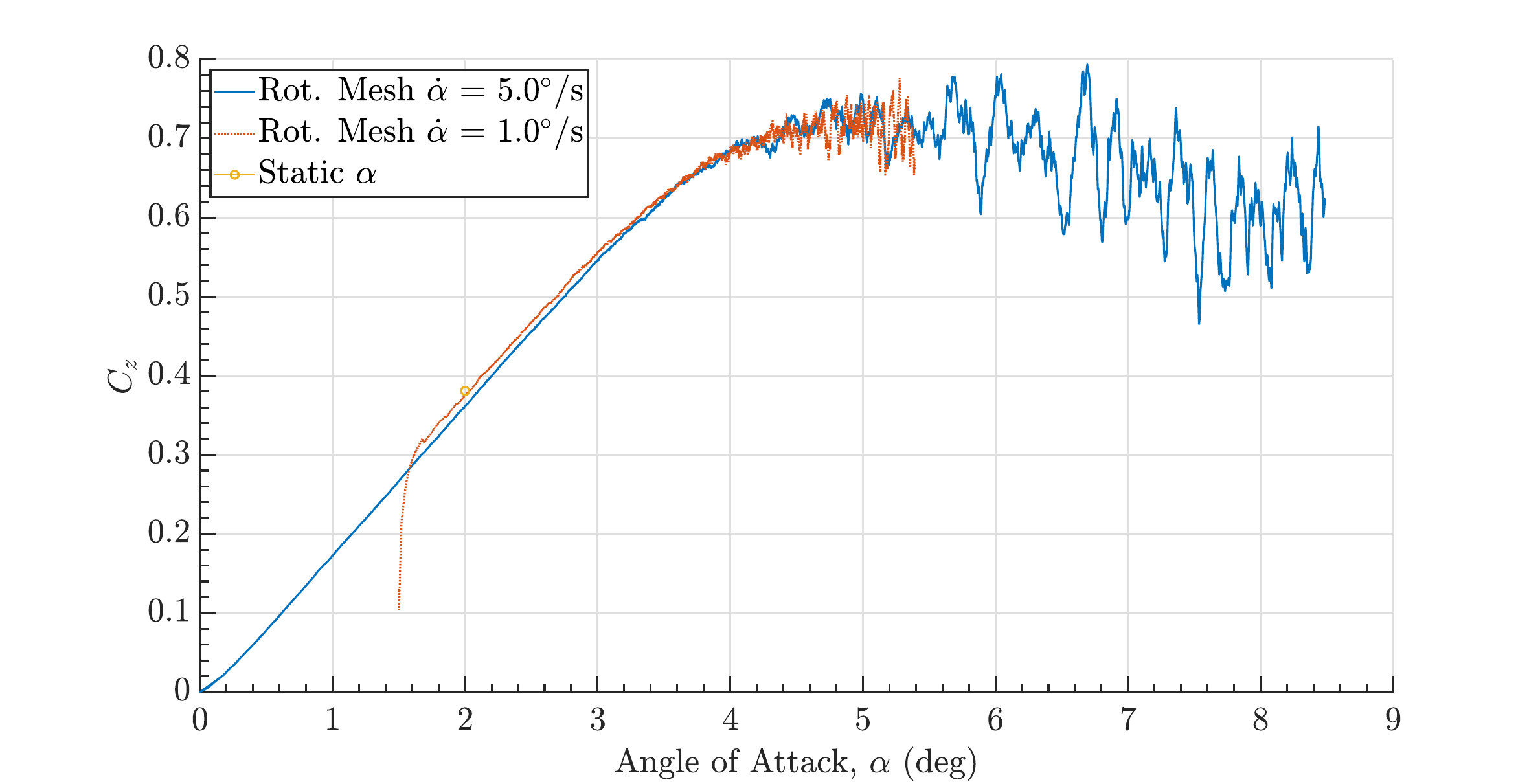}
\caption{Time history of the normal force component for the rotating airfoil cases used to identify the onset of buffet.  \label{fig:rotating_cz}}
\end{center}
\end{figure}

A rotating mesh version of the charLES solver was used to identify the onset of the initial buffet in the NACA 0012 flow under the same conditions. Rotating simulations were pursued in order to better replicate the experimental approach, which dynamically rotated the NACA 0012 airfoil at a range of Mach numbers to determine the critical alpha at which initial buffet in onset as a function of Mach number. Contours of the Mach number through a center plane of the flow are shown in Fig. \ref{fig:0012_rotating_mach} from a simulation that dynamically rotated between 1 and 5 degrees angle of attack. In this image, a shock wave and the associated shock-induced separation bubble are clearly visible as the region in which the Mach number abruptly drops. The flow is from left to right. For the rotating cases, the airfoil was started from a low angle of attack and rotated slowly through a range of alpha's of interest until significant force oscillations were observed, just as was done in the experiment. The time history of the normal force for these rotating simulations is shown in Fig. \ref{fig:rotating_cz}. Two rotation rates $1^{\circ}$ per second and $5^{\circ}$ per second were chosen to understand the sensitivity of the predictions to the rotation rate. In experiments, the rotation rates are typically significantly lower than in simulations (usually only $0.1^{\circ}$ per second), but this was not achievable presently due to the prohibitively long temporal integration window needed to match the low experimental rotation rates. Despite this, low sensitivity to the rotation rate was observed between the two rates that were considered, with normal forces computed using both rates agreeing well with one another and also with the static alpha case at $\alpha = 2.0^{\circ}$ (the case in Fig. \ref{fig:mcdevitt_static_alpha}). This result shows the rotation rates do not play a major role in governing the onset of initial buffet (over the range of rates studied) in this problem, and static angle of attack calculations can, therefore, be used to identify the onset of unsteadiness instead of rotating simulations, which significantly simplifies the computational task.

\begin{figure}[!ht]
\begin{center}
\includegraphics[width=0.6\textwidth]{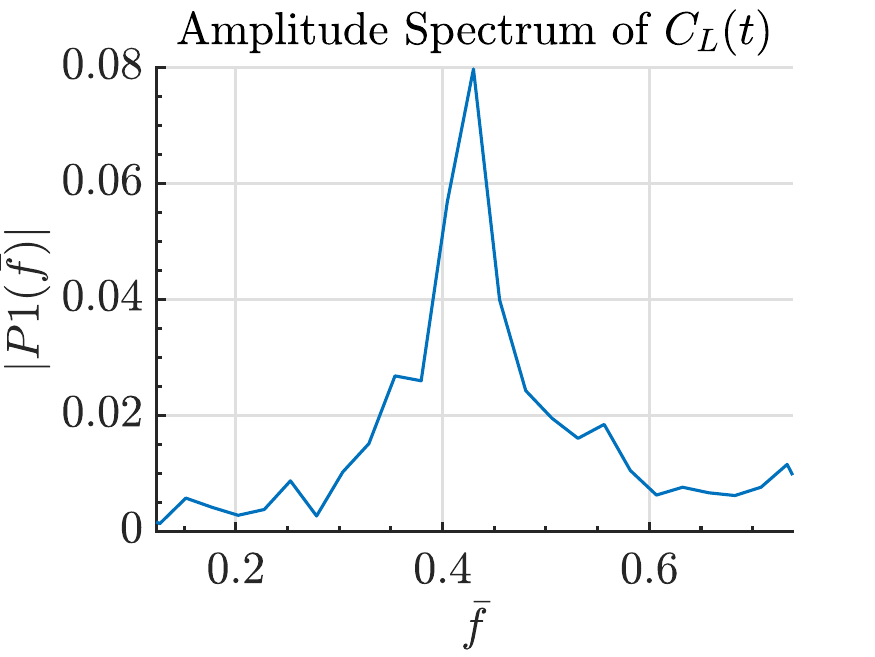}
\caption{Amplitude spectrum of the lift force oscillations at $\alpha = 5^{\circ}$ for the spanwise-periodic NACA 0012 simulations. Buffet is identified at the point where unsteady oscillations in the lift force are at least $10\%$ about the mean. The mean lift at this condition is $\approx 0.6$, and therefore, buffet is said to be achieved here. ${f}$ is a dimensional frequency used to construct the non-dimensional frequency, $\Bar{f}$ defined in Ref. \cite{mcdevitt1985static}. \label{fig:0012_amplitude_spectrum}}
\end{center}
\end{figure}

Significant force oscillations began to appear when the angle of attack varied between 4 and 5 degrees in the rotating mesh simulations at a Mach number of 0.75 (as shown in Fig. \ref{fig:rotating_cz}). Further investigations into these set points were carried out by performing fixed alpha simulations at $0.25^{\circ}$ intervals between $\alpha = 4-5^{\circ}$, and it was observed that $\alpha = 5^{\circ}$ was the minimum angle at which the fluctuations in lift varied by at least 10\%  about the mean (the experiments predicted the minimum angle for buffet as $\alpha = 4^{\circ}$). Figure \ref{fig:0012_amplitude_spectrum} shows the amplitude spectrum in the non-dimensional frequency domain associated with the time history of the lift force at this angle. It is observed that the oscillations are predominantly present at low frequencies ($f \approx 17.5$ Hz or $\Bar{f} \approx 0.43$). Table \ref{tab:0012_freq} shows the non-dimensional shock oscillation frequencies associated with buffet, with non-dimensionalization defined in Ref. \cite{mcdevitt1985static}. LES predicts shock oscillations that are delayed by $1^{\circ}$ and at a slightly lower frequency (about $10\%$) than in the experiment. The delay in the shock oscillation onset is consistent with the conclusions of Fig. \ref{fig:crm_forces_moment} in which subfigure (c) showed a slight delay in the pitching moment break that is associated with shock-induced separation and is a precursor to buffet onset. Overall, we observe that state-of-the-art LES methods do provide value in the prediction of buffet onset due to their time-accurate nature, but quantitative measures such as the alpha of initial buffet onset or the precise value of shock oscillation frequency once buffet is achieved are subject to some errors. Accurate prediction of shock location before the onset of buffet can be thought of as a necessary but insufficient condition for accurate prediction of transonic buffet.

\begin{table}[!ht]
\begin{center}
    \caption{Non-dimensional shock oscillation frequencies of the transonic 0012. Experimental frequencies are from Ref. \cite{mcdevitt1985static}. The LES uncertainty is a statistical measure of uncertainty associated with the bin width used in the FFT. The non-dimenstional frequency is defined as $\Bar{f} = 2\pi f c/U_{\infty}$ \citep{mcdevitt1985static}.}
    \begin{tabular}{cccl}
        \cline{1-3}
        \cline{1-3}
        \textbf{\begin{tabular}[c]{@{}c@{}}Case\end{tabular}} & \textbf{\begin{tabular}[c]{@{}c@{}}Alpha\end{tabular}} &  \textbf{$\bar{f}$} & \\ \cline{1-3}
        Experiment & 4  & 0.47 \\ 
        LES        & 5  & 0.43$\pm$0.025\\  \cline{1-3}\cline{1-3}
    \end{tabular}
    \label{tab:0012_freq}
\end{center}
\end{table}


\section{Conclusions} 

\label{sec:summary}

In this paper, the LES methodology with the dynamic Smagorinsky subgrid-scale model and the equilibrium wall model was used to study the flow over the transonic NASA CRM, a benchmark flow for drag prediction and CFD validation in compressible flow regimes. These calculations are among the first forays into the use of LES for such complex engineering applications in this flow regime. Sensitivities were explored around elements of the experimental setup, such as the use of an array of small cylindrical trip dots lining the leading edge of the wing at a constant spanline of $10\%$ and the inclusion of the sting mounting apparatus at the tail of the aircraft. It was also found that while HCP mesh solutions exhibited a non-monotonic grid convergence, the use of stranded meshes in the boundary layer region served to rectify this issue, even for identical cell sizes to their hexagonal close-packed counterparts. This highlighted the benefits of wall-aligned meshing approaches for LES calculations of wall-bounded turbulent flows. Only slight further improvements in solution accuracy, such as the suppression of a spurious leading edge laminar separation bubble observed on isotropic stranded meshes, were achieved by leveraging anisotropic stranded meshes (with a 4x refinement relative to their isotropic counterparts in the wall-normal direction to support fine near-wall turbulence length scales). The cost of representative fine mesh cases was assessed and it was concluded that LES of transonic external aerodynamic flows is ready for routine use in industry, especially when running on next-generation GPU machines, though the question of how best to promote laminar-to-turbulent transition remains. Some measures of the boundary layer were computed based on velocity profiles at selected stations along the wing, and non-dimensional parameters such as the momentum thickness-based Reynolds number and shape factors were reported in an appendix as an aid to turbulence modelers looking for representative canonical problems for aircraft flows at this Reynolds number. Finally, initial explorations into the predictive capabilities of LES with regards to the onset of initial buffet were made with promising preliminary results, though further investigation is needed regarding the requirements for prediction of the precise angle of attack and shock oscillation frequency at buffet onset.

\section*{Acknowledgments} 

S.T.B., R.A., and P.M. acknowledge funding from NASA Grant Number 80NSSC20M0201 and Boeing Research \& Technology. This research used resources from the Oak Ridge Leadership Computing Facility, which is a DOE Office of Science User Facility supported under Contract DE-AC05-00OR22725, and from another Director's Discretionary Allocation award program. The authors acknowledge helpful discussions with Adam Clark at Boeing and with Michael P. Whitmore at Stanford University. 

\appendix
\section{Sensitivity of Integrated Loads to Aircraft Span}
\label{sec:aaooendix_span}
\begin{figure}[!ht]
\centering
    \subfloat[\label{fig:fwd_trip}]{\includegraphics[width=0.488\textwidth]{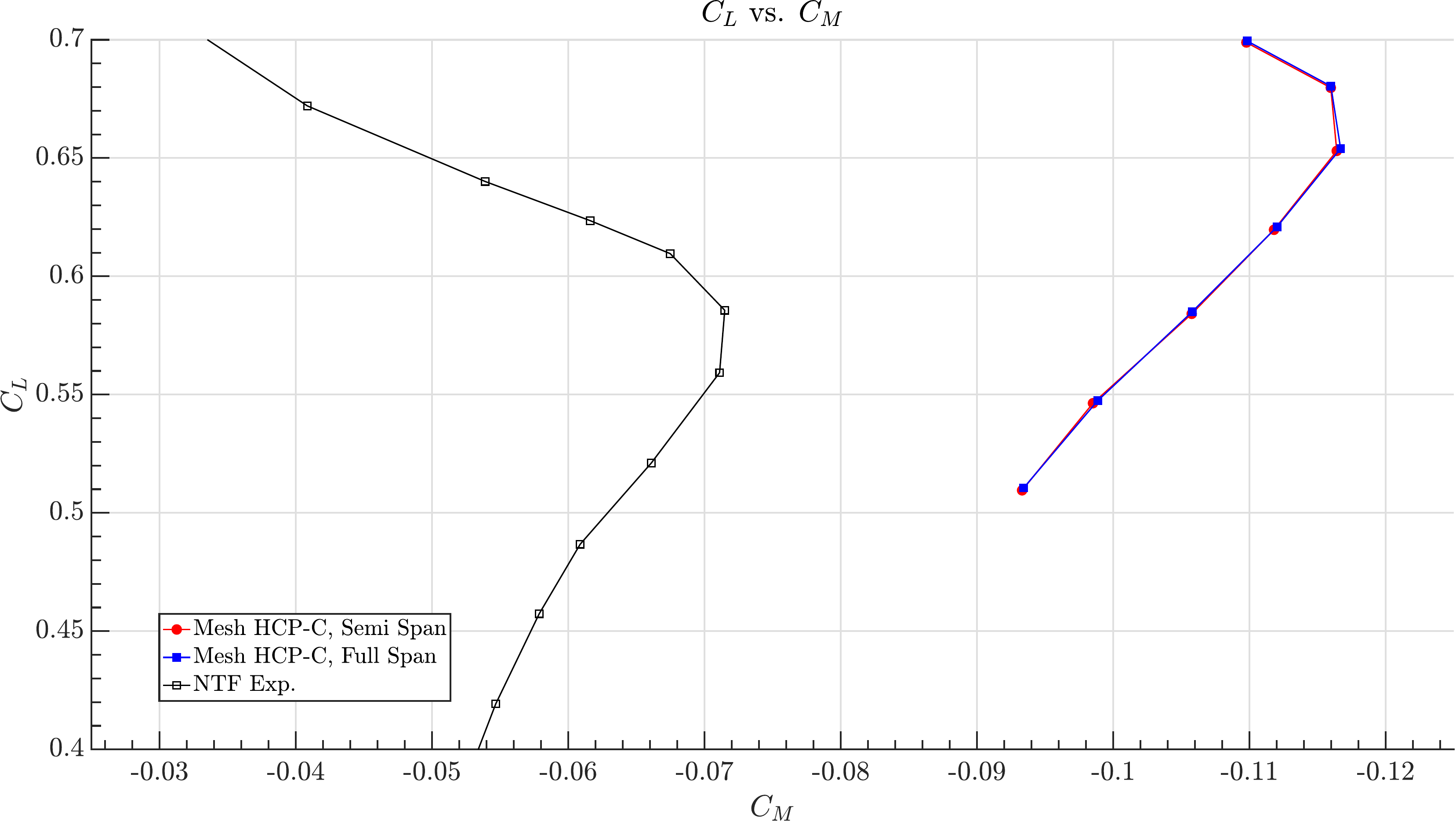}}
    \centering
    \subfloat[\label{fig:aft_trip}]{\includegraphics[width=0.495\textwidth]{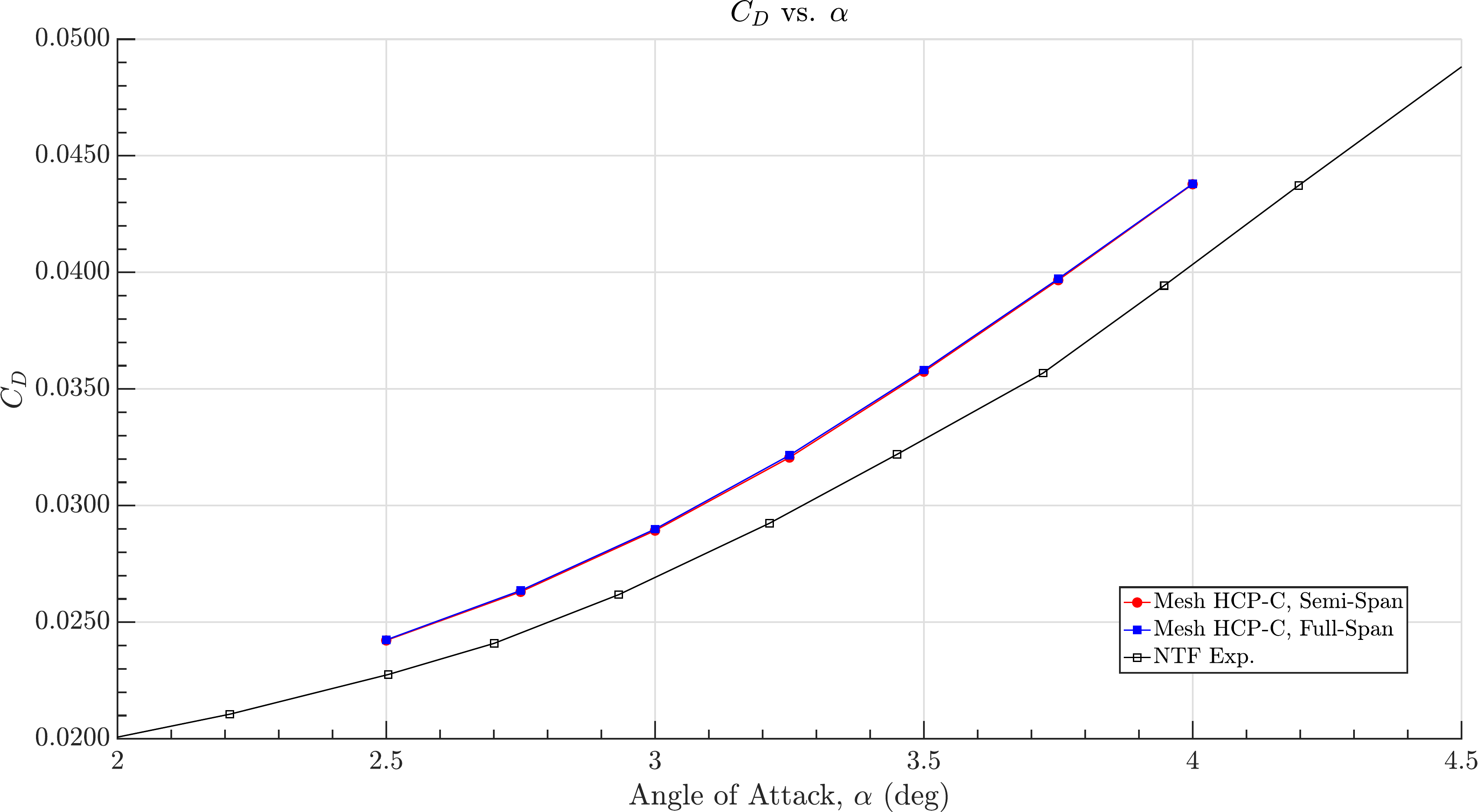}}
    \caption{Comparison of the predicted integrated loads between 
    the semi-span and the full-span transonic Common Research Model aircraft. Subfigure (a) presents the lift-moment polar, and Subfigure (b) presents the variation in total drag as a function of the angle of attack. 
    \label{fig:halffull}}
\end{figure}

In this appendix, we demonstrate that the present results are largely insensitive to the aircraft model being semi-span or full-span. Physically, the no shear stress and no penetration boundary conditions which are applied on the symmetry plane of a semi-span simulation are only strictly valid in the mean, not instantaneously. For a full-span model, the velocity fluctuations normal to the symmetry plane are finite instantaneously. This assumption was assessed by means of running a half span model with a symmetry plane on the center (x-y) plane of the simulation and compared against a full span simulation on the same grid, which was mirrored across the symmetry plane. Figure \ref{fig:halffull} presents a comparison of the integrated loads ($C_L, \, C_M$, and $C_D$) across the angle of attack sweep on the coarse grid (HCP-C). It is apparent that there is negligible sensitivity in these integrated loads to the choice of the span, and thus, we perform our studies on a semi-span model throughout this article.  

\section{Numerical Tripping Investigations}
\label{sec:appendix_a_tripping}

\begin{figure}[!ht]
\centering
    \subfloat[\label{fig:fwd_trip}]{\includegraphics[width=0.488\textwidth]{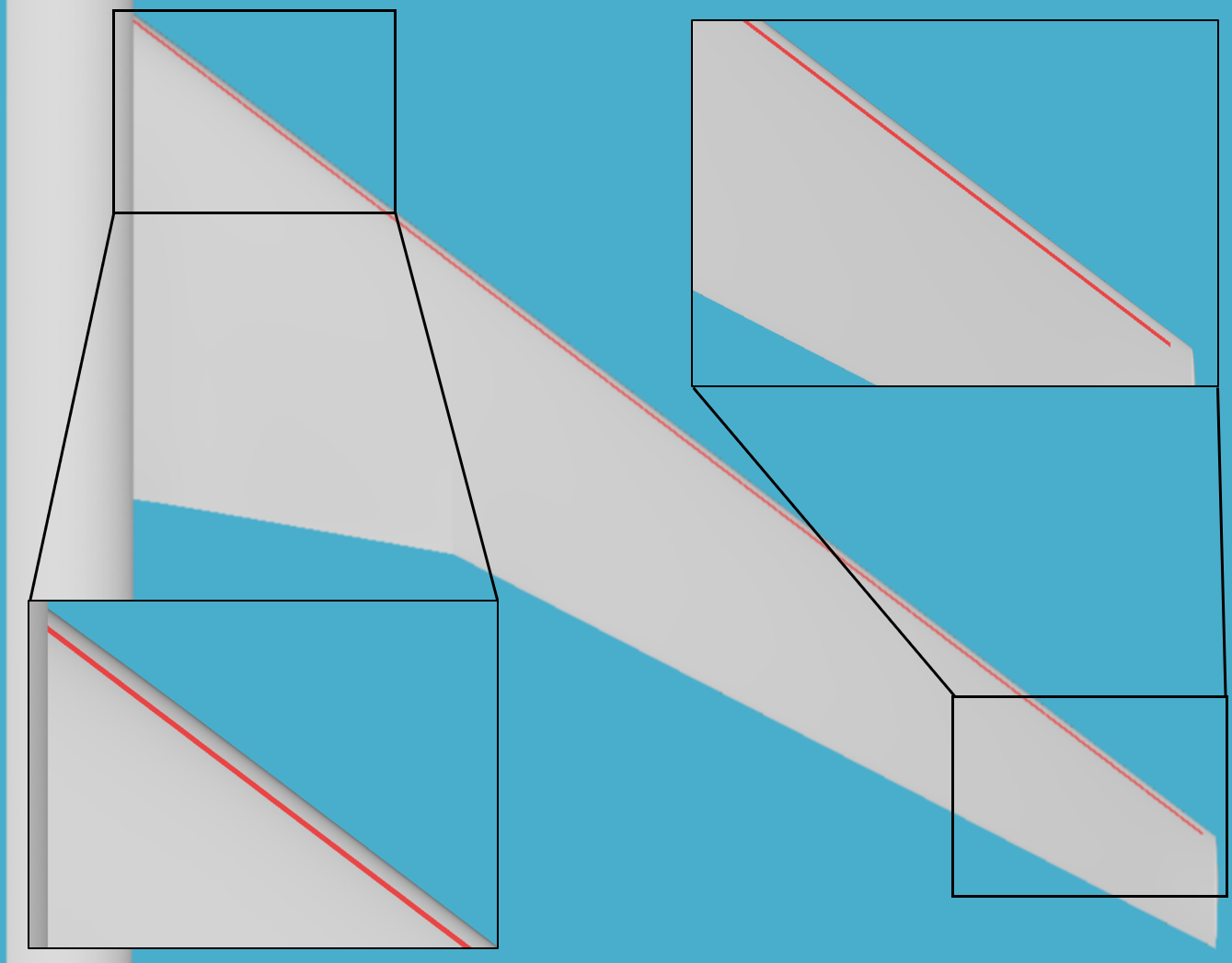}}
    \centering
    \subfloat[\label{fig:aft_trip}]{\includegraphics[width=0.495\textwidth]{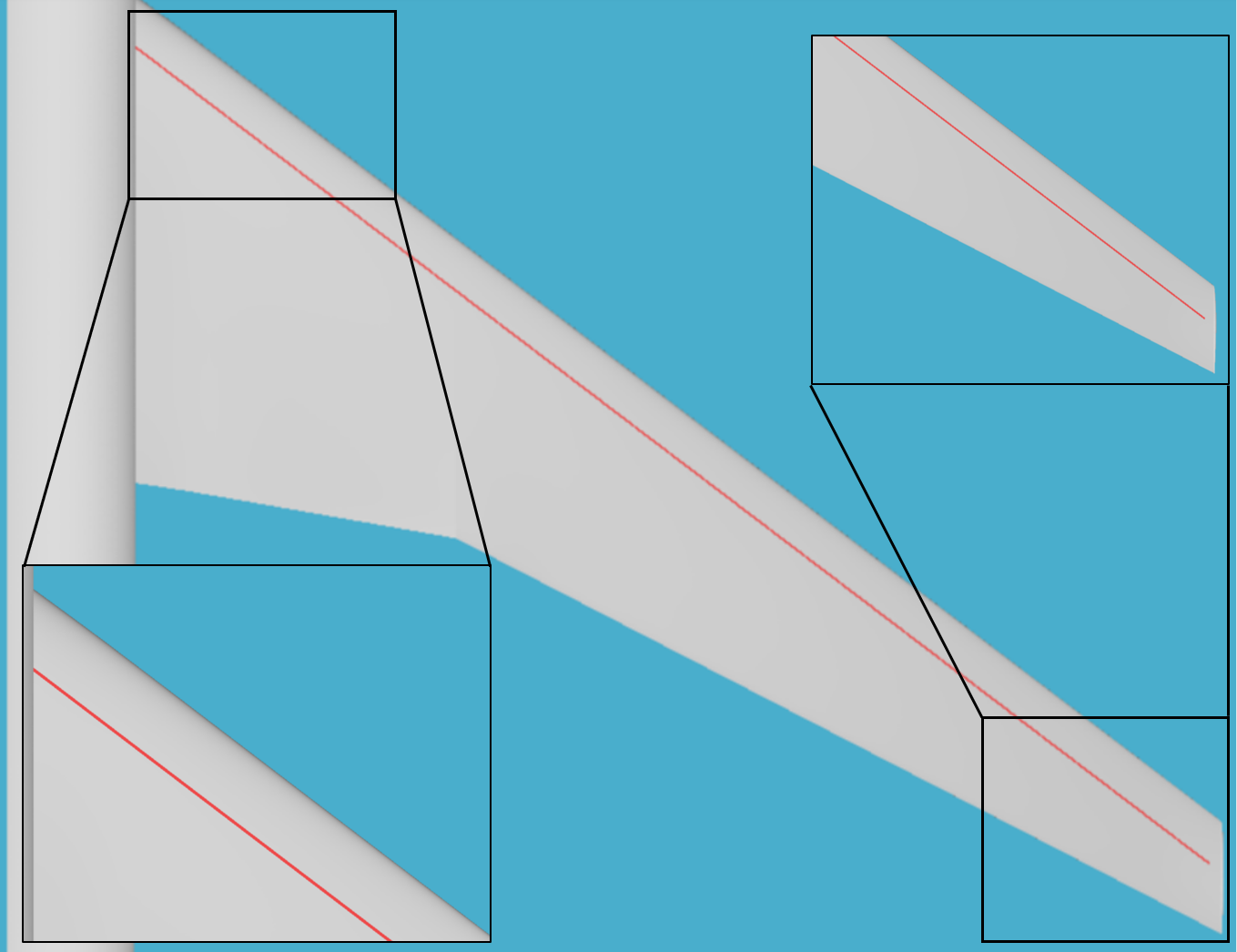}}
    \caption{Image of (a) the forward trip position and (b) the aft trip position. The inset graphics show zoomed views of the wing/juncture and wingtip regions, respectively.  \label{fig:crm_num_trip_geom}}
\end{figure}

The purpose of this appendix is to provide additional detail on the numerical tripping exercises discussed in Section \ref{sec:trip_dots_crm}. A simple sinusoidal (in time) numerical tripping approach was employed in which details of the leading edge boundary layer shown in Table \ref{tab:velprof} along with the local $u_{\tau}$ as measured from the local skin friction were used to construct a tripping amplitude ($A = u_{\tau}$) and frequency ($f = u_{\tau}/\delta_{99}$). A representative leading edge $\delta_{99}$ was chosen as $1 cm$, while a representative $u_{\tau}$ was found to be $17.5$ m/s. These parameters were applied as a sinusoidal blowing/suction boundary condition with 0 mean transpiration along a thin trip line whose chordwise extent was $0.5\% * MAC$ and extended from the wing root to the tip, as shown in Figure \ref{fig:crm_num_trip_geom}. The thickness of the numerical trip line approximately matches the trip dot diameter from \citep{rivers2010experimental,evans2020test}. In order to understand the sensitivity to the location at which the numerical tripping was applied, two locations were chosen, one that we refer to as ``forward'' and another as ``aft''. The forward location is at approx. $1\%$ x/c while the aft location is at approx. $10\%$ x/c at the wing root. The aft location more closely resembles the location of the geometric trip dots installed in the experiment. The trip line moves further aft in chord towards the wingtip because of the wing bending and because it was split on the geometry using a fixed plane that does not account for the aeroelastic deformation of the wing.  

\begin{figure}[!ht]
    \subfloat[\label{fig:crm_forces_moment_num_trip_a}]{\includegraphics[width=0.495\textwidth]{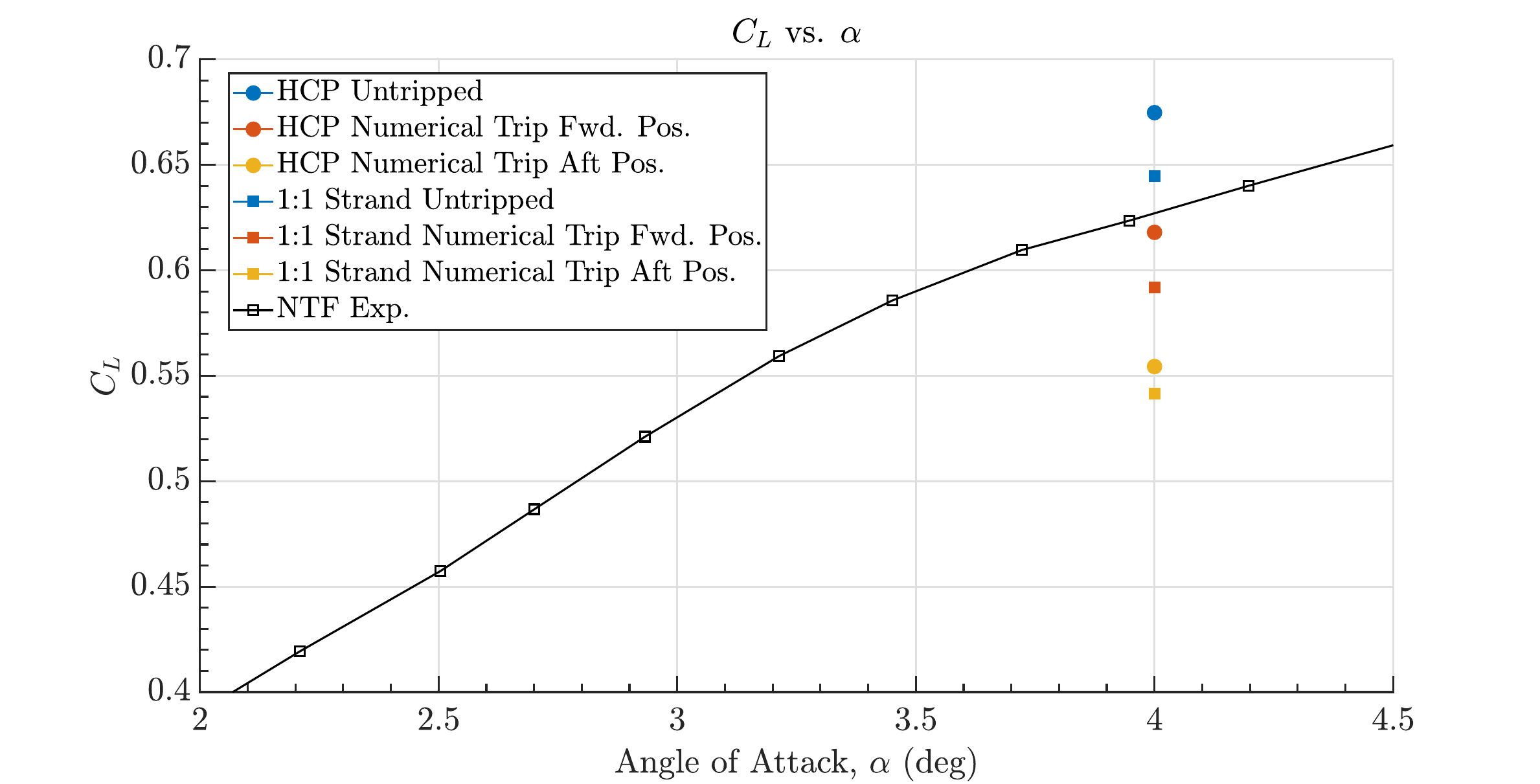}}
    \subfloat[\label{fig:crm_forces_moment_num_trip_b}]{\includegraphics[width=0.495\textwidth]{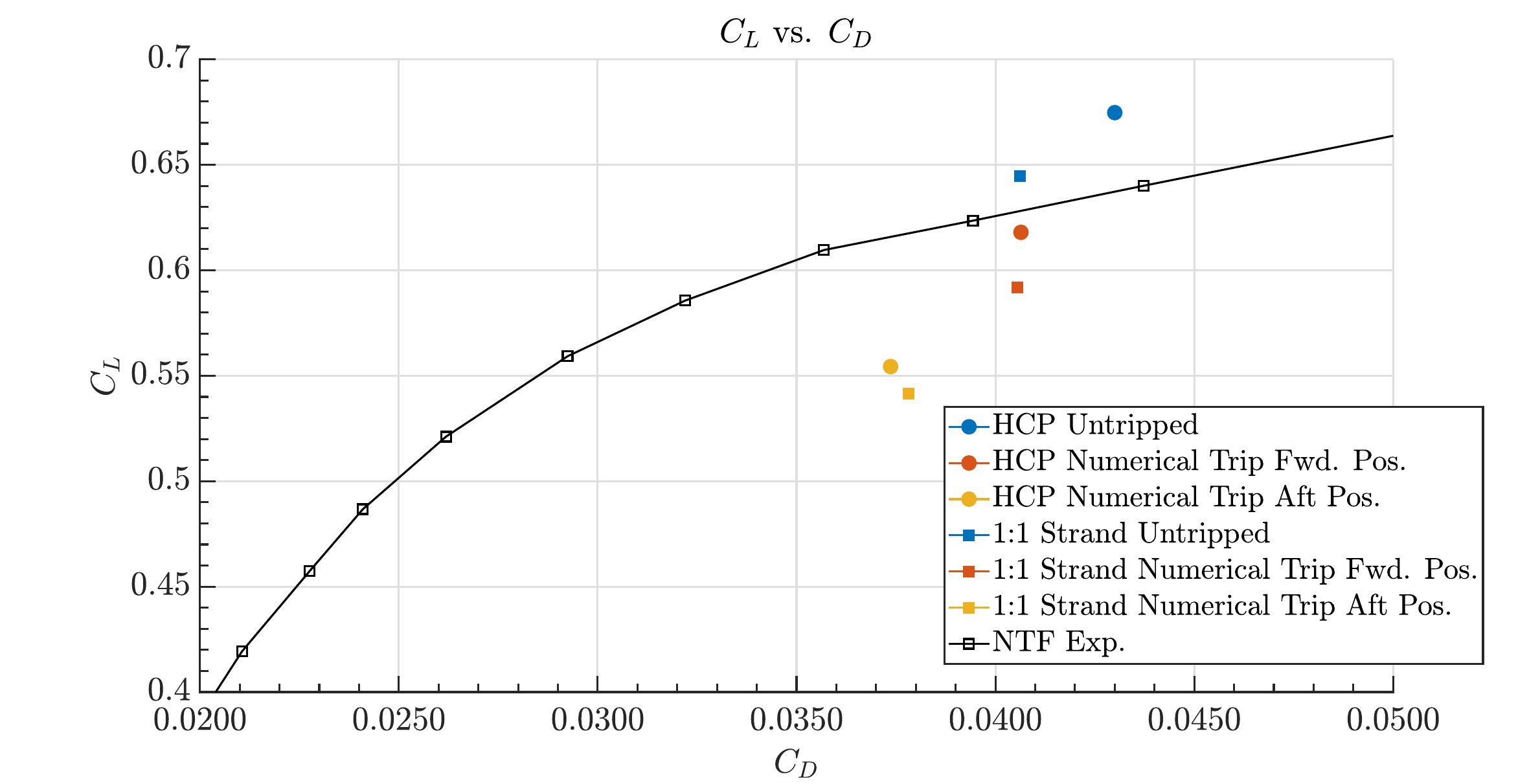}}\\ 
    \centering
    \subfloat[\label{fig:crm_forces_moment_num_trip_c}]{\includegraphics[width=0.495\textwidth]{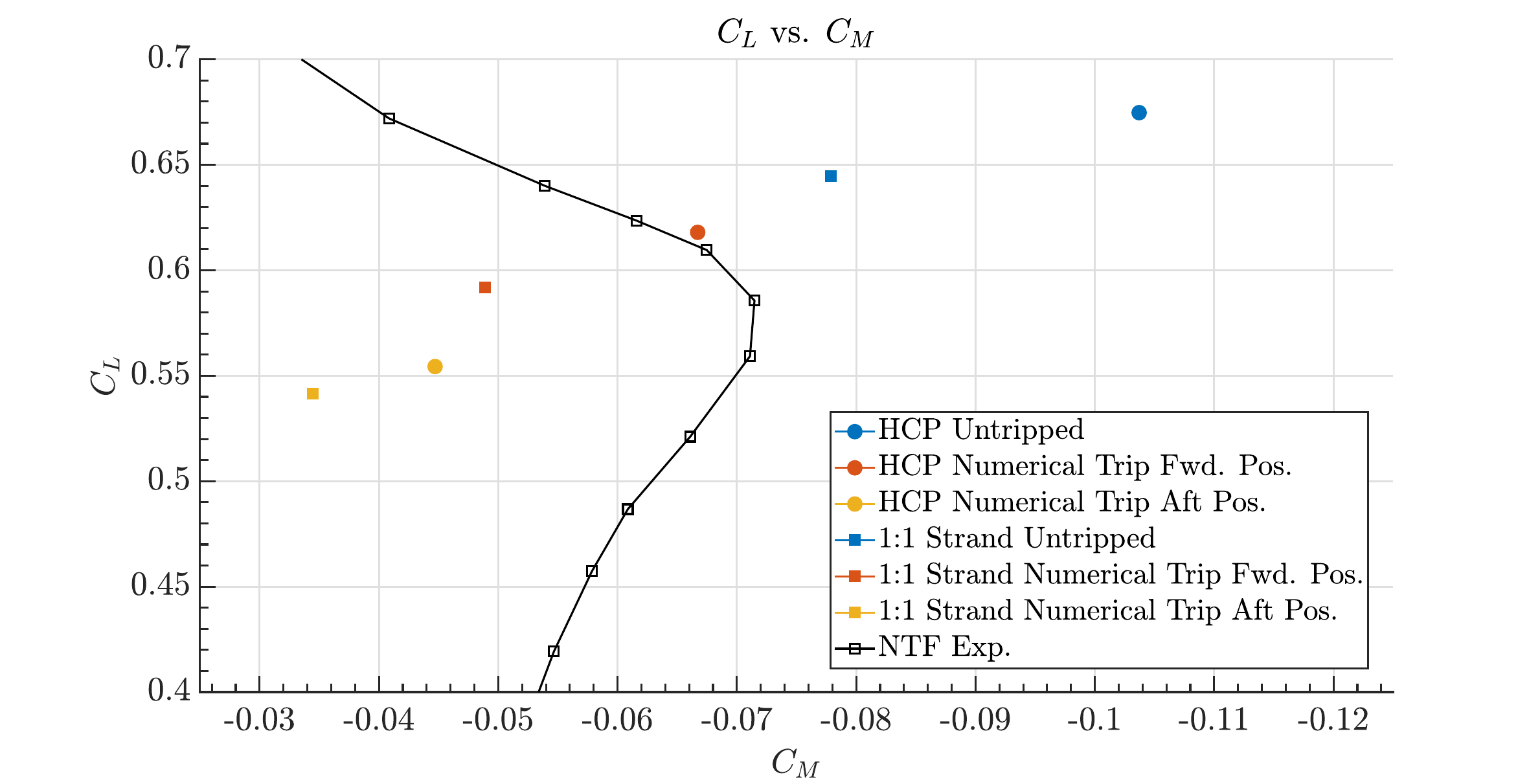}}
    \caption{Time-averaged forces/moments from the finest HCP-F and isotropic strand S-IF mesh LES simulations of the transonic CRM, including (a) the lift, (b) the drag polar, and (c) the pitching moment. The comparison includes untripped calculations as well as cases that were tripped at both the forward and aft trip positions shown in Figure \ref{fig:crm_num_trip_geom}. \label{fig:crm_forces_moment_num_trip}}
\end{figure}

\begin{figure}[!ht]
    \subfloat[\label{fig:crm_cp_num_trip_strand_vs_hcp_a}]{\includegraphics[width=0.495\textwidth]{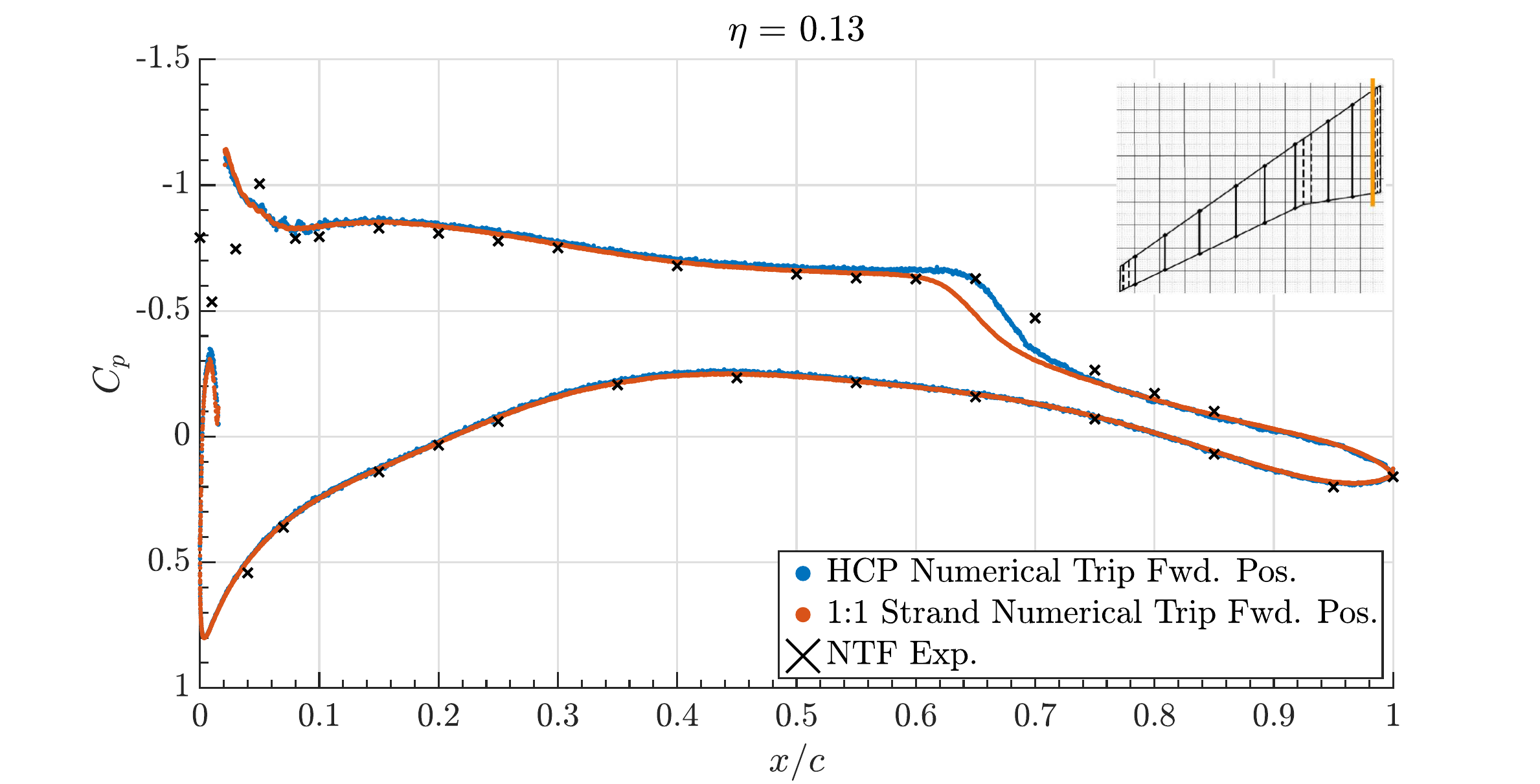}}
    \subfloat[\label{fig:crm_cp_num_trip_strand_vs_hcp_b}]{\includegraphics[width=0.495\textwidth]{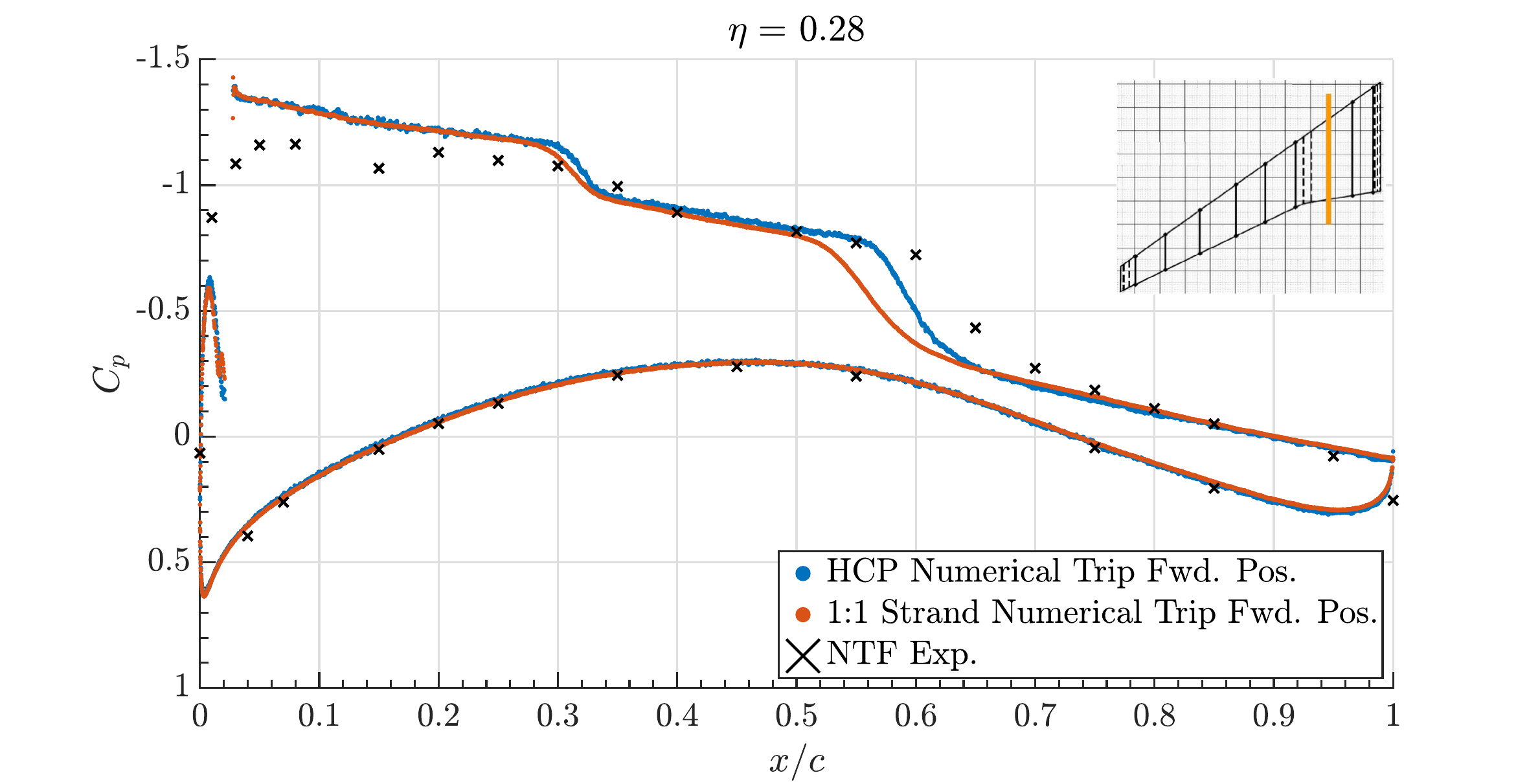}}\\ 
    \centering
    \subfloat[\label{fig:crm_cp_num_trip_strand_vs_hcp_c}]{\includegraphics[width=0.495\textwidth]{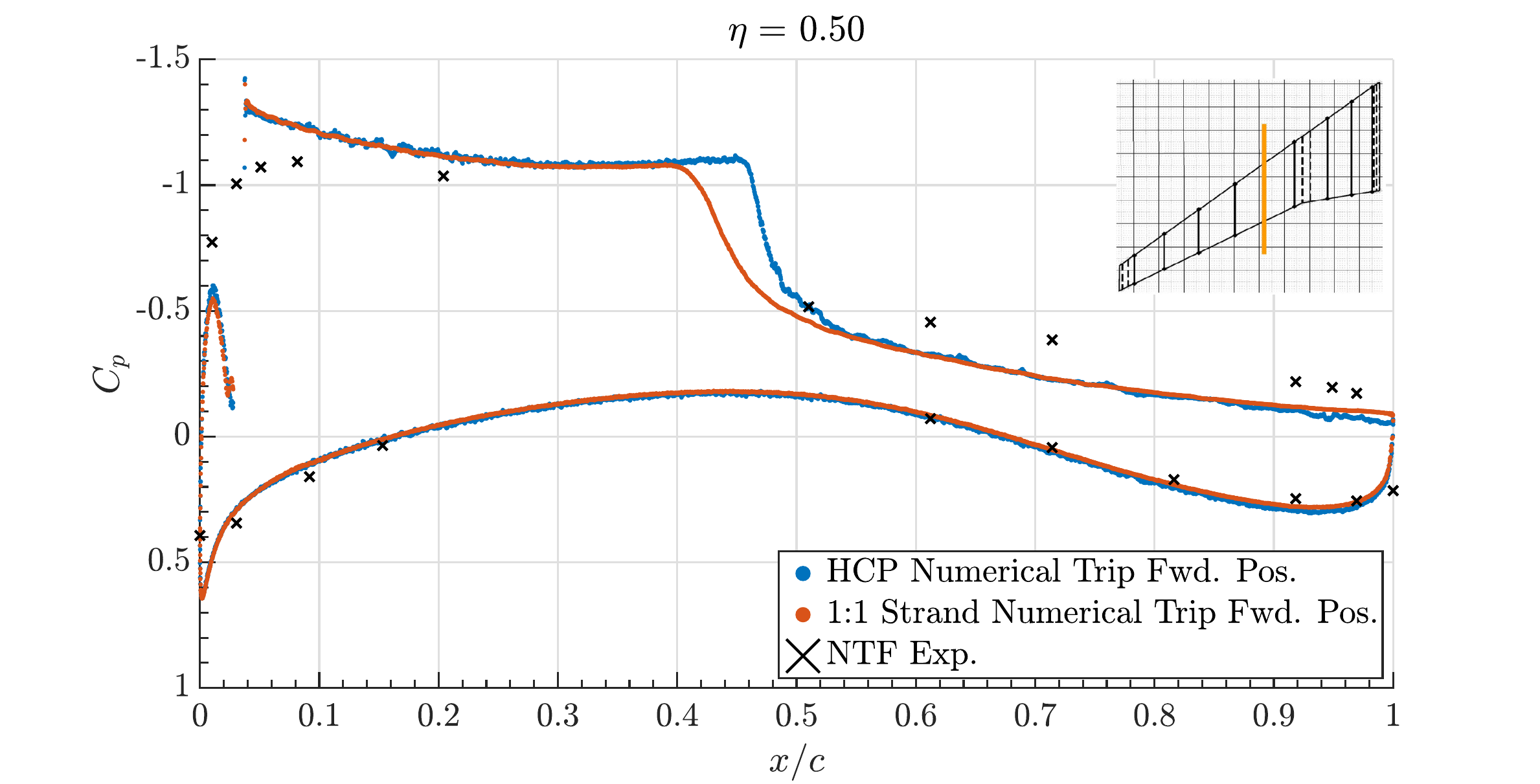}}
    \subfloat[\label{fig:crm_cp_num_trip_strand_vs_hcp_d}]{\includegraphics[width=0.495\textwidth]{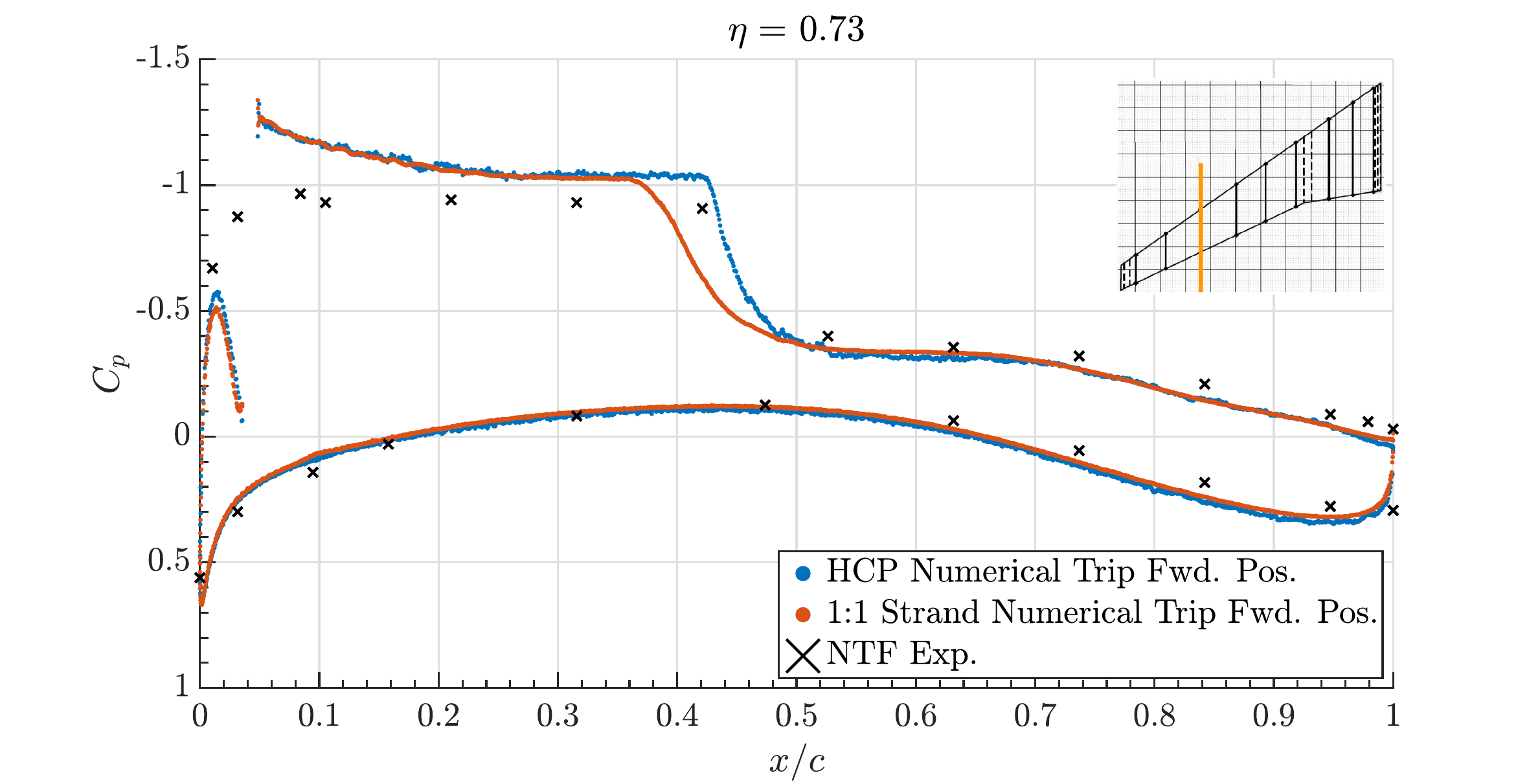}}
    \caption{Average static pressure measurements for the transonic CRM at an angle of attack of $\alpha = 4.0^{\circ}$ at four different stations along the span of the wing, ranging from (a-b) inboard to (c) mid-span to (d) outboard. The comparison includes results computed on HCP-F and S-IF meshes that were numerically tripped at the forward trip position. \label{fig:crm_cp_num_trip_strand_vs_hcp}}
\end{figure}

The numerical tripping experiments were conducted on the HCP-F and S-IF meshes in order to understand whether anchoring transition between the two mesh topologies could help to isolate the reason for the difference in accuracy observed in the two cases that were reported in Section \ref{sec:topology}, even when nominally the same near-wall resolution was employed in the two cases. The force/moment data in Figure \ref{fig:crm_forces_moment_num_trip} and the pressure data in Figure \ref{fig:crm_cp_num_trip_strand_vs_hcp} show that anchoring the transition between the grid topologies is not enough to explain the differences observed in the accuracy with which the quantities of interest are predicted for this case, as two calculations with transition anchored at the forward numerical trip location can still exhibit differences in shock location of up to $5\%$, which translates to substantial changes in lift and pitching moment. Indeed, we see that even for a relatively modest forcing amplitude ($17.5$ m/s is approx. $5\%$ of $U_{\infty}$), the changes in QOI's can be extreme, highlighting the importance of, and sensitivity to, laminar-to-turbulent transition in this flow regime. Many flow/transition control techniques exist in the literature \citep{lehmkuhl2020active}, and our recommendation is that these be further explored, especially for transonic cruise aircraft configurations simulated with LES, though further investigations (such as forcing amplitude/frequency parameter sweeps) were out of the scope of this study.

\section{Boundary Layer Details}
\label{sec:bl_details}

\begin{figure}[!ht]
    \subfloat[\label{fig:eta_pt2_slice}]{\includegraphics[width=0.495\textwidth]{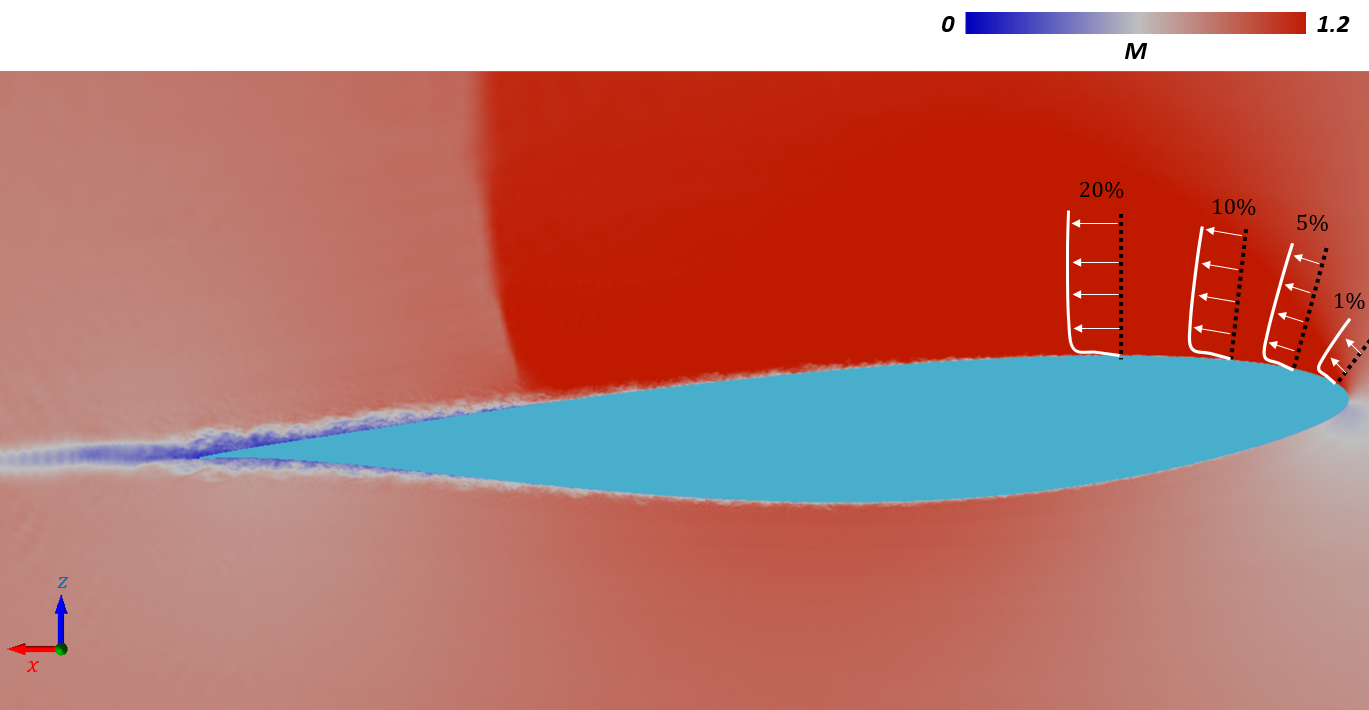}}
    \subfloat[\label{fig:eta_pt5_slice}]{\includegraphics[width=0.495\textwidth]{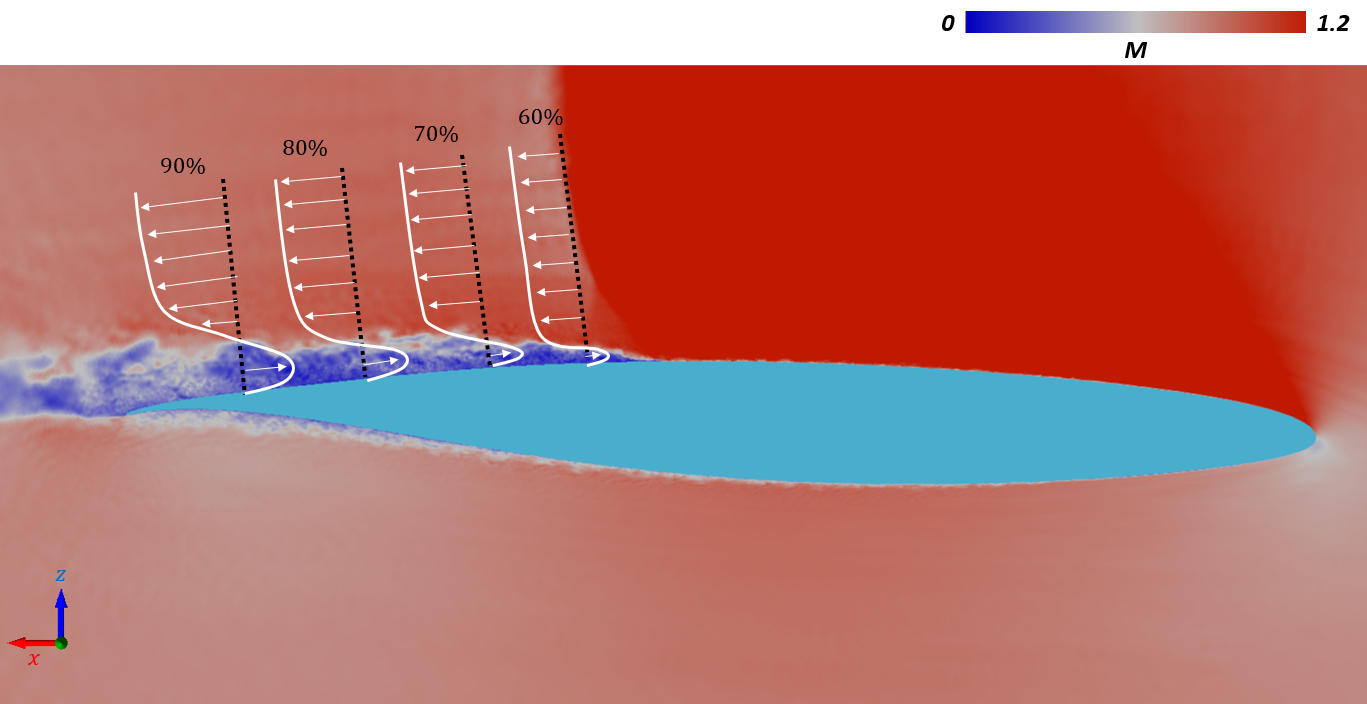}}\\ 
    \centering
    \subfloat[\label{fig:cf_bl_slice}]{\includegraphics[width=0.495\textwidth]{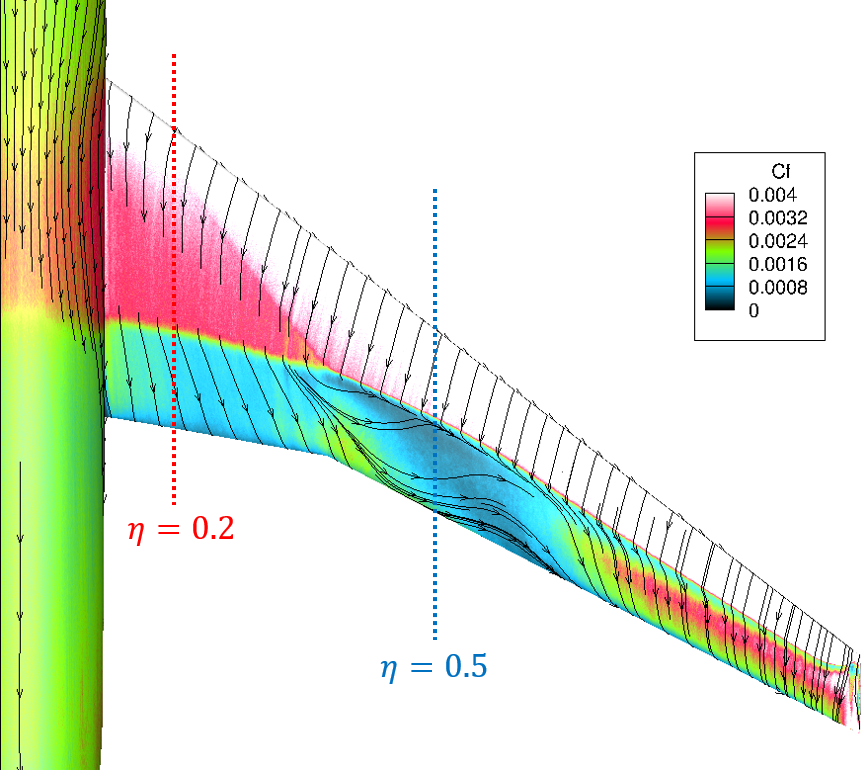}}

    \caption{Instantaneous slices of Mach number at semispan fraction (a) $\eta=0.2$ and (b) $\eta=0.5$ showing the location at which the wall-normal boundary layer probes shown in Figure \ref{fig:crm_velprof} are extracted and at which the boundary layer statistics are computed in Table \ref{tab:velprof}. Subfigure (c) shows the skin friction streamlines from the simulation on Mesh S-AM from which these boundary layer data were obtained. \label{fig:bl_slicing_locations}}
\end{figure}

\begin{figure}[!ht]
\begin{center}
\includegraphics[width=1\textwidth]{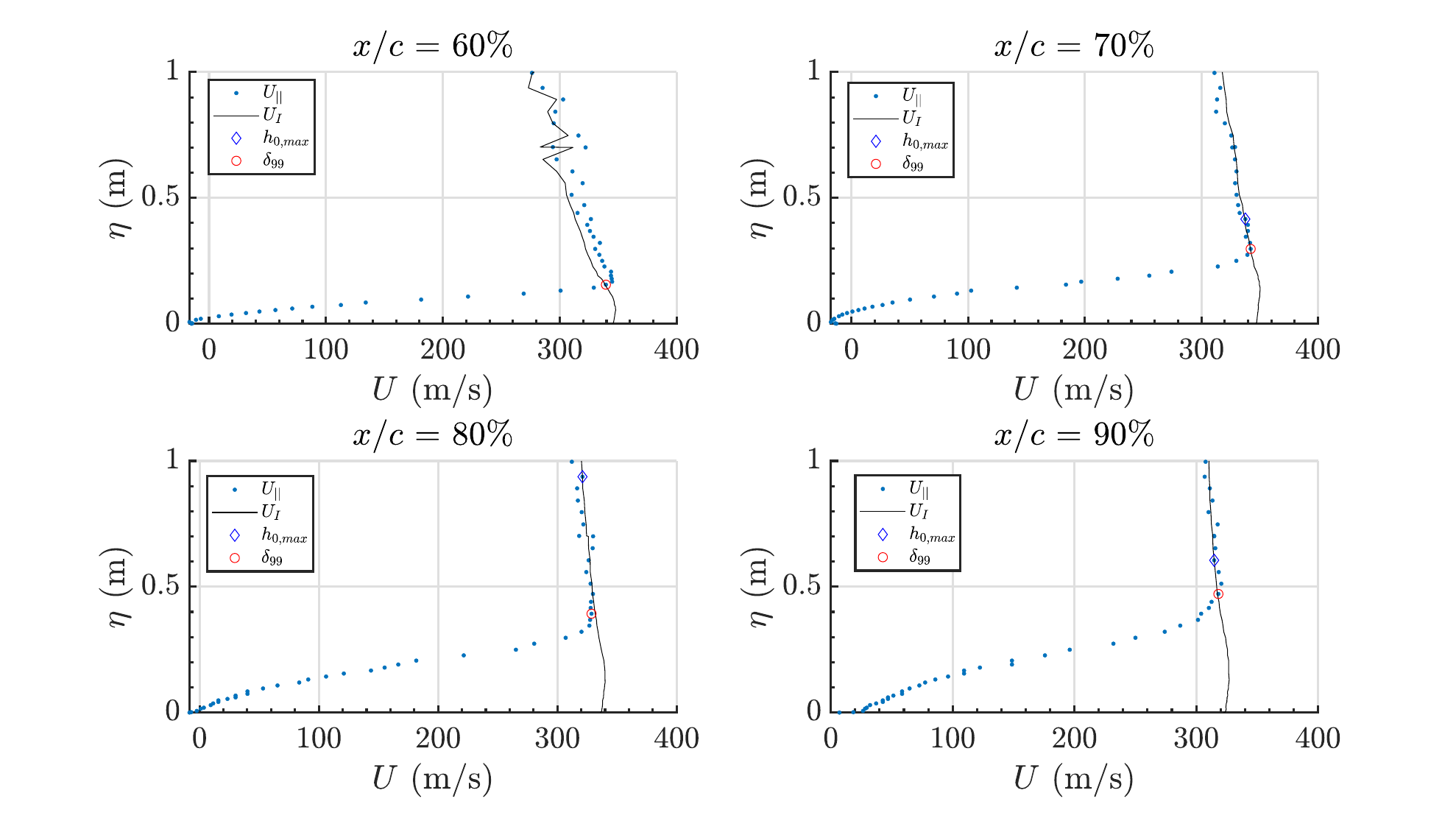}
\caption{Wall-parallel velocity profiles at mid-span ($50\%$ of the semi-span) within the separation bubble. The boundary-layer edge is determined using the compressible method of \cite{griffin2021general} for flows in non-equilibrium. \label{fig:crm_velprof}}
\end{center}
\end{figure}

Although the engineering quantities of interest for the flow over an aircraft are integrated forces and moments, it is important to quantify the ranges of Reynolds numbers, boundary layer thicknesses, and shape factors that characterize the flow over the wing. This is especially beneficial for future LES modeling efforts, in selecting canonical flows that may exhibit some of the flow features observed on the aircraft. For this reason, wall-normal velocity profiles from an anisotropic strand calculation (Mesh S-AM) at an angle of attack, $4^{\circ}$, are examined. Note that these simulations were performed at full geometric scale, but that the Reynolds number was matched to the experiments conducted in the NTF wind tunnel (an order of magnitude smaller than the true flight Reynolds numbers), so the data present a somewhat artificial test case of a full aircraft flying at wind tunnel Reynolds numbers. Still, detailed boundary layer data are not readily available even for aircraft in the wind tunnel flow regime. The velocity stations for this analysis were sampled at the leading edge of an inboard wing station and within the separation bubble at mid-span. Different spanwise locations were selected such that the development of the leading edge boundary layer ($\eta = 0.2$) could be understood independently of the influence from a large downstream separation bubble ($\eta = 0.5$). The wall-parallel velocity profiles ($U_{||}$) are shown in Fig. \ref{fig:crm_velprof}. The maximum stagnation enthalpy is calculated to establish the far-field conditions, and the boundary layer edge is determined following the method of Griffin, Fu, and Moin \cite{griffin2021general}. Table \ref{tab:velprof} summarizes the integrated boundary layer metrics such as the momentum Reynolds number and shape factor from the sampled velocity profiles in Fig. \ref{fig:crm_velprof}. It is apparent from this data that the flow over an aircraft wing experiences a wide range of Reynolds numbers based on momentum thickness ($Re_{\theta}$) and shape factors $(H)$, and suitable canonical cases must be chosen that represent some of these non-dimensional boundary layer measures. For instance, different canonical cases may be chosen to represent the thin leading-edge flow or the thick separated flow near the trailing edge. Finally, even for the finest grids considered in this work, the thin leading-edge boundary layers are marginally resolved with at most 5 grid points within the boundary layer. This points to another important necessity of future LES modeling efforts: the need to predict quantities of interest with $\mathcal{O}(5)$ points within the leading-edge boundary layer. Some initial progress has been made using sensor-aided augmentation to the equilibrium wall model \cite{agrawal2022sensor} or using machine-learning-based methods \cite{ling2022wall} in this area. However, more investigations on flows over complex geometries involving compressibility effects, such as those encountered on the transonic CRM, are required to fully assess the predictive capabilities of these advancements.

\begin{table}[!ht]
\begin{center}
    \caption{Boundary-layer characteristics for the transonic CRM flow both near the leading edge (spanwise location 0.2) and within the separation bubble (spanwise location 0.5). A wide range of Reynolds numbers and shape factors are observed as the boundary layer thickens and experiences shock-induced separation.}
    \begin{tabular}{ccccccc}
        \textbf{\begin{tabular}[c]{@{}c@{}}Spanwise\\Location\end{tabular}} & \textbf{\begin{tabular}[c]{@{}c@{}}Chordwise\\Location (\%)\end{tabular}} &  \textbf{$\delta_{99}$ (cm)} & \textbf{$\delta^{*}$ (cm)} & 
        \textbf{\begin{tabular}[c]{@{}c@{}}Shape\\Factor (H)\end{tabular}} &  
        \textbf{$Re_{\theta}$} & 
        \textbf{$\frac{\delta}{\Delta}$}
        \\ \cline{1-7} 
        0.2 & 1  & 0.76 & 0.12 & 1.86 & 472 & 2 \\ 
        0.2 & 5  & 1.24 & 0.30 & 2.50 & 657 & 3 \\ 
        0.2 & 10 & 1.25 & 0.46 & 2.44 & 729 & 4 \\ 
        0.2 & 20 & 2.62 & 0.75 & 2.34 & 1,505 & 5 \\ 
        0.5 & 60 & 15.5 & 9.7  & 6.33 & 10,558 & 20 \\ 
        0.5 & 70 & 29.7 & 16.6 & 6.67 & 18,006 & 28 \\ 
        0.5 & 80 & 39.3 & 19.6 & 4.88 & 28,648 & 29 \\ 
        0.5 & 90 & 47.1 & 21.9 & 3.80 & 41,382 & 30 \\
    \end{tabular}
    \label{tab:velprof}
\end{center}
\end{table}

\end{document}